\documentclass[aps,prd,superscriptaddress,notitlepage,showpacs,showkeys,10pt]{revtex4-1}

\usepackage{graphicx}
\usepackage{amsmath}
\usepackage{amssymb}
\usepackage{color}
\usepackage{bm}

\newcommand{\sign}[1]{\mbox{sign}\left({#1}\right)}

\definecolor{purple}{rgb}{0.8,0,0.6}
\definecolor{darkgreen}{rgb}{0.00,0.6,0.00}

\begin{document}
\title{Chiral magnetic plasmons in anomalous relativistic matter}
\date{\today}

\author{E.~V.~Gorbar}
\affiliation{Department of Physics, Taras Shevchenko National Kiev University, Kiev, 03680, Ukraine}
\affiliation{Bogolyubov Institute for Theoretical Physics, Kiev, 03680, Ukraine}

\author{V.~A.~Miransky}
\affiliation{Department of Applied Mathematics, Western University, London, Ontario N6A 5B7, Canada}
\affiliation{Department of Physics and Astronomy, Western University, London, Ontario N6A 3K7, Canada}

\author{I. A.~Shovkovy}
\affiliation{College of Integrative Sciences and Arts, Arizona State University, Mesa, Arizona 85212, USA}
\affiliation{Department of Physics, Arizona State University, Tempe, Arizona 85287, USA}

\author{P.~O.~Sukhachov}
\affiliation{Department of Applied Mathematics, Western University, London, Ontario N6A 5B7, Canada}

\begin{abstract}
The chiral plasmon modes of relativistic matter in background magnetic and strain-induced
pseudomagnetic fields are studied in detail using the consistent chiral kinetic theory. The results reveal
a number of anomalous features of these chiral magnetic and pseudomagnetic plasmons that could be used to identify them in experiment.
In a system with nonzero electric (chiral) chemical potential, the background magnetic (pseudomagnetic)
fields not only modify the values of the plasmon frequencies in the long-wavelength limit, but also affect
the qualitative dependence on the wave vector. Similar modifications can be also induced by the
chiral shift parameter in Weyl materials. Interestingly, even in the absence of the chiral shift and external
fields, the chiral chemical potential alone leads to a splitting of plasmon energies at linear order in the
wave vector.
\end{abstract}



\maketitle

\section{Introduction}
\label{sec:introduction}

The study of macroscopic implications of quantum anomalies attracted a lot of attention in
recent years. Previously, these phenomena were usually associated with high-energy and
astroparticle physics, e.g., in the context of relativistic heavy-ion collisions \cite{Kharzeev:2008-Nucl,Kharzeev:2016},
degenerate states of dense matter in compact stars \cite{Kouveliotou:1999}, and plasmas
in the early Universe \cite{Kronberg, Durrer}. However, the situation drastically changed
with the rise of Dirac and Weyl materials, whose low-energy quasiparticle excitations have
a relativistic-like spectrum. The studies of the basic properties of these novel materials attracted
a significant attention of a broad scientific community, strengthening the cross-disciplinary links between
such different fields as condensed matter and high energy physics. One of the examples of the
anomalous properties widely debated in both fields is the celebrated chiral magnetic effect (CME) \cite{Kharzeev:2008},
which has been observed indirectly in Dirac and Weyl semimetals \cite{Kim:2013dia,Li-Kharzeev:2016,Xiong:2015nna,Huang:2015eia,Arnold:2016},
as well as in heavy-ion collisions (for a review, see Ref.~\cite{Kharzeev:2016}). Note,
however, that the interpretation of the heavy-ion experiments in terms of the CME is still
ambiguous \cite{CMS:2016,Belmont:2016oqp}. The realization of a three-dimensional
(3D) Dirac semimetal phase in A$_3$Bi ($\mathrm{A=Na,K,Rb}$) and Cd$_3$As$_2$ compounds
was predicted \cite{Weng,Wang} and confirmed experimentally \cite{Borisenko,Neupane,Liu} only
a few years ago. The Weyl semimetal phase was first predicted theoretically to be realized in pyrochlore
iridates \cite{Savrasov}. It was discovered later in such compounds as $\mathrm{TaAs}$, $\mathrm{TaP}$,
$\mathrm{NbAs}$, $\mathrm{NbP}$, $\mathrm{Mo_xW_{1-x}Te}$, and $\mathrm{YbMnBi_2}$ \cite{Tong,Bian,Qian,Huang:2015eia,Belopolski,Cava}.

It is important to note that Dirac and Weyl materials not only mimic the properties of high-energy
relativistic matter, but provide an opportunity for studying novel quantum effects. For example, as shown in
Refs.~\cite{Zhou:2012ix,Zubkov:2015,Cortijo:2016yph,Cortijo:2016,Grushin-Vishwanath:2016,Pikulin:2016},
static mechanical strains applied to Weyl semimetals generate pseudomagnetic fields. As in
graphene, the corresponding effective gauge fields capture the corrections to the kinetic energy of
quasiparticles caused by unequal modifications of hopping parameters in a strained crystal.
Unlike an ordinary magnetic field $\mathbf{B}$, a pseudomagnetic
field $\mathbf{B}_5$ is felt by charge carriers of opposite chiralities as if they had opposite charges.
By recalling that, according to the Nielsen--Ninomiya theorem \cite{Nielsen-Ninomiya}, Weyl nodes always come
in pairs of opposite chiralities, we conclude that the pseudomagnetic field does not break the time-reversal
symmetry in Weyl materials by itself. (Unless the material is a $\mathbb{Z}_2$ Weyl semimetal \cite{Gorbar:2014sja},
the time reversal symmetry is broken by the chiral shift parameter, which defines the separation between
the Weyl nodes in the momentum space.) Of course, this is expected in the case of static deformations.
The characteristic strengths of the pseudomagnetic field in Dirac and Weyl materials are much smaller
than in graphene and range from about $B_5\approx0.3~\mbox{T}$, when a static torsion is applied to
a nanowire of Cd$_3$As$_2$ \cite{Pikulin:2016}, to approximately $B_5\approx15~\mbox{T}$, when a
thin film of Cd$_3$As$_2$ is bent \cite{Liu-Pikulin:2016}.

The transport properties of Dirac and Weyl semimetals under strain are studied in Refs.~\cite{Zhou:2012ix,Cortijo:2016,Pikulin:2016,Grushin-Vishwanath:2016}.
According to Ref.~\cite{Cortijo:2016}, the application of a strain in Weyl semimetals shifts the locations
 of the Weyl nodes and leads to an experimentally measurable CME current decaying exponentially
with time. It was demonstrated in Ref.~\cite{Pikulin:2016}
that pseudoelectromagnetic fields give rise to new and unusual manifestations of the chiral anomaly \cite{ABJ},
which can be observed by such conventional experimental probes as the electrical transport, ultrasonic attenuation,
and electromagnetic field emission. By using a semiclassical approach, it was shown in Ref.~\cite{Grushin-Vishwanath:2016}
that the pseudomagnetic field $\mathbf{B}_5$ contributes to the conductivity as $\sigma\sim B_5^2$. Moreover, it
was argued that topologically protected Fermi arcs are ``secretly" zero pseudo-Landau levels due to $\mathbf{B}_5$,
which is present at the boundary of the system even when a strain is absent.

Traditionally, one of the most powerful methods of mapping out the material's Fermi-surface relies on the
quantum oscillations of the density of states (DOS) with the period proportional to $1/B$. The relativistic-like nature
of the charge carriers in Weyl materials manifests itself in the phase shift and the quadratic dependence of
the DOS on the quasiparticle energy \cite{Ashby:2013}. Another distinctive feature of the Weyl materials is the unusual
type of quantum oscillations that involve both bulk and surface (Fermi arcs) states \cite{Potter:2014,Moll:2016,oscillations}.
Recently, a different mechanism for quantum oscillations in Dirac and Weyl semimetals without magnetic
fields was proposed in Ref.~\cite{Liu-Pikulin:2016}. This mechanism relies on strain-induced pseudomagnetic field $\mathbf{B}_5$ and,
similarly to the ordinary quantum oscillations in Weyl semimetals \cite{Ashby:2013}, manifests itself as oscillations of the DOS periodic in
$1/B_5$. As emphasized in Ref.~\cite{Liu-Pikulin:2016}, the great tunability of the pseudomagnetic field
provides a new and convenient experimental basis for the study of strain induced gauge fields in Weyl materials.

Electromagnetic collective excitations are very important characteristics in plasmas \cite{Krall,Landau:t10}.
The chiral chemical potential $\mu_5$, which controls the imbalance between the number densities
of right- and left-handed fermions, provides a new ingredient in relativistic plasmas absent in nonrelativistic
ones. Its role in the spectrum of collective modes started to be investigated only recently
\cite{Kharzeev,Akamatsu:2013,Stephanov:2015}. In Ref.~\cite{Akamatsu:2013}, it was shown that in the absence
of a magnetic field the chiral chemical potential splits the frequencies of the degenerate plasma modes, leaving
the nondegenerate mode intact. The authors of Ref.~\cite{Kharzeev} showed that the triangle anomaly implies the
existence of a new type of collective excitations that stems from the coupling between the density waves of the
chiral and electric charges, which was called the chiral magnetic wave (CMW). Furthermore, it was shown that 3D and 2D topologically nontrivial
materials host unusual chiral plasmonic modes confined to their surfaces \cite{Zyuzin,Hofmann} or edges \cite{Song}.
In view of the relativistic-like quasiparticle dispersion in Dirac and Weyl materials, the main properties
of their electromagnetic collective excitations should be qualitatively the same as in relativistic
plasmas. In addition, the presence of pseudomagnetic fields may result in novel features in the
spectrum of collective excitations. As we will show in this paper, this is indeed the case. If experimentally observed,
these properties of plasmon modes could allow one to extract important information about the physics of strained
Weyl materials.

In relation to the strain-induced pseudomagnetic fields, recently, the issue of the local electric charge nonconservation,
$\partial_{\mu}j^{\mu}\sim(\mathbf{E}\cdot\mathbf{B}_5)$, came to attention in Refs.~\cite{Grushin-Vishwanath:2016,Pikulin:2016}
in the framework of chiral kinetic theory. This rather striking feature was interpreted as a pumping of the electric charge between
the bulk and the boundary of the system. Moreover, it was even suggested that only the global charge conservation is respected in such
systems \cite{Grushin-Vishwanath:2016,Pikulin:2016}. In quantum field theory, where the same difficulty was previously
encountered too, it was proposed to correct the definition of the electric current by including the Chern-Simons contribution
\cite{Landsteiner:2013sja}, also known as the Bardeen-Zumino polynomial \cite{Bardeen}. Such a correction restores the
local conservation of the electric charge in the presence of generic electromagnetic and pseudoelectromagnetic fields.
One of the consequences of the corrected definition is the vanishing CME in an equilibrated plasma
\cite{Landsteiner:2013sja, Landsteiner:2016}.
Of course, testing the local charge (non-)conservation unambiguously may not be easy via an electron
transport, which is in essence a global probe. On the other hand, collective excitations provide an indirect,
but effective means to probe local processes in Weyl materials. As emphasized in our recent paper
\cite{Gorbar:2016ygi}, where the consistent chiral kinetic theory was advocated, the topological Chern-Simons
correction also introduces the dependence on the chiral shift parameter, which describes the anomalous quantum
Hall effect \cite{Burkov:2011ene,Grushin-AHE,Goswami} expected in Weyl materials, as well as modifies
the dispersion relation of the helicon mode \cite{Pellegrino}. In passing, let us note that the effect of a
constant Chern-Simons term in electrodynamics was studied in Ref.~\cite{Carroll:1989vb},
where it was found to produce birefringence of light in vacuum.

In this paper, by using the consistent chiral kinetic theory, we rigorously study the properties of the chiral
magnetic plasmons, as well as novel chiral pseudomagnetic plasmons. We argue that the frequency
of the latter is qualitatively modified by the pseudomagnetic field. Moreover, the chiral nature of both
plasmons is manifested in the oscillations of electric, as well as chiral charge densities. Similarly to
the background magnetic or pseudomagnetic field, the chiral shift $\mathbf{b}$ lifts the degeneracy
of the plasmon modes, providing an effective means of experimentally resolve the issue of the consistent
versus covariant implementations of the chiral anomaly in the chiral kinetic theory.

The paper is organized as follows. The consistent chiral kinetic theory in the presence of electromagnetic and
pseudoelectromagnetic fields is described in Sec.~\ref{sec:CKT}. The general analysis of collective excitations
in constant background magnetic and pseudomagnetic fields is presented in Sec.~\ref{sec:collective-B}. The
resulting polarization vector in the limit of weak (pseudo)magnetic fields is obtained in Sec.~\ref{sec:collective-B-small}.
The cases of plasmon modes propagating along and perpendicular to the external
magnetic field are studied in Secs.~\ref{sec:collective-B-long} and \ref{sec:collective-B-arb-tr}, respectively.
The discussion and summary of the main results relevant for experiments are given in
Sec.~\ref{sec:Summary-Discussions}. Some useful technical results and formulas are presented in Appendices
\ref{sec:App-ref} through \ref{sec:App-Ai}.

\section{The consistent chiral kinetic theory}
\label{sec:CKT}

The chiral kinetic theory aims at the description of the time evolution of one-particle distribution functions
$f_{\lambda}(t,\mathbf{p},\mathbf{r})$ for the right- ($\lambda=+$) and left-handed ($\lambda=-$) fermions.
In order to simplify the notations, in the following the arguments of $f_{\lambda}$ will not be shown explicitly.
The general form of the kinetic equations for chiral fermions reads
\begin{equation}
\partial_t f_{\lambda}+\dot{\mathbf{p}}\cdot \partial_{\mathbf{p}} f_{\lambda}+\dot{\mathbf{r}}\cdot \partial_\mathbf{r}f_{\lambda}=I_{\rm coll}(f_{\lambda}),
\label{CKT-kinetic-equations-general}
\end{equation}
where $I_{\rm coll}(f_{\lambda})$ is the collision integral. When the deviations of the distribution function
from its equilibrium value are small, one may use the relaxation time approximation for the collision integral, i.e.,
$I_{\rm coll}=-(f_{\lambda}-f^{\rm (eq)}_{\lambda})/\tau$, where $\tau$ is the relaxation time,
and
\begin{equation}
f^{\rm (eq)}_{\lambda} =\frac{1}{e^{(\epsilon_{\mathbf{p}}-\mu_{\lambda})/T}+1},
\label{CKT-equilibrium-function}
\end{equation}
is the equilibrium distribution function. Here, $\mu_{\lambda}=\mu+\lambda\mu_5$ is the effective chemical
potential for the right- ($\lambda=+$) and left-handed ($\lambda=-$) fermions, $\mu$ is the electric
chemical potential, $\mu_5$ is the chiral chemical potential, and $\epsilon_{\mathbf{p}}$ is the fermion
dispersion relation, see Eq.~(\ref{CKT-epsilon_p}) below. (Note that we use the units, in which
the Boltzmann constant is $k_B=1$.) The distribution function for antiparticles (holes)
$\bar{f}^{\rm (eq)}_{\lambda}$ is obtained by replacing $\mu_{\lambda}\to - \mu_{\lambda}$.

In this study, we are interested in chiral matter in background electromagnetic and pseudoelectromagnetic
fields. It is convenient to introduce the following effective electric and magnetic fields that act on fermions of
given chirality:
\begin{eqnarray}
\mathbf{E}_{\lambda}=\mathbf{E}+\lambda\mathbf{E}_{5}, \qquad \mathbf{B}_{\lambda}=\mathbf{B}+\lambda\mathbf{B}_{5}.
\label{CKT-fields}
\end{eqnarray}
From a physics viewpoint, the pseudoelectromagnetic fields $\mathbf{E}_{5}$ and $\mathbf{B}_{5}$ stem from
strains in Weyl materials \cite{Cortijo:2016yph,Liu-Pikulin:2016,Pikulin:2016}. The pseudoelectric field $\mathbf{E}_{5}$
is obtained by dynamically deforming the sample. The static pseudomagnetic field $\mathbf{B}_{5}$ can be generated by
applying static torsion or bending.

In the presence of a weak effective magnetic field $\mathbf{B}_{\lambda}$, $\hbar|e\mathbf{B}_{\lambda}|/(c p^2) \ll 1$,
the fermion energy $\epsilon_{\mathbf{p}}$ is given by \cite{Son}
\begin{equation}
\epsilon_{\mathbf{p}}=v_Fp - \lambda e\frac{v_F\hbar}{c} \frac{(\mathbf{B}_{\lambda}\cdot\mathbf{p})}{2p^2}
= v_Fp\left(1 - \frac{e}{c}(\mathbf{B}_{\lambda}\cdot \mathbf{\Omega}_{\lambda})\right),
\label{CKT-epsilon_p}
\end{equation}
where $v_F$ is the Fermi velocity, $c$ is the speed of light, $e$ is an electric charge ($e<0$ for the electron),
\begin{equation}
\mathbf{\Omega}_{\lambda} =\lambda \hbar\frac{\hat{\mathbf{p}}}{2p^2}
\end{equation}
is the Berry curvature \cite{Berry:1984}, $p\equiv|\mathbf{p}|$, and $\hat{\mathbf{p}}=\mathbf{p}/p$.
Note that for antiparticles one should replace
$\mathbf{\Omega}_{\lambda}\to-\mathbf{\Omega}_{\lambda}$. The Berry curvature is a crucial ingredient
in the chiral kinetic theory because it captures the nontrivial topology of massless chiral fermions and
allows to reproduce the chiral anomaly in the presence of external electromagnetic fields (for a clear
exposition, see, e.g., Ref.~\cite{Stephanov}).

Henceforth, we will consider only static deformations and, therefore, the external pseudoelectric field $\mathbf{E}_{5}$
will be absent. By making use of Eq.~(\ref{CKT-epsilon_p}), we find that the quasiparticle velocity $\mathbf{v}$ equals
\begin{equation}
\mathbf{v}= \partial_\mathbf{p}\epsilon_{\mathbf{p}}
=v_F\hat{\mathbf{p}} \left[1+2\frac{e}{c} \left(\mathbf{B}_{\lambda} \cdot \mathbf{\Omega}_{\lambda}\right) \right]
- \frac{e v_F}{c}\mathbf{B}_{\lambda}\left(\hat{\mathbf{p}} \cdot \mathbf{\Omega}_{\lambda}\right).
\label{CKT-v}
\end{equation}
The Berry curvature modifies the semiclassical equations of motion for $\dot{\mathbf{r}}$ and $\dot{\mathbf{p}}$
\cite{Niu,Xiao,Duval} leading to the following equation of the chiral kinetic theory \cite{Son,Stephanov,Son-Spivak}
in the collisionless limit:
\begin{equation}
\partial_tf_{\lambda}+\frac{1}{1+\frac{e}{c}(\mathbf{B}_{\lambda}\cdot\mathbf{\Omega}_{\lambda})}
\left[\Big(e\tilde{\mathbf{E}}_{\lambda}
+\frac{e}{c}(\mathbf{v}\times \mathbf{B}_{\lambda})
+\frac{e^2}{c}(\tilde{\mathbf{E}}_{\lambda}\cdot\mathbf{B}_{\lambda})\mathbf{\Omega}_{\lambda}\Big)\cdot
\partial_\mathbf{p}f_{\lambda}
+\Big(\mathbf{v}+e(\tilde{\mathbf{E}}_{\lambda}\times\mathbf{\Omega}_{\lambda})
+\frac{e}{c}(\mathbf{v}\cdot\mathbf{\Omega}_{\lambda})\mathbf{B}_{\lambda}\Big)\cdot \partial_\mathbf{r}f_{\lambda}\right]=0,
\label{CKT-kinetic-equation}
\end{equation}
where $\tilde{\mathbf{E}}_{\lambda} = \mathbf{E}_{\lambda}-(1/e)\partial_\mathbf{r}\epsilon_{\mathbf{p}}$ and the factor
$1/[1+e(\mathbf{B}_{\lambda}\cdot\mathbf{\Omega}_{\lambda})/c]$ accounts for the correct definition of the phase-space
volume that satisfies the Liouville's theorem \cite{Xiao,Duval}. Let us note that the use of collisionless limit is
justified in the current study if the frequency of plasmon modes is much larger than $1/\tau$, where $\tau$
is the shortest relaxation time of the system. In this connection, let us mention that there can exist different relaxation
mechanisms, including the ones that change the chirality of quasiparticles, e.g., see Ref.~\cite{Stephanov:2015}. The
situation is further complicated by the fact that, in general, the relaxation time may depend on the model parameters,
types of impurities, and background fields, e.g., see Refs.~\cite{GoswamiPixley,SpivakAndreev}. Here, however, we
assume that the sample is sufficiently clean and the collisionless limit is valid.

The evolution of the electromagnetic fields $\mathbf{E}$ and $\mathbf{B}$ is determined by the Maxwell equations
\begin{eqnarray}
\bm{\nabla}\cdot\mathbf{E} = 4\pi \rho, &\qquad & \bm{\nabla}\times\mathbf{E} = -\frac{1}{c} \partial_t\mathbf{B},
\label{CKT-Poisson-vector-scalar-Maxwell-equation1} \\
\bm{\nabla} \cdot\mathbf{B} = 0, &\qquad &   \bm{\nabla}\times\mathbf{B} = \frac{4\pi}{c}\mathbf{j}+\frac{1}{c} \partial_t\mathbf{E}.
\label{CKT-Poisson-vector-scalar-Maxwell-equation2}
\end{eqnarray}
In the first equation, the electric charge density contains the contributions of both left- and right-handed particles,
$\rho=\sum_{\lambda=\pm}\rho_{\lambda}$, where
\begin{equation}
\rho_{\lambda}=\sum_{\rm p, a}e\int\frac{d^3p}{(2\pi \hbar)^3}
\left[1+\frac{e}{c}(\mathbf{B}_{\lambda}\cdot\mathbf{\Omega}_{\lambda})\right]f_{\lambda}.
\label{CKT-charge-density}
\end{equation}
Here $\sum_{\rm p,a}$ denotes the summation over particles and antiparticles contributions.
It is important to note that the pseudoelectromagnetic fields $\mathbf{E}_5$ and $\mathbf{B}_5$ are not governed by the Maxwell
equations. Instead, they are defined by deformations in strained Weyl materials. By using the Maxwell equations
(\ref{CKT-Poisson-vector-scalar-Maxwell-equation1}) and (\ref{CKT-Poisson-vector-scalar-Maxwell-equation2}), together
with Eq.~(\ref{CKT-charge-density}), one can derive the following continuity equations:
\begin{eqnarray}
\partial_t\rho_{\lambda} +\partial_\mathbf{r}\cdot \mathbf{j}_{\lambda} &=& \lambda
\frac{e^3}{4\pi^2 \hbar^2 c} ({\mathbf{E}}_{\lambda}\cdot\mathbf{B}_{\lambda}),
\label{CKT-dn/dt-2}
\end{eqnarray}
where fermion currents are \cite{Son-Spivak,Stephanov,Son}
\begin{eqnarray}
\mathbf{j}_{\lambda} &=& -\sum_{\rm p, a}e\int\frac{d^3p}{(2\pi \hbar)^3}\left(\epsilon_{\mathbf{p}}\partial_{\mathbf{p}}
+\frac{e}{c}\epsilon_{\mathbf{p}}\mathbf{B}_{\lambda}(\mathbf{\Omega}_{\lambda}\cdot\partial_{\mathbf{p}})
+\epsilon_{\mathbf{p}}\mathbf{\Omega}_{\lambda}\times\partial_{\mathbf{r}} - e\mathbf{E}_{\lambda}\times\mathbf{\Omega}_{\lambda}\right)
f_{\lambda}\nonumber\\
&=&\sum_{\rm p, a}e\int\frac{d^3p}{(2\pi \hbar)^3}\left[\mathbf{v}
+\frac{e}{c} \epsilon_{\mathbf{p}} \mathbf{B}_{\lambda} (\partial_{\mathbf{p}}\cdot \mathbf{\Omega}_{\lambda})
+\frac{e}{c}(\mathbf{v}\cdot\mathbf{\Omega}_{\lambda}) \mathbf{B}_{\lambda}
+e(\tilde{\mathbf{E}}_{\lambda}\times\mathbf{\Omega}_{\lambda})\right]\,f_{\lambda}
+\sum_{\rm p, a}e\partial_{\mathbf{r}}\times \int\frac{d^3p}{(2\pi \hbar)^3} f_{\lambda}\epsilon_{\mathbf{p}}\mathbf{\Omega}_{\lambda},
\label{CKT-electric-current-b}
\end{eqnarray}
and the total electric current density is $\mathbf{j}=\sum_{\lambda=\pm}\mathbf{j}_{\lambda}$.
Note that we integrated by parts in Eq.~(\ref{CKT-electric-current-b}). By using the fermion energy
in Eq.~(\ref{CKT-epsilon_p}) and the identity $(\partial_{\mathbf{p}}\cdot \mathbf{\Omega}_{\lambda})=2\pi\hbar \delta^{(3)} (\mathbf{p})$,
it is easy to check that the second term in the square brackets of Eq.~(\ref{CKT-electric-current-b}) vanishes.
The last term in Eq.~(\ref{CKT-electric-current-b}) is the magnetization current
\begin{equation}
\mathbf{j}_{\lambda}^{\rm (curl)}\equiv c\, \partial_{\mathbf{r}}\times \bm{\mathcal{M}}_{\lambda},
\label{CKT-electric-current-curl}
\end{equation}
where
\begin{equation}
 \bm{\mathcal{M}}_{\lambda}\equiv \sum_{\rm p, a}\frac{e}{c}\int\frac{d^3p}{(2\pi \hbar)^3} f_{\lambda}\epsilon_{\mathbf{p}}\mathbf{\Omega}_{\lambda}
\label{CKT-electric-current-curl-m}
\end{equation}
is the effective magnetization. Since $\mathbf{j}_{\lambda}^{\rm (curl)}$ is a total curl, it does not modify
the continuity equation. Nevertheless, this term plays an important role in the Maxwell equations and,
consequently, in the spectrum of collective excitations.

The continuity equation (\ref{CKT-dn/dt-2}) can be rewritten equivalently in terms of the electric and chiral currents, i.e.,
\begin{eqnarray}
\label{CKT-dn/dt-n5}
\partial_t\rho_5 +\partial_\mathbf{r}\cdot\mathbf{j}_5 &=& \frac{e^3}{2\pi^2 \hbar^2 c}
\Big[(\mathbf{E}\cdot\mathbf{B}) +(\mathbf{E}_{5}\cdot\mathbf{B}_{5})\Big],\\
\label{CKT-dn/dt-n}
\partial_t\rho +\partial_\mathbf{r}\cdot \mathbf{j} &=& \frac{e^3}{2\pi^2 \hbar^2 c}
\Big[(\mathbf{E}\cdot\mathbf{B}_{5}) +(\mathbf{E}_{5}\cdot\mathbf{B})\Big].
\end{eqnarray}
The first equation is related to the celebrated chiral anomaly \cite{ABJ} and expresses the nonconservation
of the chiral charge in the presence of electromagnetic or pseudoelectromagnetic fields. From a physics viewpoint,
this nonconservation can be understood as a pumping of the chiral charge between the Weyl nodes of opposite
chiralities.

The second equation describes the anomalous local nonconservation of the {\em electric} charge in
electromagnetic and pseudoelectromagnetic fields. Roughly speaking, it implies that the electric
charge can be literarily created out of nothing or annihilated into nothing in certain local processes.
Clearly, this poses a very serious problem and signifies some incompleteness of the framework. It was
suggested in Refs.~\cite{Pikulin:2016,Grushin-Vishwanath:2016} that the local nonconservation of the
electric charge in Eq.~(\ref{CKT-dn/dt-n}) describes a pumping of the electric charge between the
bulk and the boundary of the system and, therefore, there is no violation of the global electric charge conservation.
As we discussed in Ref.~\cite{Gorbar:2016ygi}, the conservation of the electric charge should be
enforced locally by using the consistent definition of the electric current.

It is worth noting that Eqs.~(\ref{CKT-dn/dt-n5}) and (\ref{CKT-dn/dt-n}) are known in quantum field theory as the
{\it covariant} anomaly relations. They originate from the fermionic sector of the theory, in which left-
and right-handed fermions are treated symmetrically. However, such a treatment is inconsistent with
the gauge symmetry. It was proposed in Ref.~\cite{Landsteiner:2013sja} (for a clear exposition, see also
Ref.~\cite{Landsteiner:2016}) that the correct physical currents, satisfying the local conservation of the
electric charge, are the {\it consistent} currents. As we argued in Ref.~\cite{Gorbar:2016ygi}, the same
should apply to the chiral kinetic theory. Thus, the correct formulation of the theory is the {\em
consistent chiral kinetic theory} that utilizes the consistent definition of the electric current. The additional
topological contribution to the current density reads \cite{Bardeen,Landsteiner:2013sja,Landsteiner:2016,Gorbar:2016ygi}:
\begin{equation}
\delta j^{\mu} =  \frac{e^3}{4\pi^2 \hbar^2 c} \epsilon^{\mu \nu \rho \lambda} A_{\nu}^5 F_{\rho \lambda},
\label{consistent-def-0}
\end{equation}
where $A_{\nu}^5=b_{\nu}+\tilde{A}^5_{\nu}$ is the axial potential. Unlike the electromagnetic potential
$A_{\nu}$, the axial potential is an observable quantity. Indeed, in Weyl materials, $b_0$ and $\mathbf{b}$ correspond
to energy and momentum-space separations between the Weyl nodes, respectively. Strain-induced axial
(or, equivalently, pseudoelectromagnetic) fields are described by $\tilde{A}_{\nu}^5$, which is directly
related to the deformation tensor \cite{Zubkov:2015,Cortijo:2016yph,Cortijo:2016,Grushin-Vishwanath:2016,Pikulin:2016,Liu-Pikulin:2016}.
As is easy to check, the consistent electric current, i.e.,
\begin{equation}
J^{\nu} \equiv (c\rho+c\delta \rho, \mathbf{j}+\delta \mathbf{j}),
\label{consistent-4-current}
\end{equation}
is nonanomalous, $\partial_{\nu} J^{\nu}=0$, and, therefore, the electric charge
is locally conserved. For the sake of clarity, let us rewrite Eq.~(\ref{consistent-def-0}) in components
\begin{eqnarray}
\delta \rho &=&\frac{e^3}{2\pi^2 \hbar^2c^2}\,(\mathbf{A}^5\cdot\mathbf{B}),
\label{consistent-charge-density}
 \\
\delta \mathbf{j} &=&\frac{e^3}{2\pi^2 \hbar^2 c}\,A^5_0 \,\mathbf{B} - \frac{e^3}{2\pi^2 \hbar^2 c}\,(\mathbf{A}^5\times\mathbf{E}).
\label{consistent-current-density}
\end{eqnarray}
In order to obtain nonzero pseudomagnetic field $\mathbf{B}_{5}$, the axial vector potential $\tilde{\mathbf{A}}^5$ should
depend on coordinates. Such a dependence greatly complicates analytical calculations. In the following, therefore, we will assume
that the field $\mathbf{B}_{5}$ is weak or, in other words, that $\tilde{\mathbf{A}}^5$ is negligible compared to the chiral
shift $\mathbf{b}$. Then, it is justified to replace $A_0^5\to b_0$ and $\mathbf{A}^5\to\mathbf{b}$ in
Eqs.~(\ref{consistent-charge-density}) and (\ref{consistent-current-density}).

It is worth noting that the topological correction (\ref{consistent-def-0}) plays an important role even without
pseudoelectromagnetic fields. Indeed, the first term in Eq.~(\ref{consistent-current-density}) with $b_0=-\mu_5/e$
is exactly what is necessary to cancel the CME current in the equilibrium state \cite{Landsteiner:2016}. Furthermore, the second term in
$\delta \mathbf{j}$ describes the anomalous Hall effect in Weyl materials \cite{Burkov:2011ene, Grushin-AHE, Goswami},
\begin{eqnarray}
\mathbf{j}_{\rm AHE} = - \frac{e^3}{2\pi^2 \hbar^2 c}\,(\mathbf{b}\times\mathbf{E}),
\label{CKT-AHE}
\end{eqnarray}
in the framework of the semiclassical kinetic theory \cite{Gorbar:2016ygi}.

\section{Collective modes: general consideration}
\label{sec:collective-B}

By making use of the consistent chiral kinetic theory, let us determine the spectrum of the high-frequency
plasmon excitations in a Weyl material in the presence of a constant background field
$\mathbf{B}_{0,\lambda}\equiv \mathbf{B}_0+\lambda \mathbf{B}_{0,5}$. Here, $\mathbf{B}_0$
is an ordinary magnetic field and $\mathbf{B}_{0,5}$ is a strain-induced pseudomagnetic field.
For the sake of simplicity, we will assume that $\mathbf{B}_{0}\parallel\mathbf{B}_{0,5}$.
In addition to the external fields $\mathbf{B}_0$ and $\mathbf{B}_{0,5}$, collective
modes in Weyl materials will induce also weak oscillating electromagnetic fields
$\mathbf{E}^{\prime}$ and $\mathbf{B}^{\prime}$. Further, these fields could drive
the dynamical deformations of the Weyl material, which in turn generate
oscillating pseudomagnetic fields $\mathbf{E}_5^{\prime}$ and $\mathbf{B}_5^{\prime}$.
However, the latter will be extremely weak and will be neglected in the following.

Our consideration of the electromagnetic collective modes will use the standard approach
of physical kinetics \cite{Krall,Landau:t10}, but generalized to account for
the Berry curvature, the pseudomagnetic field, and the topological current correction.
As usual, we seek $\mathbf{E}^{\prime}$ and $\mathbf{B}^{\prime}$ in the form of
plain waves:
\begin{equation}
\mathbf{E}^{\prime} = \mathbf{E} e^{-i\omega t+i\mathbf{k}\cdot\mathbf{r}} ,
\qquad
\mathbf{B}^{\prime} = \mathbf{B} e^{-i\omega t+i\mathbf{k}\cdot\mathbf{r}} ,
\end{equation}
with frequency $\omega$ and wave vector $\mathbf{k}$. The Maxwell's equations
(\ref{CKT-Poisson-vector-scalar-Maxwell-equation1}) and (\ref{CKT-Poisson-vector-scalar-Maxwell-equation2}) imply
that $\mathbf{B}^{\prime} = c(\mathbf{k}\times \mathbf{E}^{\prime})/\omega$ and
\begin{eqnarray}
\mathbf{k}\left(\mathbf{k}\cdot \mathbf{E}^{\prime} \right) - k^2 \mathbf{E}^{\prime}
=-\frac{\omega^2}{c^2}\left(\mathbf{E}^{\prime}+4\pi \mathbf{P}^{\prime}\right),
\label{collective-B-tensor-spectrum-EM-0}
\end{eqnarray}
where $\mathbf{P}^{\prime}=\mathbf{P}e^{-i\omega t+i\mathbf{k}\cdot\mathbf{r}}$ denotes the polarization vector. By introducing the electric
susceptibility tensor $\chi^{mn}$ (where $m,n=1,2,3$ denote spatial components), the polarization
vector can be given in the following form:
\begin{equation}
P^{\prime m}= i\frac{J^{\prime m}}{\omega}=\chi^{mn}E^{\prime n},
\label{polarization-tensor}
\end{equation}
where $J^{\prime m}$ is an oscillating part of current (\ref{consistent-4-current}).
Then, from Eq.~(\ref{collective-B-tensor-spectrum-EM-0}), we obtain
\begin{eqnarray}
\left( n_0^2 \omega^2- c^2k^2 \right)\delta^{mn}E^n= -c^2 k^m k^nE^n -4\pi\omega^2\chi^{mn}E^n,
\label{collective-B-tensor-spectrum-EM-1}
\end{eqnarray}
where we also restored the refractive index $n_0$. In the case of Dirac semimetal Cd$_3$As$_2$,
the latter is $n_0\approx6$ \cite{Freyland}. In order to simplify our analysis here, we will neglect the dependence of the refractive
index on the frequency. Further, let us note that while our analytical results and main qualitative conclusions should be valid for generic Weyl or Dirac materials,
in our numerical calculations we will use the value of the Fermi velocity of Cd$_3$As$_2$ \cite{Neupane}, i.e.,
$v_F\approx9.8~\mbox{eV\AA}\approx1.5 \times10^{8}~\mbox{cm/s}$.

Equation (\ref{collective-B-tensor-spectrum-EM-1}) admits non-trivial solutions only if the corresponding determinant vanishes, i.e.,
\begin{equation}
\mbox{det}\left[\left( n_0^2 \omega^2- c^2k^2 \right)\delta^{mn} + c^2 k^m k^n + 4\pi\omega^2\chi^{mn}\right]=0.
\label{collective-B-tensor-dispersion-relation-general}
\end{equation}
In essence, this is the characteristic equation that determines the dispersion relations of collective modes.

\begin{figure}[!ht]
\begin{center}
\includegraphics[width=0.45\textwidth]{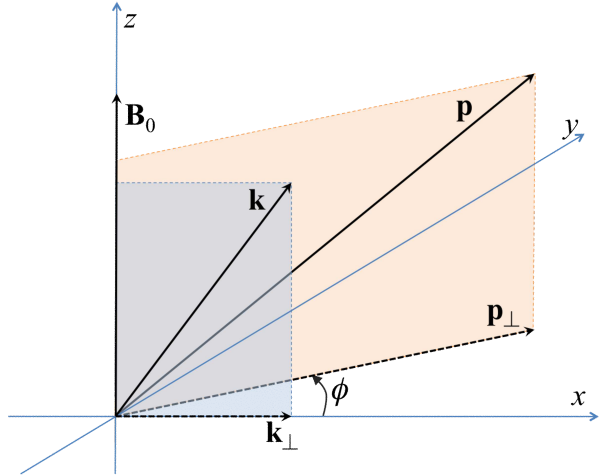}
\caption{The relative orientation of all relevant vectors used in the
analysis of collective modes in an external magnetic field.}
\label{fig:Illustration}
\end{center}
\end{figure}

In order to determine the susceptibility tensor $\chi^{mn}$ in the chiral matter at hand, we use the
consistent chiral kinetic theory. As usual, we choose the ansatz for the distribution function in the form
$f_{\lambda}=f_{\lambda}^{\rm (eq)}+\delta f_{\lambda}$, where
$f_{\lambda}^{\rm (eq)}$ is the equilibrium distribution function given in Eq.~(\ref{CKT-equilibrium-function}),
and
\begin{equation}
\delta f_{\lambda} = f_{\lambda}^{(1)} e^{-i\omega t+i\mathbf{k}\cdot\mathbf{r}}
\end{equation}
is a perturbation induced by the oscillating $\mathbf{E}^{\prime}$ and $\mathbf{B}^{\prime}$ fields. Taking into account that
$\partial_t f_{\lambda}^{\rm (eq)} =0$ and $\partial_\mathbf{r} f_{\lambda}^{\rm (eq)} =0$, it is useful to check that in
the {\em zeroth} order of perturbation theory, the chiral kinetic equation reduces to the equation
\begin{equation}
\left(\hat{\mathbf{p}}\times\mathbf{B}_{0,\lambda}\right)
\cdot \partial_\mathbf{p} f_{\lambda}^{\rm (eq)} =0.
\label{collective-B-kinetic-equation_zeroth}
\end{equation}
Obviously, Eq.~(\ref{collective-B-kinetic-equation_zeroth}) is automatically satisfied for any distribution function which
depends on $|\mathbf{p}|$ and $(\mathbf{p}\cdot \mathbf{B}_{0,\lambda})$ including the equilibrium distribution function
(\ref{CKT-equilibrium-function}) with the dispersion law (\ref{CKT-epsilon_p}). In the {\em first} order of perturbation theory, the chiral kinetic equation (\ref{CKT-kinetic-equation}) gives
\begin{equation}
i\left[ (1+\kappa_{\lambda}) \omega
- (\mathbf{v}\cdot \mathbf{k})-\frac{e}{c}(\mathbf{v}\cdot\mathbf{\Omega}_{\lambda})(\mathbf{B}_{0,\lambda}\cdot \mathbf{k}) \right]
f_{\lambda}^{(1)}
- \frac{e}{c}(\mathbf{v}\times \mathbf{B}_{0,\lambda}) \cdot \partial_\mathbf{p}f_{\lambda}^{(1)}
=e\Big[ (\tilde{\mathbf{E}}\cdot\mathbf{v})
+\frac{e}{c}(\mathbf{v}\cdot\mathbf{\Omega}_{\lambda}) (\tilde{\mathbf{E}}\cdot\mathbf{B}_{0,\lambda})\Big]
\frac{\partial f_{\lambda}^{\rm (eq)}}{\partial \epsilon_{\mathbf{p}}},
\label{collective-B-kinetic-equation-13}
\end{equation}
where we used the shorthand notations
\begin{eqnarray}
\kappa_{\lambda} &\equiv& \frac{e}{c} (\mathbf{\Omega}_{\lambda} \cdot \mathbf{B}_{0,\lambda})
= \lambda \hbar\frac{e\left(\hat{\mathbf{p}}\cdot\mathbf{B}_{0,\lambda}\right)}{2c p^2}, \\
\tilde{\mathbf{E}}&=& \mathbf{E}
+i\frac{\lambda v_F \hbar}{2\omega p} \mathbf{k} \left(\hat{\mathbf{p}}
\cdot[\mathbf{k}\times\mathbf{E}]\right).
\end{eqnarray}
Here we took into account an oscillating magnetic field $\mathbf{B}^{\prime}$ in the energy dispersion, i.e.,
\begin{equation}
\epsilon_{\mathbf{p}} = v_Fp-\frac{\lambda \hbar v_F (e\mathbf{B}_{0,\lambda}\cdot\hat{\mathbf{p}})}{2cp}
-\frac{\lambda \hbar v_F (e\mathbf{B}^{\prime}\cdot\hat{\mathbf{p}})}{2cp}.
\label{collective-B-total-epsilon}
\end{equation}
By making use of the cylindrical coordinates (with the $z$-axis pointing along the magnetic field
$\mathbf{B}_{0}$ and $\phi$ being the azimuthal angle of momentum $\mathbf{p}$), we can render
Eq.~(\ref{collective-B-kinetic-equation-13}) in the following form:
\begin{equation}
\frac{v_F}{c}(1+2\kappa_{\lambda})\frac{ eB_{0,\lambda}}{p}\frac{\partial f_{\lambda}^{(1)}}{\partial \phi}
+i\left[ (1+\kappa_{\lambda})  \omega
- v_F (1+2\kappa_{\lambda})(\hat{\mathbf{p}} \cdot \mathbf{k}) \right]  f_{\lambda}^{(1)}
= ev_F(1+2\kappa_{\lambda})(\hat{\mathbf{p}} \cdot \tilde{\mathbf{E}}) \frac{\partial f_{\lambda}^{\rm (eq)}}{\partial \epsilon_{\mathbf{p}}},
\label{collective-B-eq-general-0}
\end{equation}
where terms quadratic in $B_{0,\lambda}$ were omitted. To this order, the last equation is
equivalent to
\begin{equation}
\frac{\partial f_{\lambda}^{(1)}}{\partial \phi}
+i\left[ \frac{c p \omega/v_F-c(\mathbf{p}\cdot \mathbf{k})}{ eB_{0,\lambda}}
-  \frac{ \lambda  \hbar \omega  (\hat{\mathbf{p}}\cdot\mathbf{B}_{0,\lambda})}{2v_F p B_{0,\lambda}}
 \right]  f_{\lambda}^{(1)}
= \frac{c(\mathbf{p}\cdot \tilde{\mathbf{E}})}{B_{0,\lambda}}
\frac{\partial f_{\lambda}^{\rm (eq)}}{\partial \epsilon_{\mathbf{p}}}.
\label{collective-B-eq-general-1}
\end{equation}
This equation has the same form as its counterpart in nonrelativistic plasma
(see, e.g., Ref.~\cite{Landau:t10}), i.e.,
\begin{equation}
\frac{\partial f_{\lambda}^{(1)}}{\partial \phi} +i (a_1+a_2\cos\phi)f_{\lambda}^{(1)}
=Q(\phi),
\label{collective-B-eq-general-2}
\end{equation}
where
\begin{equation}
Q(\phi) = a_3 \cos(\phi_E-\phi) + a_4 +  a_5\frac{p_\parallel k_\parallel
+ p_\perp k_\perp \cos\phi}{p^2} \left[E_{\perp}p_{\perp}k_{\parallel}\sin{(\phi-\phi_E)} +E_{ \perp}p_{\parallel}k_{\perp}\sin{(\phi_E)}
-E_{\parallel}k_{\perp}p_{\perp}\sin{(\phi)}\right],
\end{equation}
and
\begin{eqnarray}
a_1 &=& \frac{c p \omega/v_F - c p_\parallel k_\parallel}{eB_{0,\lambda}} - \frac{\lambda \hbar \omega  p_\parallel }{2v_F p^2}, \\
a_2 &=&  - \frac{c p_\perp k_\perp }{eB_{0,\lambda}}, \\
a_3 &=&  \frac{c p_\perp E_{\perp}}{B_{0,\lambda}}
\frac{\partial f_{\lambda}^{\rm (eq)}}{\partial \epsilon_{\mathbf{p}}}, \\
a_4 &=& \frac{c p_\parallel E_{\parallel} }{B_{0,\lambda}}
\frac{\partial f_{\lambda}^{\rm (eq)}}{\partial \epsilon_{\mathbf{p}}}, \\
a_5 &=& \frac{i\lambda \hbar v_F}{2\omega B_{0,\lambda}} \frac{\partial f_{\lambda}^{\rm (eq)}}{\partial \epsilon_{\mathbf{p}}}.
\end{eqnarray}
Here $\phi_{E}$ denotes the azimuthal angle of $\mathbf{E}$,
which, similarly to $\phi$, is measured from the $\mathbf{k}_{\perp}$ direction in the plane perpendicular to the magnetic field. The relative
orientation of all relevant vectors is visualized in Fig.~\ref{fig:Illustration}.

The general solution to Eq.~(\ref{collective-B-eq-general-2}) reads
\begin{equation}
f_{\lambda}^{(1)} (\phi)  = C_1 e^{-i(a_1 \phi+ a_2\sin\phi)}
+e^{-i a_2 \sin\phi}
\int_{0}^{\phi-C_0} e^{i[a_2 \sin(\phi-\tau)-a_1 \tau]}
Q(\phi-\tau)  d\tau.
\label{general-solution}
\end{equation}
For $f^{(1)}_{\lambda}(\phi)$ to be a periodic function of $\phi$, one should set $C_1=0$ and
$C_0=\sign{eB_{0,\lambda}}\infty $. The last condition ensures the finiteness of integral at
$\omega\to\omega+i0$, which describes a gradual turning on of the oscillating fields. It is worth noting that
for $\tilde{\mathbf{E}}=0$ the  kinetic equation (\ref{collective-B-kinetic-equation-13}) reduces
to the homogeneous equation considered in Ref.~\cite{Stephanov:2015}, where the effects of
dynamical electromagnetism were not taken into account. The solution of this equation, which
describes the CMW, was obtained by averaging it over momentum $\mathbf{p}$ in Ref.~\cite{Stephanov:2015}.
If such an averaging is not performed, then, as we showed above, a homogeneous solution
to Eq.~(\ref{collective-B-eq-general-2}) is given by the first term in Eq.~(\ref{general-solution}).
However, this is not a valid solution since it is not periodic in $\phi$. Thus, we
conclude that there is no solutions without fluctuating electric field $\mathbf{E}^{\prime}$ and,
consequently, the effects of the dynamical electromagnetism are always present in the collective
modes, including the CMW. As will be shown in Sec.~\ref{sec:collective-B-long}, the CMW is,
in fact, the longitudinal chiral magnetic plasmon.

The correct solution to Eq.~(\ref{collective-B-eq-general-2}) is due to the inhomogeneous term
proportional to the electric field and is given by
\begin{equation}
f_{\lambda}^{(1)} (\phi)  =\int_{0}^{-s_B\infty}d\tau \left\{ \frac{c (\mathbf{p} \cdot \mathbf{E}_{\tau})}{B_{0,\lambda}}
+i\frac{\lambda \hbar v_F c}{2\omega B_{0,\lambda}} \left(\hat{\mathbf{p}}
\cdot [\mathbf{K}_{\tau}^{(1)}\times \mathbf{E}_{\tau}]\right)
(\hat{\mathbf{p}}\cdot\mathbf{K}_{\tau}^{(1)}) \right\}
\frac{\partial f_{\lambda}^{\rm (eq)}}{\partial \epsilon_{\mathbf{p}}}
e^{-i \frac{\left( p \Omega_{\tau}- c\mathbf{p} \cdot \mathbf{K}_{\tau}\right)}{eB_{0,\lambda}} },
\label{collective-B-f1}
\end{equation}
where $s_B=\sign{eB_{0,\lambda}}$ and we introduced the following variables:
\begin{eqnarray}
\label{collective-B-Omega-tau-be}
\Omega_{\tau} &=& \tau c \omega/v_F, \quad \mathbf{E}_{\tau}= \left( \mathbf{E}_{\perp\tau} , E_\parallel \right)
=\mathbf{E}\cos\tau + (\hat{\mathbf{z}} \times \mathbf{E} )\sin\tau
+\hat{\mathbf{z}}(\mathbf{E}\cdot \hat{\mathbf{z}}) (1-\cos\tau), \\
\mathbf{K}_{\tau} &=& \left(2 \mathbf{K}_{\perp\tau} \sin{\left(\frac{\tau}{2}\right)} ,
\tau k_\parallel +\lambda \hbar\frac{\omega eB_{0,\lambda}}{2c v_F p^2}\tau \right)
=\mathbf{k}\sin\tau+(\hat{\mathbf{z}} \times \mathbf{k} )(1-\cos\tau)
+\hat{\mathbf{z}}(\mathbf{k}\cdot \hat{\mathbf{z}}) (\tau-\sin\tau)
+\frac{\lambda \hbar \omega\tau }{2c v_F p^2} e\mathbf{B}_{0,\lambda} ,\\
\mathbf{K}_{\tau}^{(1)} &=& \left(\mathbf{K}_{\perp\tau}^{(1)}, k_{\parallel}\right) = \mathbf{k} \cos{\tau} +(\hat{\mathbf{z}}
\times\mathbf{k})\sin{\tau} +\hat{\mathbf{z}}(\mathbf{k}\cdot\hat{\mathbf{z}}) (1-\cos{\tau}).
\label{collective-B-Omega-tau-ee}
\end{eqnarray}
The vectors $\mathbf{K}_{\perp\tau} = k_\perp \left\{\cos\left(\tau/2\right), \sin\left(\tau/2\right)\right\}$,
$\mathbf{E}_{\perp\tau} = E _\perp \left\{\cos \left(\tau+\phi_E\right),\sin \left(\tau+\phi_E\right)\right\}$, and
$\mathbf{K}_{\perp\tau}^{(1)} = k_\perp \left\{\cos\tau, \sin\tau\right\}$
are obtained from $\mathbf{k}_\perp$ and $\mathbf{E}_\perp$ by rotating them about the direction
of the effective magnetic field ($\hat{\mathbf{z}}\equiv \mathbf{B}_{0,\lambda}/B_{0,\lambda}$) by the angles $\tau/2$ and $\tau$,
respectively. Note that
\begin{eqnarray}
\label{collective-B-K2-tau-be}
K_{\tau}^2 &=& \tau^2 \left(k_\parallel +\lambda\hbar \frac{\omega eB_{0,\lambda}}{2cv_Fp^2} \right)^2+4k_\perp^2 \sin^2\frac{\tau}{2},\\
(\mathbf{K}_{\tau} \cdot \mathbf{E}_{\tau} )
&=& (\mathbf{k}\cdot \mathbf{E}) \sin\tau +(\mathbf{k}\cdot\hat{\mathbf{z}})(\hat{\mathbf{z}}\cdot\mathbf{E})(\tau-\sin\tau)
+(\hat{\mathbf{z}}\cdot [\mathbf{E}\times \mathbf{k}])(1-\cos\tau) +(\hat{\mathbf{z}}\cdot \mathbf{E})
\frac{\lambda \hbar\omega\tau }{2cv_Fp^2} eB_{0,\lambda}.
\label{collective-B-K2-tau-ee}
\end{eqnarray}
Further, the polarization vector (\ref{polarization-tensor}) has the following explicit form:
\begin{eqnarray}
\mathbf{P} &=& \sum_{\rm p,a} \sum_{\lambda} \frac{ie}{\omega}
\int\frac{d^3p}{(2\pi \hbar)^3} \left\{ e(\tilde{\mathbf{E}}\times\mathbf{\Omega}_{\lambda})
+\frac{e}{\omega}(\mathbf{v}\cdot\mathbf{\Omega}_{\lambda})\left( \mathbf{k} \times \mathbf{E} \right) +\frac{e}{c}(\mathbf{\delta v}
\cdot\mathbf{\Omega}_{\lambda})\mathbf{B}_{0,\lambda}\right\} f_{\lambda}^{\rm (eq)} \nonumber\\
&+&  \sum_{\rm p,a} \sum_{\lambda}\frac{\lambda e^2 \hbar v_F}{2\omega^2}\int\frac{d^3p}{(2\pi \hbar)^3} \frac{1}{p} f_{\lambda}^{\rm (eq)} [\mathbf{k}\times
\mathbf{\Omega}_{\lambda}] \left(\hat{\mathbf{p}}\cdot[\mathbf{k}\times\mathbf{E}]\right) +\sum_{\rm p,a} \sum_{\lambda} \frac{ie}{\omega} \int\frac{d^3p}{(2\pi \hbar)^3} v_F\hat{\mathbf{p}}
\left[1+2\frac{e}{c}(\mathbf{\Omega}_{\lambda}\cdot\mathbf{B}_{0,\lambda}) \right] f^{(1)}_{\lambda}\nonumber\\
&-&\sum_{\rm p,a} \sum_{\lambda}\frac{e}{\omega} \int\frac{d^3p}{(2\pi \hbar)^3}
\epsilon_{\mathbf{p}} f^{(1)}_{\lambda} (\mathbf{k}\times\mathbf{\Omega}_{\lambda}) -i \frac{e^3}{2\pi^2 \omega c \hbar^2}(\mathbf{b}\times \mathbf{E}) + i \frac{e^3 b_0}{2\pi^2\omega^2 \hbar^2} (\mathbf{k}\times \mathbf{E}).
\label{collective-B-polarization}
\end{eqnarray}
Here the second and fourth terms stem from the magnetization current $\mathbf{j}^{\rm (curl)}$ given by Eq.~(\ref{CKT-electric-current-curl}).
The last two terms in Eq.~(\ref{collective-B-polarization}) originate from the topological correction $\delta \mathbf{j}$ given by Eq.~(\ref{consistent-current-density}).
Note also that the correction to the velocity $\delta\mathbf{v}$ stems from the oscillating magnetic field in the energy dispersion
(\ref{collective-B-total-epsilon}) and is equal to
\begin{equation}
\delta\mathbf{v} = \frac{2ev_F}{c} \hat{\mathbf{p}}\left(\mathbf{B}\cdot\mathbf{\Omega}_{\lambda}\right)
-\frac{ev_F}{c}\mathbf{B}\left(\hat{\mathbf{p}}\cdot\mathbf{\Omega}_{\lambda}\right)
=\frac{\lambda \hbar ev_F}{2\omega p^2} \Big\{2\hat{\mathbf{p}}\left(\hat{\mathbf{p}}\cdot[\mathbf{k}\times\mathbf{E}]\right)
-[\mathbf{k}\times\mathbf{E}]\Big\}.
\end{equation}

\begin{figure}[!ht]
\begin{center}
\includegraphics[width=0.45\textwidth]{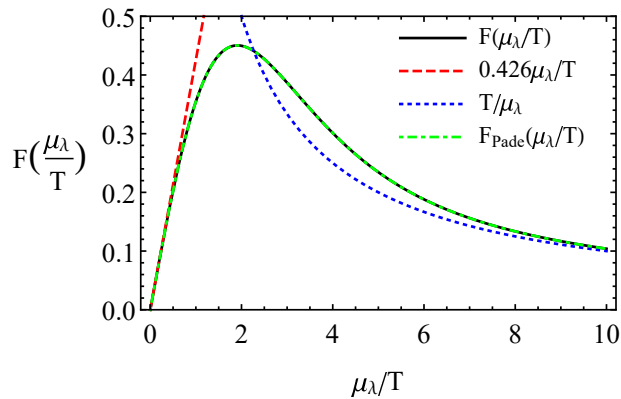}
\caption{The function $F\left(\mu_\lambda/T\right)$, its Pad\'{e} approximant, and two of its asymptotes.}
\label{fig:F}
\end{center}
\end{figure}

Let us calculate contributions to the polarization vector $\mathbf{P}$ up to the linear order in $B_{0,\lambda}$.
Using formulas in Appendix~\ref{sec:App-ref}, we find that the first two terms in Eq.~(\ref{collective-B-polarization})
can be presented in the following form (for the simplicity of presentation, we will omit the contribution of antiparticles
and the summation over chiralities):
\begin{eqnarray}
\label{collective-B-X1-a}
 \mathbf{X}_1^{\rm  (a)} &=& \frac{ie^2}{\omega}\int\frac{d^3p}{(2\pi \hbar)^3} (\mathbf{E}\times\mathbf{\Omega}_{\lambda}) f_{\lambda}^{\rm (eq)}
\simeq -i\frac{e^3\hbar^2 v_F (\mathbf{E}\times\mathbf{B}_{0,\lambda})}{12 c \omega}\int\frac{d^3p}{(2\pi \hbar)^3}\frac{1}{p^3}
\frac{\partial f^{(0)}_{\lambda}}{\partial \epsilon_{\mathbf{p}}} +O\left(B_{0,\lambda}^2\right), \\
\label{collective-B-X1-b}
 \mathbf{X}_1^{\rm  (b)} &=& -\frac{e^2}{\omega}\int\frac{d^3p}{(2\pi \hbar)^3}  \frac{\lambda \hbar v_F }{2\omega p} (\mathbf{E} \cdot[\mathbf{k}
 \times\hat{\mathbf{p}}])(\mathbf{k}\times\mathbf{\Omega}_{\lambda})f_{\lambda}^{\rm (eq)} \simeq -\frac{v_Fe^2}{24\pi^2 \hbar\omega^2 } \int\frac{dp}{p} f^{(0)}_{\lambda} \left[\mathbf{k}\times(\mathbf{k}\times\mathbf{E})\right] +O\left(B_{0,\lambda}^2\right),\\
\label{collective-B-X1-c}
 \mathbf{X}_1^{\rm  (c)} &=& \frac{ie^2}{\omega}\int\frac{d^3p}{(2\pi \hbar)^3} \frac{(\mathbf{v}\cdot\mathbf{\Omega}_{\lambda})}{\omega} \left( \mathbf{k} \times
 \mathbf{E} \right)  f_{\lambda}^{\rm (eq)} \simeq  i\left( \mathbf{k} \times \mathbf{E} \right)  \frac{\lambda e^2 T}{4\pi^2\hbar^2 \omega^2} \ln
 \left(1+e^{\mu_\lambda/T}\right) +O\left(B_{0,\lambda}^2\right),\\
\label{collective-B-X1-d}
\mathbf{X}_1^{\rm  (d)} &=& \frac{ie^2}{\omega} \int\frac{d^3p}{(2\pi \hbar)^3} \frac{(\mathbf{\delta v}\cdot\mathbf{\Omega}_{\lambda})\mathbf{B}_{0,\lambda}}{c}
f_{\lambda}^{\rm (eq)} \simeq i\int\frac{d^3p}{(2\pi \hbar)^3} f^{(0)}_{\lambda} \frac{e^3 \hbar^2 v_F \mathbf{B}_{0,\lambda}}
{4c\omega^2 p^4} \left(\hat{\mathbf{p}}\cdot[\mathbf{k}\times\mathbf{E}]\right) =O\left(B_{0,\lambda}^2\right),\\
\label{collective-B-X1-e}
\mathbf{X}_1^{\rm  (e)}&=&\frac{e^2\hbar^2 v_F}{4\omega^2} \int\frac{d^3p}{(2\pi \hbar)^3} \frac{1}{p^3} [\mathbf{k}
\times\hat{\mathbf{p}}]\left(\hat{\mathbf{p}}\cdot[\mathbf{k}\times\mathbf{E}] \right) \left[ f^{(0)}_{\lambda} -\frac{\lambda \hbar v_F
e (\mathbf{B}_{0,\lambda}\cdot\hat{\mathbf{p}})}{2cp}\left(\frac{\partial f^{(0)}_{\lambda}}{\partial \epsilon_{\mathbf{p}}} \right)\right]
\nonumber\\
&\simeq& \frac{v_F e^2}{24\pi^2 \hbar\omega^2} \int\frac{dp}{p} f^{(0)}_{\lambda} \left[\mathbf{k} \times(\mathbf{k}\times\mathbf{E})\right] +O\left(B_{0,\lambda}^2\right),
\end{eqnarray}
where the equilibrium distribution function was expanded as
\begin{equation}
f_{\lambda}^{\rm (eq)}\simeq\left(f_{\lambda}^{(0)} - \frac{\lambda \hbar v_F e(\mathbf{B}_{0, \lambda}\cdot \hat{\mathbf{p}})}{2cp}
\frac{\partial f^{(0)}_{\lambda}}{\partial \epsilon_{\mathbf{p}}} \right) + O(B_{0,\lambda}^2).
\end{equation}
Here $f^{(0)}_{\lambda}$ is given by Eq.~(\ref{CKT-equilibrium-function}) with $\epsilon_{\mathbf{p}}=v_Fp$.
Note that the integrals over $p$ in $\mathbf{X}_1^{(a)}$, $\mathbf{X}_1^{(b)}$, and $\mathbf{X}_1^{(e)}$ are divergent
in the infrared. However, the terms $\mathbf{X}_1^{(b)}$ and $\mathbf{X}_1^{(e)}$ cancel each other and, as we will show below,
the divergency in $\mathbf{X}_1^{(a)}$ is removed after taking into account the antiparticles contribution. Therefore, by
using formulas in Appendix~\ref{sec:App-ref} and adding the contributions of antiparticles, we obtain
\begin{eqnarray}
\label{collective-B-X1}
\mathbf{X}_1 &=& \sum_{\rm p, a} \left(\mathbf{X}_1^{\rm (a)}+\mathbf{X}_1^{\rm (b)}+\mathbf{X}_1^{\rm (c)}+\mathbf{X}_1^{\rm  (e)}\right)=
-i\frac{e^3\hbar^2v_F(\mathbf{E}\times\mathbf{B}_{0,\lambda})}{12c\omega} \int\frac{d^3p}{(2\pi \hbar)^3}\frac{1}{p^3}\left(
\frac{\partial f^{(0)}_{\lambda}}{\partial \epsilon_{\mathbf{p}}} -\frac{\partial \bar{f}^{(0)}_{\lambda}}{\partial \epsilon_{\mathbf{p}}} \right)  \nonumber\\
&+& i\frac{\lambda T e^2}{4\pi^2 \hbar^2 \omega^2} \left( \mathbf{k} \times \mathbf{E}\right) \left[\ln
\left(1+e^{\mu_\lambda/T}\right) -\ln\left(1+e^{-\mu_\lambda/T}\right)\right] =i\frac{e^3v_F(\mathbf{E}\times\mathbf{B}_{0,\lambda})}{24\pi^2 \hbar \omega cT} F\left(\frac{\mu_\lambda}{T}\right)
 + i\frac{\lambda \mu_\lambda e^2}{4\pi^2 \hbar^2 \omega^2} \left( \mathbf{k} \times \mathbf{E} \right),
\end{eqnarray}
where the function
\begin{eqnarray}
\label{collective-B-F1-def}
F\left(\nu_\lambda \right) \equiv -T\int\frac{dp}{p}\left( \frac{\partial f^{(0)}_{\lambda}}{\partial \epsilon_{\mathbf{p}}}
-\frac{\partial \bar{f}^{(0)}_{\lambda}}{\partial \epsilon_{\mathbf{p}}} \right)
\end{eqnarray}
can be easily computed numerically and $\nu_{\lambda}\equiv\mu_{\lambda}/T$. Its high- and low-temperature
asymptotes equal
$F\left(\nu_\lambda\right)\simeq 7 \zeta(3) \nu_\lambda/(2\pi^2)\approx 0.426
\nu_\lambda$ for $T\to \infty$ and
$F\left(\nu_\lambda\right)\simeq \nu_\lambda^{-1}$ for $T\to 0$, respectively. The function $F\left(\nu_\lambda \right)$ could be
well approximated by the Pad\'e approximant of order [5/6], i.e.,
\begin{eqnarray}
\label{collective-B-F1-Pade}
F\left(\nu_\lambda \right) \simeq
\frac{7\zeta(3)}{2\pi^2}  \frac{\nu_\lambda +0.03533\nu_\lambda^3 +0.0007432\nu_\lambda^5}
{1+0.2290\nu_\lambda^2+0.01567 \nu_\lambda^4+0.0003098\nu_\lambda^6}.
\end{eqnarray}
We plot function $F\left(\nu_\lambda\right)$ together with its asymptotes
and the Pad\'e approximant (\ref{collective-B-F1-Pade}) in Fig.~\ref{fig:F}.

The third and fourth terms in Eq.~(\ref{collective-B-polarization}) contain $f^{(1)}_{\lambda}$,
which is expressed through the integral over $\tau$, and are difficult to calculate in the general case.
In the next section, we will consider them in the limit of small $\mathbf{B}_{0,\lambda}$, where the main
contribution to the integrals over $\tau$ comes from the region of small $\tau$.

\section{Polarization vector in the limit of small background field}
\label{sec:collective-B-small}

In the limit of small $\mathbf{B}_{0,\lambda}$, the main contribution to the integral over
$\tau$ in Eq.~(\ref{collective-B-f1}) comes from the region
of small $\tau$. Therefore, one should expand $\sin\tau$ and $\cos\tau$ in Eqs.~(\ref{collective-B-Omega-tau-be})
through (\ref{collective-B-K2-tau-ee}) in powers of $\tau$, use integrals (\ref{App-int-tau-be}) through
(\ref{App-int-tau-ee}), and keep the terms that contribute linearly in the
effective magnetic field $\mathbf{B}_{0,\lambda}$ to the polarization vector.

The third term in Eq.~(\ref{collective-B-polarization}) contains the following integral:
\begin{eqnarray}
\mathbf{X}_2 &=& \frac{ie}{\omega}\int\frac{d^3p}{(2\pi \hbar)^3} v_F\hat{\mathbf{p}}
\left[1+2\frac{e}{c}(\mathbf{\Omega}_{\lambda}\cdot\mathbf{B}_{0,\lambda}) \right] f_{\lambda}^{(1)}
 = \frac{ie}{\omega}\int_{0}^{-s_B\infty }d\tau \int\frac{d^3p}{(2\pi \hbar)^3} v_F\hat{\mathbf{p}}
\left[1+\frac{\lambda \hbar  e(\hat{\mathbf{p}}\cdot \mathbf{B}_{0,\lambda})}{cp^2} \right] \nonumber\\
 &\times&\left\{ \frac{c \mathbf{p} \cdot \mathbf{E}_{\tau}}{B_{0,\lambda}} +i\frac{\lambda \hbar v_F c}{2\omega B_{0,\lambda}}
 \left(\hat{\mathbf{p}}\cdot [\mathbf{K}_{\tau}^{(1)}\times \mathbf{E}_{\tau}]\right) (\hat{\mathbf{p}}\cdot\mathbf{K}_{\tau}^{(1)}) \right\}
\left[\frac{\partial f^{(0)}_{\lambda}}{\partial \epsilon_{\mathbf{p}}} -\frac{\lambda \hbar v_F e (\mathbf{B}_{0,\lambda}\cdot
\hat{\mathbf{p}})}{2cp} \frac{\partial^2 f^{(0)}_{\lambda}}{\partial \epsilon_{\mathbf{p}}^2} \right]e^{-i p\frac{\Omega_{\tau}- c
\hat{\mathbf{p}} \cdot \mathbf{K}_{\tau}}{eB_{0,\lambda}}} \nonumber\\
&=&\mathbf{X}_{2}^{\rm (a)}+\mathbf{X}_{2}^{\rm (b)}+\mathbf{X}_{2}^{\rm (c)}+\mathbf{X}_{2}^{\rm (d)}.
\label{collective-B-X2}
\end{eqnarray}
The expressions for coefficients $\mathbf{X}_{2}^{\rm (a)}$ through $\mathbf{X}_{2}^{\rm (d)}$ are rather cumbersome. Therefore, we
present them in Appendix~\ref{sec:App-X-j}.

The fourth term in Eq.~(\ref{collective-B-polarization}) can be rewritten as follows:
\begin{eqnarray}
\mathbf{X}_{3}&=&-\frac{e\lambda\hbar v_F}{\omega}\int\frac{d^3p}{(2\pi \hbar)^3} \frac{(\mathbf{k}\times\hat{\mathbf{p}})}{2p}
\left[1-\frac{\lambda \hbar e (\hat{\mathbf{p}}\cdot\mathbf{B}_{0,\lambda})}{2 p^2c}\right] \int_{0}^{-s_B\infty}d\tau
\left\{\frac{c \mathbf{p} \cdot \mathbf{E}_{\tau}}{B_{0,\lambda}} +i\frac{\lambda \hbar v_F c}{2\omega B_{0,\lambda}}
\left(\hat{\mathbf{p}}\cdot [\mathbf{K}_{\tau}^{(1)}\times \mathbf{E}_{\tau}]\right) (\hat{\mathbf{p}}\cdot\mathbf{K}_{\tau}^{(1)}) \right\}
\nonumber\\
&\times&\left[\frac{\partial f^{(0)}_{\lambda}}{\partial \epsilon_{\mathbf{p}}} -\frac{\lambda \hbar v_F e (\mathbf{B}_{0,\lambda}\cdot
\hat{\mathbf{p}})}{2cp} \frac{\partial^2 f^{(0)}_{\lambda}}{\partial \epsilon_{\mathbf{p}}^2}\right]
e^{-i p\frac{\Omega_{\tau}- c\hat{\mathbf{p}} \cdot \mathbf{K}_{\tau}}{ eB_{0,\lambda}}} = \mathbf{X}_{3}^{\rm (a)}+\mathbf{X}_{3}^{\rm (b)}+\mathbf{X}_{3}^{\rm (c)}+\mathbf{X}_{3}^{\rm (d)},
\label{collective-B-magnetization}
\end{eqnarray}
where the explicit expressions for $\mathbf{X}_{3}^{\rm (a)}$ through $\mathbf{X}_{3}^{\rm (d)}$ are also given in
Appendix~\ref{sec:App-X-j}.

It worth noting that the contributions $\mathbf{X}_{3}^{\rm (c)}$ and $\mathbf{X}_{3}^{\rm (d)}$ contain
a divergent integral of the following type:
\begin{equation}
\int_{\Lambda_{\rm IR}}^{\infty}\frac{dp}{p^2} \frac{\partial f^{(0)}_{\lambda}}{\partial \epsilon_{\mathbf{p}}} = v_F\Bigg\{ -\frac{1}{v_Fp}
\frac{\partial f^{(0)}_{\lambda}}{\partial \epsilon_{\mathbf{p}}} \Big|^{\infty}_{\Lambda_{\rm IR}} +\int_{\Lambda_{\rm IR}}^{\infty}\frac{dp}{p}
\frac{\partial^2 f^{(0)}_{\lambda}}{\partial \epsilon_{\mathbf{p}}^2} \Bigg\} = -\frac{1}{4T\Lambda_{\rm IR} \cosh^2{\left(\frac{\mu_{\lambda}}{2T}\right)}}
+v_F\int_{\Lambda_{\rm IR}}^{\infty}\frac{dp}{p} \frac{\partial^2 f^{(0)}_{\lambda}}{\partial \epsilon_{\mathbf{p}}^2}.
\label{collective-B-tensor-Lambda-IR}
\end{equation}
Here we introduced an infrared cutoff $\Lambda_{\rm IR}=C\sqrt{\hbar |eB_{0,\lambda}|/c}$ with a numerical
constant $C$ of order unity. Such a cutoff has a transparent physical meaning: it separates the phase space
of large momenta, where the semiclassical description is valid, from the infrared region $p\lesssim \Lambda_{\rm IR}$,
where such a description fails (for details, see also Ref.~\cite{Stephanov}). In our numerical calculations, we will use
$C=1$. After adding the contribution of antiparticles, one can set $\Lambda_{\rm IR}$ to zero in the last term of
Eq.~(\ref{collective-B-tensor-Lambda-IR}). Indeed, then the corresponding integral is no longer divergent in
the infrared and can be expressed in terms of the derivative of function $F(\nu_\lambda)$ with respect to $\nu_\lambda$.

By combining all contributions to the polarization vector (\ref{collective-B-polarization}), we arrive at the final
result in the following form:
\begin{eqnarray}
\mathbf{P} &=& A_0 \mathbf{E} +A_1 [\mathbf{E}\times \hat{\mathbf{z}}] +A_2[\mathbf{E}\times \hat{\mathbf{k}}]
+A_3\hat{\mathbf{k}}(\hat{\mathbf{k}}\cdot\mathbf{E}) +A_4\hat{\mathbf{k}}(\hat{\mathbf{z}}\cdot\mathbf{E})
+A_5\hat{\mathbf{z}}(\hat{\mathbf{k}}\cdot
\mathbf{E}) +A_6\mathbf{E}(\hat{\mathbf{z}}\cdot\hat{\mathbf{k}}) \nonumber\\
&+&
A_7[\hat{\mathbf{k}}\times\hat{\mathbf{z}}](\hat{\mathbf{k}}\cdot\mathbf{E}) +A_{8}\hat{\mathbf{k}} (\hat{\mathbf{z}}\cdot[\mathbf{E}\times
\hat{\mathbf{k}}]) +A_{9}\hat{\mathbf{k}}(\hat{\mathbf{k}}\cdot\hat{\mathbf{z}})(\hat{\mathbf{k}}\cdot\mathbf{E})+A_{10}(\hat{\mathbf{k}}\cdot
\hat{\mathbf{z}})[\hat{\mathbf{k}}\times\mathbf{E}] + A_{11} [\mathbf{b}\times \mathbf{E}].
\label{collective-B-tensor-P}
\end{eqnarray}
The explicit expressions for all coefficients $A_i$, with $i=\overline{0,11}$, are given in Appendix~\ref{sec:App-Ai}.
It is worth noting that the second term in Eq.~(\ref{collective-B-tensor-P}) is related to the usual Faraday rotation, as well as
its anomalous counterpart. Indeed, as one can check from its definition in Eq.~(\ref{collective-B-tensor-A1}) in Appendix~\ref{sec:App-Ai},
the coefficient $A_1$ in front of this term is proportional to a magnetic or pseudomagnetic field. Also, the last
term in polarization vector (\ref{collective-B-tensor-P}) captures the anomalous Faraday effect of Weyl materials,
i.e., the rotation of polarization in the absence of a background magnetic field \cite{Kargarian}.
In terms of the electric susceptibility tensor, we obtain
\begin{eqnarray}
\chi^{mn} &=& A_0 \delta^{mn} +A_1 \varepsilon^{mn3} +A_2 \varepsilon^{mnl}\hat{\mathbf{k}}^l
+A_3\hat{\mathbf{k}}^m\hat{\mathbf{k}}^n +A_4\delta^{n3}\hat{\mathbf{k}}^m +A_5\delta^{m3}\hat{\mathbf{k}}^n +A_6\delta^{mn}\hat{\mathbf{k}}^3
+A_7\varepsilon^{ml3}\hat{\mathbf{k}}^l\hat{\mathbf{k}}^n
\nonumber\\
&+&A_{8}\varepsilon^{nl3}\hat{\mathbf{k}}^l\hat{\mathbf{k}}^m
+A_{9}\hat{\mathbf{k}}^m\hat{\mathbf{k}}^n\hat{\mathbf{k}}^3 +A_{10}\varepsilon^{mln}\hat{\mathbf{k}}^l\hat{\mathbf{k}}^3 +A_{11} \varepsilon^{mln}\mathbf{b}^l.
\label{collective-B-tensor-chi}
\end{eqnarray}
By setting $\mathbf{k} = (k_\perp, 0, k_\parallel)$, where $k_\perp=k_x=k\sin{\theta}$, $k_\parallel=k_z=k\cos{\theta}$, and $\theta$ is the
angle between $\mathbf{B}_{0}$ and $\mathbf{k}$, we have in components
\begin{eqnarray}
\label{collective-B-tensor-chi-be}
\chi^{11} &=& A_0+A_3 \sin^2{\theta} +\cos{\theta}\left(A_6+\sin^2{\theta} A_9\right), \\
\chi^{22} &=& A_0+A_6\cos{\theta}, \\
\chi^{33} &=& A_0+\cos{\theta}\left(\cos{\theta} A_3 +A_4+A_5+A_6  +
\cos^2{\theta} A_9\right), \\
\chi^{12} &=& A_1 +A_2\cos{\theta} -A_8\sin^2{\theta} -A_{10}\cos^2{\theta} -A_{11}b_z, \\
\chi^{21} &=& -A_1 -A_2\cos{\theta} -A_7\sin^2{\theta} +A_{10}\cos^2{\theta} +A_{11}b_z, \\
\chi^{13} &=& (A_2 +A_4 -\cos{\theta} A_9)\sin{\theta} +A_{11}b_y , \\
\chi^{31} &=& (A_3 \cos{\theta} +A_5 +A_9\cos^2{\theta})\sin{\theta}
 -A_{11}b_y, \\
\chi^{23} &=& (A_2 -A_7 \cos{\theta} -A_{10} \cos{\theta})\sin{\theta} -A_{11}b_x, \\
\chi^{32} &=& -(A_2 +A_8 \cos{\theta} -A_{10}\cos{\theta})\sin{\theta} +A_{11}b_x.
\label{collective-B-tensor-chi-ee}
\end{eqnarray}
For an arbitrary $\theta$, the dispersion relation (\ref{collective-B-tensor-dispersion-relation-general}) is quite bulky, therefore, we present it in Appendix~\ref{sec:App-Ai} as Eq.~(\ref{collective-B-tensor-dispersion-relation-theta}).
For the sake of clarity and brevity, we consider in the next sections only two simple cases of the
collective mode propagation with respect to the external magnetic field $\mathbf{B}_{0}$: the longitudinal propagation
(i.e., $k_{\perp}=0$) and the transverse propagation (i.e., $k_{\parallel}=0$).

\section{Collective modes propagating along the magnetic field}
\label{sec:collective-B-long}

In this section, we consider the collective modes propagating parallel to the background magnetic field,
i.e., $\mathbf{k}\parallel \mathbf{B}_{0}$. Then, the dispersion equation (\ref{collective-B-tensor-dispersion-relation-general}) [or, equivalently, Eq.~(\ref{collective-B-tensor-dispersion-relation-theta})]
at $k_{\perp}=0$ is
\begin{eqnarray}
&&\left[ n_0^2 +4\pi(A_0+A_3+A_4+A_5+A_6+A_9)\right]\left[\left( n_0^2 \omega^2-c^2k^2+4\pi\omega^2(A_0+A_6)\right)^2 +16\pi^2\omega^4\left(A_1+A_2-A_{10}-A_{11}b_{\parallel} \right)^2 \right] \nonumber\\
&& +16\pi^2\omega^2 A_{11}^2 b_\perp^2\left[ n_0^2 \omega^2 -c^2k^2+4\pi\omega^2(A_0+A_6)\right] =0,
\label{consistent-B5-long-dispersion-relation-1}
\end{eqnarray}
where $b_{\parallel} \equiv b_z$ and $b_{\perp} \equiv \sqrt{b_x^2+b_y^2}$. It appears that, for an arbitrary orientation of the chiral shift $\mathbf{b}$, one cannot
separate the dispersion relations of the longitudinal ($\mathbf{E}\parallel\mathbf{k}$) and transverse ($\mathbf{E}\perp\mathbf{k}$)
modes in the last equation. However, such a separation is possible in the special case with $\mathbf{b}_\perp=0$, where
\begin{eqnarray}
\label{consistent-B5-Long-Eq1}
 n_0^2 +4\pi (A_0+A_3+A_4+A_5+A_6+A_9) &=& 0, \\
\label{consistent-B5-Long-Eq2}
 n_0^2 \omega^2-c^2k^2+4\pi\omega^2(A_0+A_6)\mp 4\pi i(A_1+A_2-A_{10}-A_{11}b_{\parallel})\omega^2 &=&0.
\end{eqnarray}
By making use of the coefficients $A_i$ in Appendix~\ref{sec:App-Ai}, we can rewrite Eq.~(\ref{consistent-B5-Long-Eq1})
in the following explicit form:
\begin{eqnarray}
1+\frac{3\Omega_{e}^2}{v_F^2k^2}\left[1 -\frac{\omega}{2v_Fk}
\ln{\left|\frac{\omega+v_Fk}{\omega-v_Fk}\right|} -\frac{2 \alpha e v_F^5k^2 (\mathbf{B}_{0,5}\cdot\mathbf{k})}{3 \pi \hbar c \omega \Omega_{e}^2 (\omega^2-v_F^2k^2)}\right]=0.
\label{consistent-B5-Long-Eq1-expl}
\end{eqnarray}
Here we introduced the shorthand notations for the fine structure constant $\alpha=e^2/(\hbar v_F n_0^2 )$ and
the Langmuir (plasma) frequency
\begin{equation}
\Omega_e \equiv \sqrt{\frac{4\alpha}{3\pi\hbar^2}\left(\mu^2+\mu_5^2 +\frac{\pi^2 T^2}{3}\right)}.
\end{equation}
The dispersion relations for the longitudinal and transverse modes are given by the solutions of
Eqs.~(\ref{consistent-B5-Long-Eq1}) and (\ref{consistent-B5-Long-Eq2}), respectively. In the limit of long
wavelengths $ck\ll \Omega_e$, as well as small $B_{0}$, $B_{0,5}$, and $b_{\parallel}$, the corresponding
analytical solutions are
\begin{eqnarray}
\label{consistent-B5-Long-Eq1-app}
\omega_l &\simeq& \Omega_e\sqrt{1 +\frac{2 \alpha e(\mathbf{B}_{0,5}\cdot\mathbf{k}) v_F^3}{\pi c\hbar \Omega_e^3} +\frac{3}{5}\left(\frac{v_Fk}{\Omega_e}\right)^2 +\frac{\alpha e(\mathbf{B}_{0,5}\cdot\mathbf{k}) v_F^3}{5\pi c\hbar \Omega_e^3} \left(\frac{v_Fk}{\Omega_e}\right)^2 +\frac{12}{175}\left(\frac{v_Fk}{\Omega_e}\right)^4 +\cdots} , \\
\label{consistent-B5-Long-Eq2-app}
\omega_{\rm tr}^{\pm} &\simeq & \sqrt{\left(\Omega_{e,B}^\pm\right)^2 + A_1^\pm v_F k  + A_2^\pm (v_F k)^2 +\cdots},
\end{eqnarray}
where
\begin{eqnarray}
\left(\Omega_{e,B}^\pm\right)^2 &\simeq & \Omega_e^2 \pm \frac{4\alpha v_F^2e}{3\pi c \hbar^2 \Omega_e}\left\{\mu B_{0}+\mu_5 B_{0,5} -\frac{3 \hbar \Omega_e^2 b_{\parallel}}{2v_F} -\frac{\hbar^2\Omega_e^2}{8T}\left[(B_{0}+B_{0,5}) F \left(\frac{\mu+\mu_5}{T}\right) +(B_{0}-B_{0,5}) F \left(\frac{\mu-\mu_5}{T}\right)\right]\right\} \nonumber\\
&+&O(B_{0}^2, B_{0,5}^2, b_{\parallel}^2),\\
A_1^\pm &\simeq & \mp \frac{2 \alpha \mu_5}{3 \hbar\pi} + \frac{2 \alpha^2 eb_{\parallel} v_F\mu_5}{3 c\hbar^2\pi^2 \Omega_e} +\frac{4\alpha^2 v_F^2 e \mu_5}{9c\hbar^3\pi^2\Omega_e^3} \Bigg\{ B_{0}\mu+B_{0,5}\mu_5  \nonumber\\
&+& \frac{\hbar^2\Omega_e^2}{8T} \left[(B_{0}+B_{0,5}) F \left(\frac{\mu+\mu_5}{T}\right) +(B_{0}-B_{0,5}) F \left(\frac{\mu-\mu_5}{T}\right) \right]\Bigg\}
+O(B_{0}^2, B_{0,5}^2, b_{\parallel}^2),\\
A_2^\pm &\simeq & \left(\frac{1}{5} +\frac{c^2}{v_F^2  n_0^2 }\right) \pm
\frac{\alpha eb_{\parallel}}{45\pi^3 c \hbar^3 \Omega_e^3 v_F} \left[5\alpha^2v_F^2 \mu_5^2 -9\pi^2\Omega_e^2 \hbar^2\left(5\frac{c^2}{ n_0^2 }-v_F^2\right)\right] \nonumber\\
 &\pm& \frac{2e\,\alpha}{45c\hbar^4\pi^3\Omega_e^5}\Bigg\{ (B_{0}\mu+B_{0,5}\mu_5) \Big[5\alpha^2v_F^2\mu_5^2+3\pi^2\hbar^2\Omega_e^2\left(3v_F^2-5\frac{c^2}{ n_0^2 }\right)\Big] +\frac{\hbar^2\Omega_e^2}{24T} \Big[5\alpha^2v_F^2\mu_5^2+9\pi^2\hbar^2\Omega_e^2\left(6v_F^2-5\frac{c^2}{ n_0^2 }\right)\Big]\nonumber\\
 &\times& \left[(B_{0}+B_{0,5}) F \left(\frac{\mu+\mu_5}{T}\right) +(B_{0}-B_{0,5}) F \left(\frac{\mu-\mu_5}{T}\right) \right] \Bigg\} +O(B_{0}^2, B_{0,5}^2, b_{\parallel}^2).
\end{eqnarray}
Here we took into account that the CME should be absent in equilibrium, i.e., set $\mu_5=-eb_0$ \cite{Landsteiner:2016}.
Let us note that the topological corrections given by the last two terms in Eq.~(\ref{collective-B-polarization}) do not
influence the dispersion law of the longitudinal mode. The general feature common for all modes under
consideration is the presence of nonzero gaps of the order of the Langmuir frequency $\Omega_e$ in their spectra.
As is clear, this is the consequence of taking dynamical electromagnetism into account.

\begin{figure}[!ht]
\begin{center}
\includegraphics[width=0.45\textwidth]{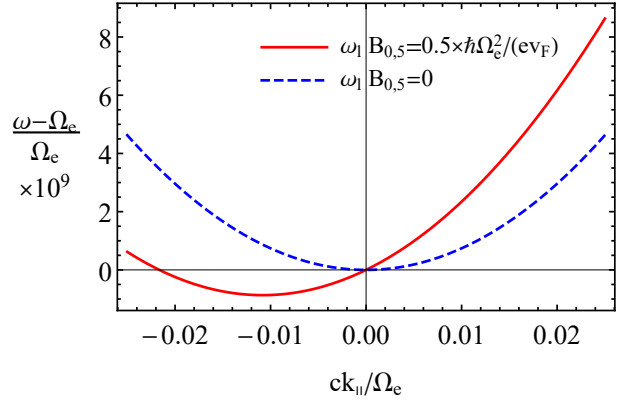}
\caption{The dispersion relation of the longitudinal collective mode given by Eq.~(\ref{consistent-B5-Long-Eq1-app}) as a function of the wave vector at $B_{0,5}=0.5\,\hbar\Omega_e^2/(v_F e)$ (red solid line) and $B_{0,5}=0$ (blue dashed line).}
\label{fig:consistent-long-bz-l}
\end{center}
\end{figure}

The dispersion relation of the longitudinal mode $\omega_l$ is shown in Fig.~\ref{fig:consistent-long-bz-l}
in the case of vanishing $B_{0,5}$, as well as $B_{0,5}= 0.5\,\hbar\Omega_e^2/(v_F e)$. As we see from the figure,
the dispersion relation of the longitudinal mode has a minimum. By making use of the analytical expression
in Eq.~(\ref{consistent-B5-Long-Eq1-app}), we find that its location is given by
$v_Fk_{\parallel}/\Omega_e \approx-5v_F^2\alpha eB_{0,5}/(3\pi c \hbar \Omega_e^2)$.
This minimum is somewhat reminiscent of the roton modes in the superfluid helium-4 \cite{Landau:1947}.
Note, however, that the minimum is global in the model at hand.

One can see from Eq.~(\ref{consistent-B5-Long-Eq1-expl}), as well as approximate expression (\ref{consistent-B5-Long-Eq1-app})
that the frequency of the longitudinal mode $\omega_{l}$ does not explicitly depend on temperature.
Moreover, it depends linearly on the pseudomagnetic field $\mathbf{B}_{0, 5}$ [see the
second and fourth terms under the square root in Eq.~(\ref{consistent-B5-Long-Eq1-app})] and does not depend explicitly on the electric and chiral chemical potentials. One might
call this type of collective excitations \emph{the chiral pseudomagnetic plasmon}.
Its origin is related to the anomalous term on the right-hand side of Eq.~(\ref{CKT-dn/dt-n}). It is worth noting that
in view of our approximation of weakly varying $\mathbf{A}^5$ the effect of $\mathbf{B}_{0, 5}$
should be small. In the general case, the proper account of space dependent $\mathbf{A}^{5}$
should be studied in more detail.

It is important to note that the CMW is, in fact, a chiral magnetic plasmon \cite{Gorbar:2016ygi}, just like the chiral pseudomagnetic
one. In the case of a nonzero wave vector, $\mathbf{k}\neq 0$, the chiral nature of
these plasmons can be seen from the oscillations of the chiral charge density (for completeness,
we also present the oscillating part of the electric charge density), i.e.,
\begin{equation}
\label{collective-B-n5}
\delta \rho_5 =
-\sin{\left(\omega t -\mathbf{k}\cdot\mathbf{r}\right)}  n_0^2 \Bigg\{\frac{2\alpha \mu \mu_5 (\mathbf{E}\cdot\mathbf{k})}{\pi^2\hbar^2 v_F^2 k^2}  \left[1-\frac{\omega}{2v_Fk}
\ln{\left|\frac{\omega+v_Fk}{\omega-v_Fk}\right|}\right]
-\frac{ \alpha v_F\omega e(\mathbf{E}\cdot \mathbf{B}_{0})}{2\pi^2 \hbar c (\omega^2-v_F^2k^2)} \Bigg\},
\end{equation}
\begin{eqnarray}
\label{collective-B-n}
\delta \rho &=&
-\sin{\left(\omega t -\mathbf{k}\cdot\mathbf{r}\right)}  n_0^2 \Bigg\{\frac{ 3 \Omega_e^2(\mathbf{E}\cdot\mathbf{k})}{4\pi v_F^2 k^2}  \left[1-\frac{\omega}{2v_Fk}
\ln{\left|\frac{\omega+v_Fk}{\omega-v_Fk}\right|}\right]
-\frac{\alpha v_F \omega e(\mathbf{E}\cdot \mathbf{B}_{0,5})}{2\pi^2 \hbar c (\omega^2-v_F^2k^2)} \Bigg\} \nonumber\\
&+&\cos{\left(\omega t -\mathbf{k}\cdot\mathbf{r}\right)}  n_0^2 \frac{e \alpha v_F (\mathbf{b}\cdot[\mathbf{k}\times\mathbf{E}])}{2\pi^2\hbar \omega c}.
\end{eqnarray}
Note that the pseudomagnetic field $\mathbf{B}_{0,5}$ leads to an additional oscillating term in the electric
charge density [see the second term in the curly brackets in Eq.~(\ref{collective-B-n})], related to the right-hand side of
Eq.~(\ref{CKT-dn/dt-n}). The last term in Eq.~(\ref{collective-B-n}) stems from the topological correction given
by Eq.~(\ref{consistent-charge-density}). It is important to emphasize the topological origin of the chiral
charge density oscillations, which manifests itself in the absence of the temperature dependence
in $\delta \rho_5$. In addition, while the first term in Eq.~(\ref{collective-B-n5}) is related to the chiral electric separation effect
\cite{Huang:2013, Jiang:2015, Pu:2014}, the second one has its origin in the chiral anomaly given by Eq.~(\ref{CKT-dn/dt-n5}).
It is worth mentioning that in the absence of the dynamical electromagnetism there are two modes of the
CMW, i.e., the right- and left-handed modes, which propagate along and against the magnetic field and have
the same amplitude. Their electric and chiral charge densities oscillate in phase and antiphase,
respectively \cite{Kharzeev:2016,Chernodub:2015}. However, it is clear that this is not the case in
our analysis, which rigorously takes dynamical electromagnetism into account.
Indeed, we see that the fluctuations of electric and chiral charge densities have different
magnitudes and depend on the magnetic and pseudomagnetic fields, as well as on the chiral shift.
In fact, even the magnitudes of the collective modes propagating along and against the pseudomagnetic field
direction (i.e., with $e\mathbf{B}_{0,5}\cdot\mathbf{k}>0$ and $e\mathbf{B}_{0,5}\cdot\mathbf{k}<0$) are
different from each other.

\begin{figure}[!ht]
\begin{minipage}[ht]{0.32\linewidth}
\center{\includegraphics[width=1.0\linewidth]{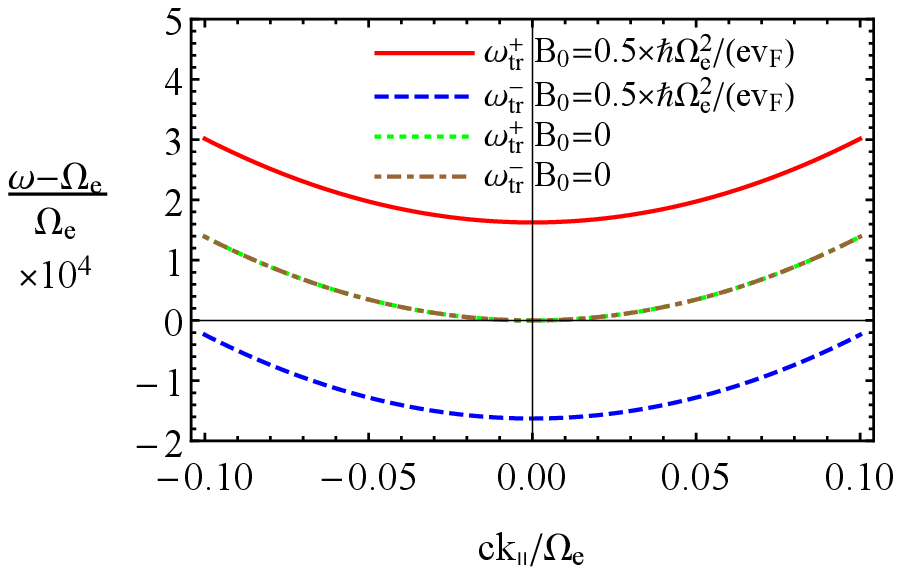} \\
{\small (a) $\mu=\sqrt{3\pi /(4 \alpha)}\hbar\Omega_e$, $\mu_5=0$, $B_{0,5}=0$, $b_{\parallel}=0$}}
\end{minipage}
\begin{minipage}[ht]{0.32\linewidth}
\center{\includegraphics[width=1.0\linewidth]{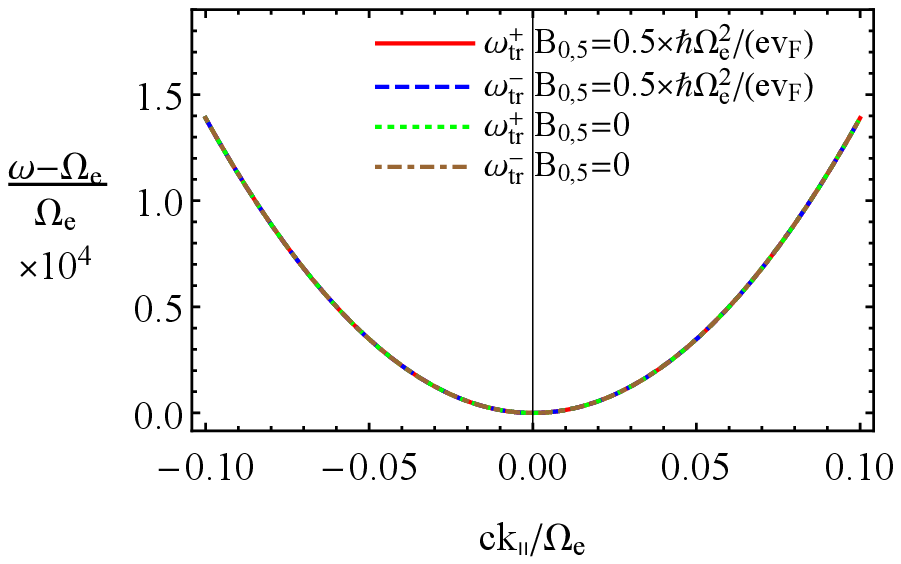} \\
{\small (b) $\mu=\sqrt{3\pi /(4 \alpha)}\hbar\Omega_e$, $\mu_5=0$, $B_0=0$, $b_{\parallel}=0$}}
\end{minipage}
\begin{minipage}[ht]{0.32\linewidth}
\center{\includegraphics[width=1.0\linewidth]{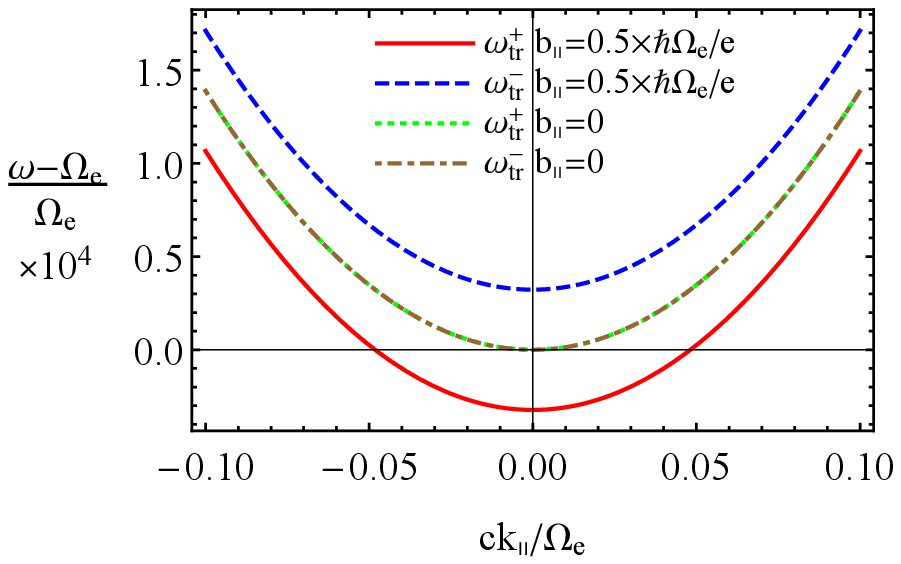} \\
{\small (c) $\mu=\sqrt{3\pi /(4 \alpha)}\hbar\Omega_e$, $\mu_5=0$, $B_0=0$, $B_{0,5}=0$}}
\end{minipage}
\begin{minipage}[ht]{0.32\linewidth}
\center{\includegraphics[width=1.0\linewidth]{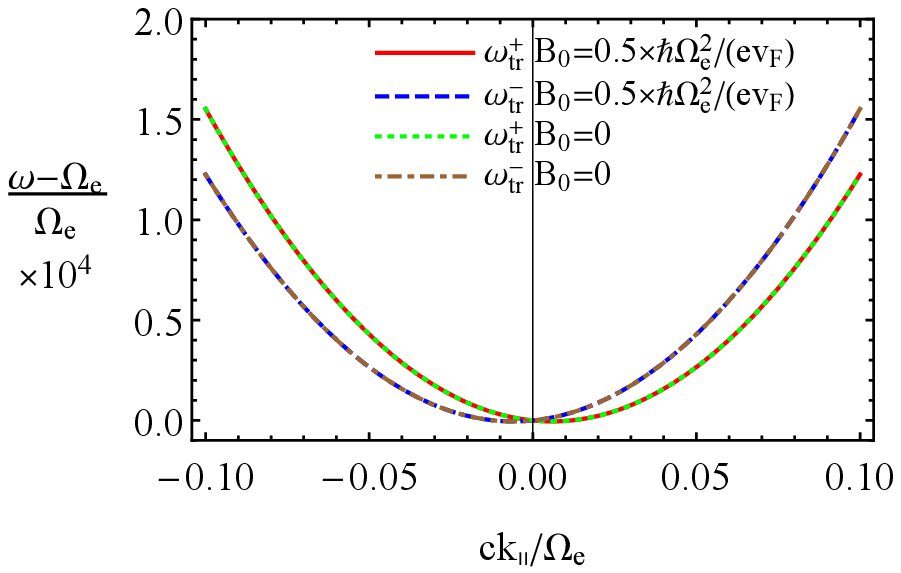} \\
{\small (d) $\mu=0$, $\mu_5=\sqrt{3\pi /(4 \alpha)}\hbar\Omega_e$, $B_{0,5}=0$, $b_{\parallel}=0$}}
\end{minipage}
\begin{minipage}[ht]{0.32\linewidth}
\center{\includegraphics[width=1.0\linewidth]{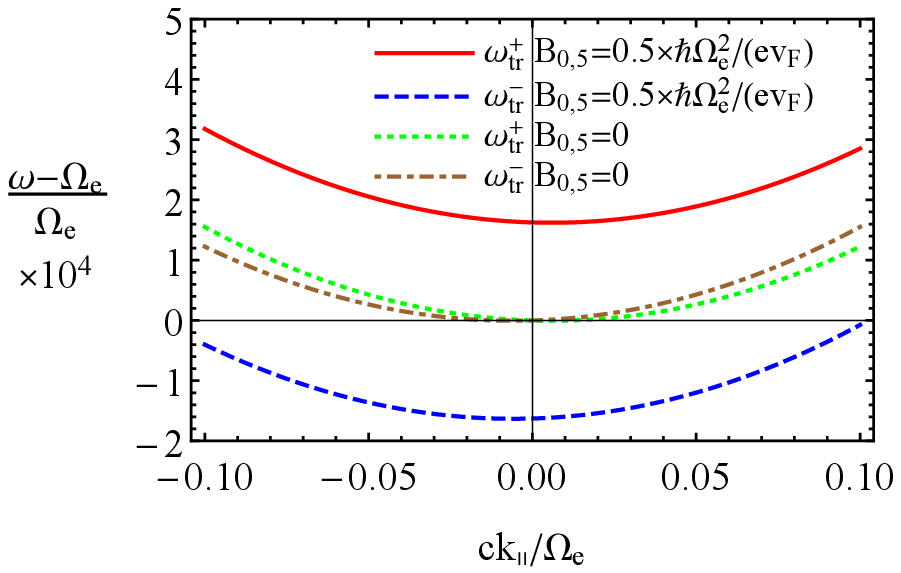} \\
{\small (e) $\mu=0$, $\mu_5=\sqrt{3\pi /(4 \alpha)}\hbar\Omega_e$, $B_0=0$, $b_{\parallel}=0$}}
\end{minipage}
\begin{minipage}[ht]{0.32\linewidth}
\center{\includegraphics[width=1.0\linewidth]{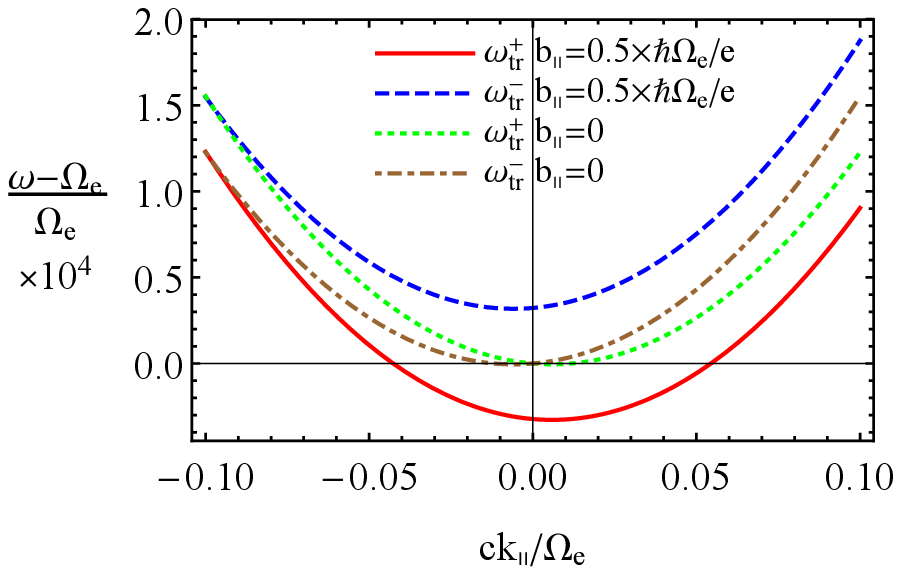} \\
{\small (f) $\mu=0$, $\mu_5=\sqrt{3\pi /(4 \alpha)}\hbar\Omega_e$, $B_0=0$, $B_{0,5}=0$}}
\end{minipage}
\caption{The dispersion relations of the transverse collective modes given by Eq.~(\ref{consistent-B5-Long-Eq2}) at different values of $\mu$, $\mu_5$, $B_0$, $B_{0,5}$, and $b_{\parallel}$. Red solid and green dotted lines correspond to $\omega_{\rm tr}^{+}$. Blue dashed and brown dot-dashed lines correspond to $\omega_{\rm tr}^{-}$. We set $eb_0=-\mu_5$ and $T=0$.}
\label{fig:consistent-long-bz-tr}
\end{figure}

The dispersion relations for the transverse modes $\omega_{\rm tr}^{\pm}$ are shown in Fig.~\ref{fig:consistent-long-bz-tr}
for several different values of $\mu$, $\mu_5$, $B_0$, $B_{0,5}$, and $b_{\parallel}$. In a general case, the degeneracy
of the two transverse modes is lifted. Moreover, as is evident from panels (a) and (e) in Fig.~\ref{fig:consistent-long-bz-tr}
[see also Eq.~(\ref{consistent-B5-Long-Eq2-app})], the presence of an external magnetic $\mathbf{B}_{0}$ (pseudomagnetic $\mathbf{B}_{0,5}$)
field together with a nonzero electric chemical $\mu$ (chiral chemical $\mu_5$) potential splits the plasma frequencies
at $\mathbf{k}=0$. In contrast, as one can see from panels (b) and (d) in Fig.~\ref{fig:consistent-long-bz-tr},
the corresponding plasma frequencies are the same at nonzero magnetic $\mathbf{B}_{0}$ (pseudomagnetic
$\mathbf{B}_{0,5}$) field and finite chiral chemical $\mu_5$ (electric chemical $\mu$) potential. This finding agrees with the
underlying physics behind the splitting of the plasma frequencies, namely, the Lorentz force and its anomalous
counterpart that require the presence of both a nonzero magnetic (pseudomagnetic) field and a finite electric
(chiral) chemical potential.  Another interesting property of the transverse modes is the dependence on
the chiral chemical potential $\mu_5$, which splits the two energies at nonzero wave vectors [see panel (d)
in Fig.~\ref{fig:consistent-long-bz-tr}]. This effect can be traced back to the CME. As one
can see from panels (c) and (f) in Fig.~\ref{fig:consistent-long-bz-tr},
the effect of the chiral shift parameter $b_{\parallel}$ is qualitatively similar to
the effect of the external magnetic $\mathbf{B}_{0}$ (pseudomagnetic $\mathbf{B}_{0,5}$) field applied to the system with nonzero
electric chemical $\mu$ (chiral chemical $\mu_5$) potential.

\begin{figure}[!ht]
\begin{center}
\includegraphics[width=0.45\textwidth]{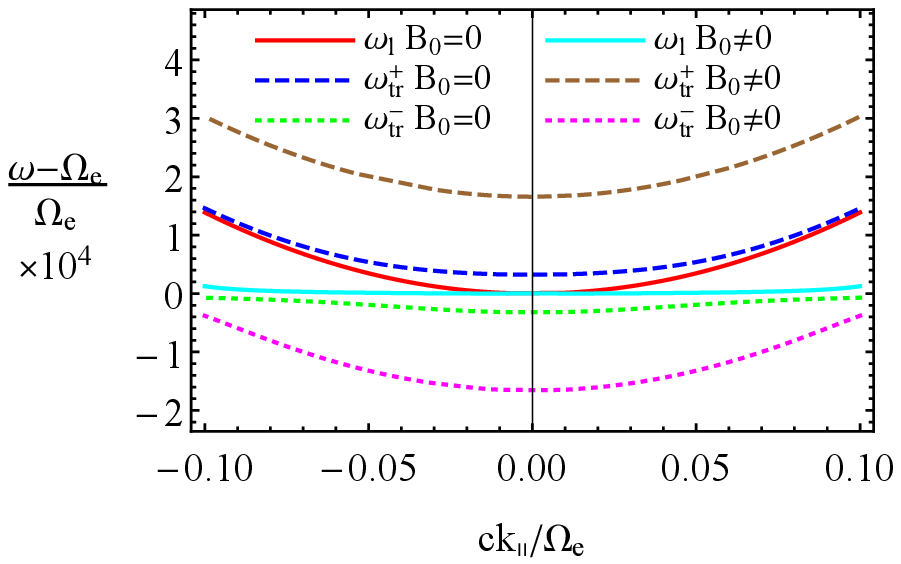}\hfill
\includegraphics[width=0.45\textwidth]{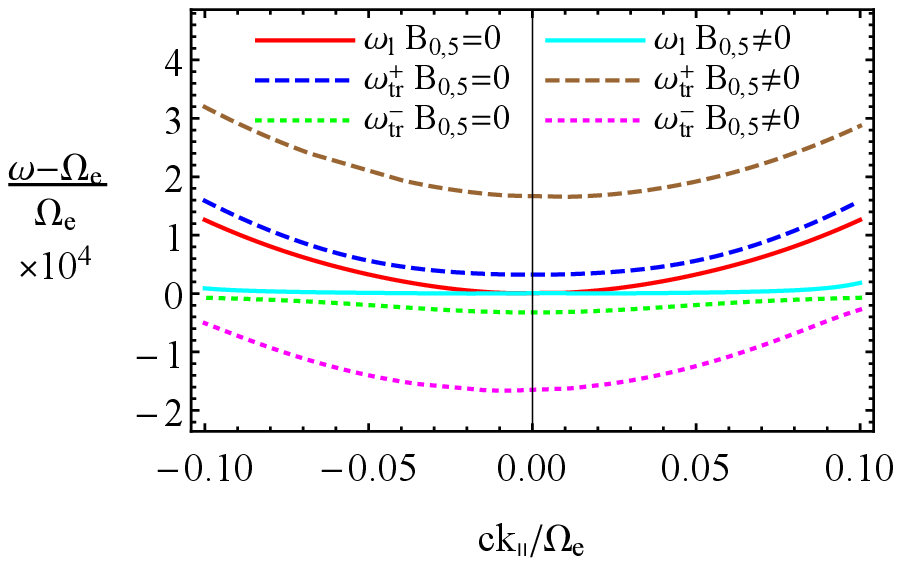}
\caption{The dispersion relations of the collective modes given by Eq.~(\ref{consistent-B5-long-dispersion-relation-1}) for
$\mu=\sqrt{3\pi /(4 \alpha)}\hbar\Omega_e$, $\mu_5=0$ (left panel) and $\mu=0$, $\mu_5=\sqrt{3\pi /(4 \alpha)}\hbar\Omega_e$
(right panel). Red (cyan) solid line corresponds to $\omega_{l}$, blue (brown) dashed one denotes $\omega_{\rm tr}^{+}$, and
green (magenta) dotted one corresponds to $\omega_{\rm tr}^{-}$ at $B_0=0$ [$B_{0}=0.5\,\hbar\Omega_e^2/(v_Fe)$] in the left
panel and $B_{0,5}=0$ [$B_{0,5}=0.5\,\hbar\Omega_e^2/(v_Fe)$] in the right panel. We set $b_\perp=0.5\,\hbar\Omega_e/e$,
$eb_0=-\mu_5$, and $T=0$.}
\label{fig:consistent-long-bx}
\end{center}
\end{figure}

Before concluding this section, it is instructive to consider the dispersion relations of the chiral plasmons
in the case when the chiral shift parameter is perpendicular to the magnetic field. Without loss of generality, we
can choose $\mathbf{b}_\perp$ along the $x$ direction and find the solutions to the spectral equation
(\ref{consistent-B5-long-dispersion-relation-1}) using numerical methods. The results are shown in
Fig.~\ref{fig:consistent-long-bx} at various values of $\mu$, $\mu_5$, $B_0$, and $B_{0,5}$.
As one can see in Fig.~\ref{fig:consistent-long-bx}, a nonzero chiral shift $b_\perp$ significantly changes
the behavior of the collective modes. In addition to lifting the degeneracy of the plasmons, $b_\perp$
mixes longitudinal and transverse modes leading to a much stronger dependence of the former on $k$
(cf. Figs.~\ref{fig:consistent-long-bz-l} and \ref{fig:consistent-long-bx}).
We also find that the magnetic (pseudomagnetic) field applied to the system with a nonzero electric (chiral)
chemical potential enhances the splitting between the longitudinal $\omega_l$ and transverse
$\omega_{\rm tr}^{\pm}$ modes.

\section{Collective modes propagating perpendicular to the magnetic field}
\label{sec:collective-B-arb-tr}

\begin{figure}[!ht]
\begin{center}
\includegraphics[width=0.45\textwidth]{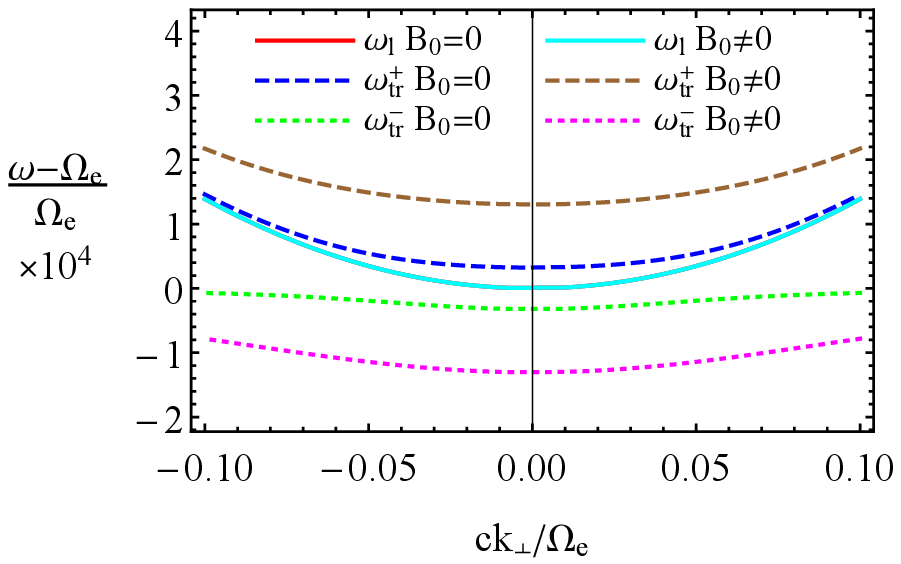}\hfill
\includegraphics[width=0.45\textwidth]{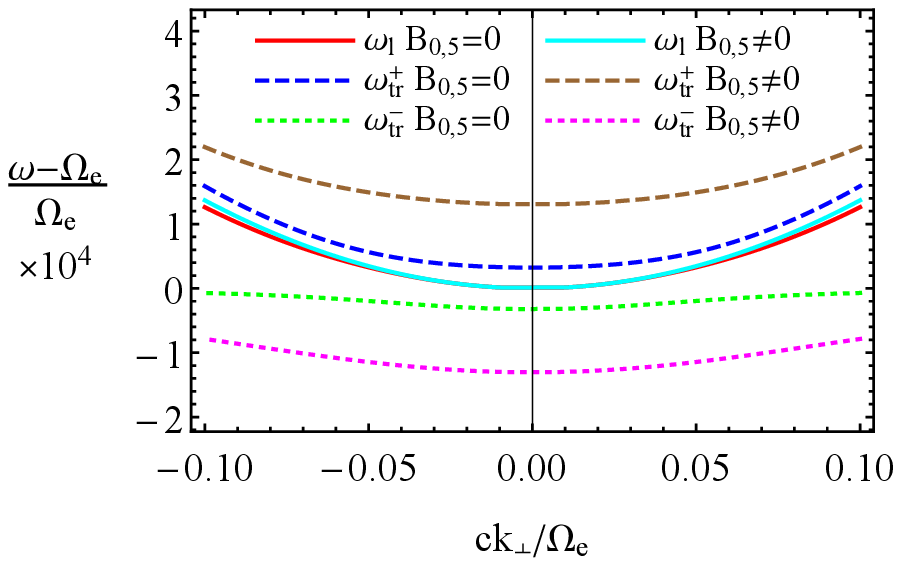}
\caption{The dispersion relations of the collective modes given by Eq.~(\ref{collective-B-tensor-dispersion-relation-theta}) at
$k_{\parallel}=0$ and $b_{\parallel}=0.5\hbar\Omega_e/e$ for $\mu=\sqrt{3\pi /(4 \alpha)}\hbar\Omega_e$, $\mu_5=0$
(left) and $\mu=0$, $\mu_5=\sqrt{3\pi /(4 \alpha)}\hbar\Omega_e$ (right). Red (cyan) solid line corresponds
to $\omega_{l}$, blue (brown) dashed one denotes $\omega_{\rm tr}^{+}$, and green (magenta) dotted one corresponds
to $\omega_{\rm tr}^{-}$ at $B_0=0$ [$B_{0}=0.5\hbar\Omega_e^2/(v_Fe)$] in the left panel and $B_{0,5}=0$
[$B_{0,5}=0.5\hbar\Omega_e^2/(v_Fe)$] in the right panel. We set $eb_0=-\mu_5$ and $T=0$.}
\label{fig:consistent-tr-bz}
\end{center}
\end{figure}

In this section we consider the dispersion relations for collective modes propagating perpendicular
to the magnetic field, i.e., $\mathbf{k}\perp\mathbf{B}_0$. The corresponding characteristic equation
is given by Eq.~(\ref{collective-B-tensor-dispersion-relation-theta}) at $k_{\parallel}=0$. In the long wavelength limit,
$\mathbf{k}=0$, the solutions for the plasma frequencies can be obtained analytically \cite{Gorbar:2016ygi}.
In the general case at $\mathbf{k}\neq 0$, however, the dispersion relations can be obtained only numerically.
The collective mode frequencies $\omega_l$ and $\omega_{\rm tr}^{\pm}$ are plotted in Fig.~\ref{fig:consistent-tr-bz}
for $b_{\parallel}=0.5\,\hbar\Omega_e/e$ and $\mathbf{b}_\perp=0$. As one can see, an external magnetic
(pseudomagnetic) field together with an electric (chiral) chemical potential increases the splitting of the modes.
Note that this finding agrees with the analytical results at $\mathbf{k}=0$ obtained in Ref.~\cite{Gorbar:2016ygi}.
It is also clear from the right panel of Fig.~\ref{fig:consistent-tr-bz} that there is a slightly noticeable effect of the
background pseudomagnetic field on the longitudinal mode $\omega_{l}$ in the system with a finite chiral
chemical potential.

\begin{figure}[!t]
\begin{center}
\includegraphics[width=0.45\textwidth]{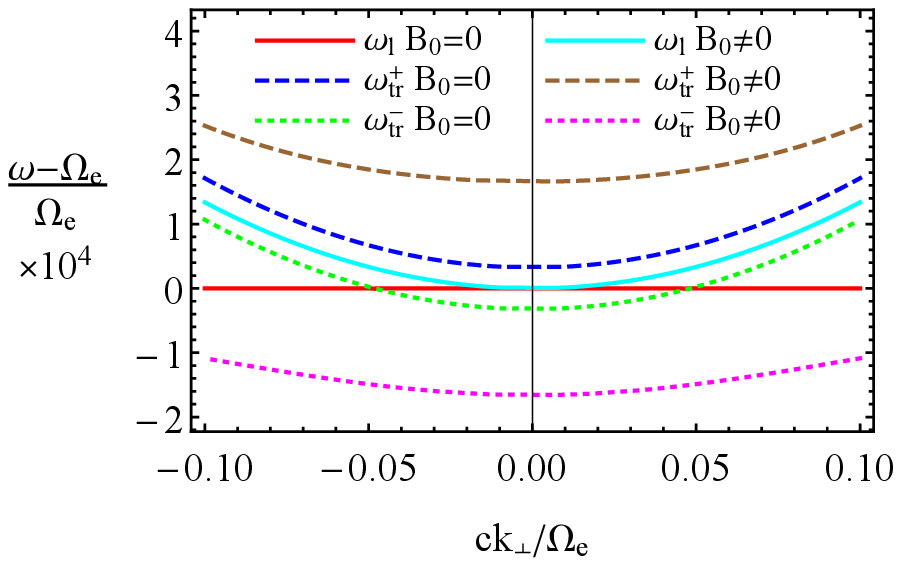}\hfill
\includegraphics[width=0.45\textwidth]{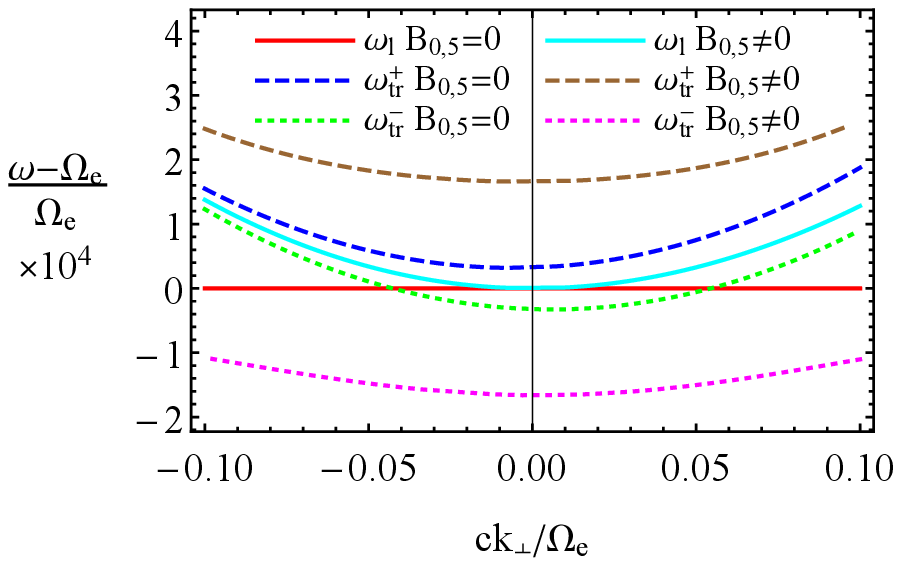}
\includegraphics[width=0.45\textwidth]{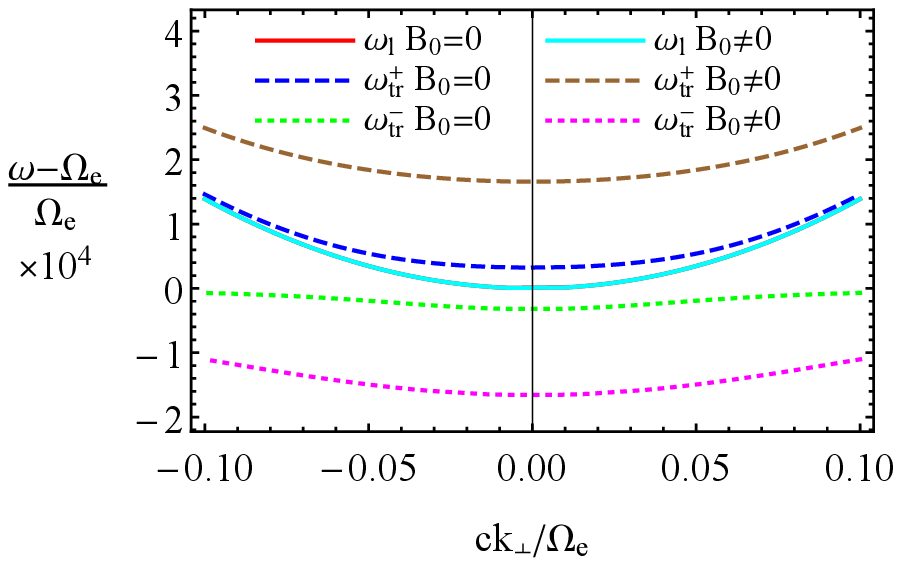}\hfill
\includegraphics[width=0.45\textwidth]{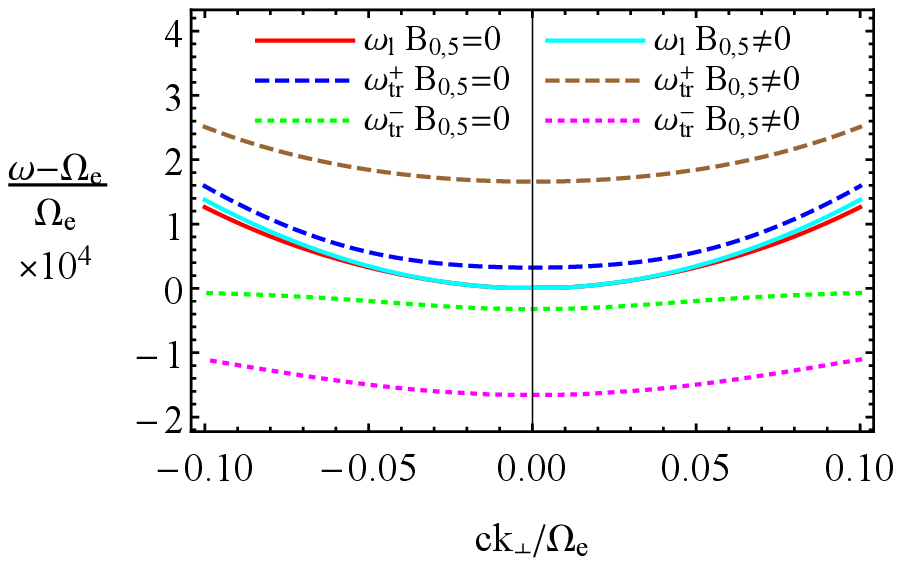}
\caption{The dispersion relations of the collective modes given by Eq.~(\ref{collective-B-tensor-dispersion-relation-theta}) at
$k_{\parallel}=0$ and $\mathbf{b}_{\perp}\neq0$ for $\mu=\sqrt{3\pi /(4 \alpha)}\hbar\Omega_e$, $\mu_5=0$ (left panel) and
$\mu=0$, $\mu_5=\sqrt{3\pi /(4 \alpha)}\hbar\Omega_e$ (right panel). Red (cyan) solid line corresponds to $\omega_{l}$,
blue (brown) dashed one denotes $\omega_{\rm tr}^{+}$, and green (magenta) dotted one corresponds to
$\omega_{\rm tr}^{-}$ at $B_0=0$ [$B_{0}=0.5\hbar\Omega_e^2/(v_Fe)$] in the left panel and $B_{0,5}=0$
[$B_{0,5}=0.5\hbar\Omega_e^2/(v_Fe)$] in the right panel. We set $b_{x}=0.5\,\hbar\Omega_e/e$ (top row),
$b_{y}=0.5\,\hbar\Omega_e/e$ (bottom row), $eb_0=-\mu_5$, and $T=0$.}
\label{fig:consistent-tr-bx-by}
\end{center}
\end{figure}

Now, let us discuss the properties of the collective modes $\omega_l$ and $\omega_{\rm tr}^{\pm}$ for two different orientations
of the chiral shift $\mathbf{b}_\perp$ with respect to the wave vector $\mathbf{k}$ (assuming that $b_\parallel =0$). In the case of
$\mathbf{k}\parallel\mathbf{b}_\perp$, the dispersion relations of the longitudinal and transverse modes are shown in the two top
panels of Fig.~\ref{fig:consistent-tr-bx-by}.
While at zero magnetic or pseudomagnetic field the frequency of the longitudinal mode $\omega_l$ is unaffected by the chiral shift
parameter, the frequencies of the two transverse modes $\omega_{\rm tr}^{\pm}$ split similarly to the case with a nonzero
chiral shift shown in Fig.~\ref{fig:consistent-long-bz-tr}.
The functional dependence of the dispersion relations on the wave vector is somewhat different when a nonzero
magnetic (left panel) or pseudomagnetic field (right panel) is present. Instead of crossing, the frequencies
$\omega_l$ and $\omega_{\rm tr}^{-}$ are repelled from each other with increasing value of $k$. We emphasize
that there is no symmetry with respect to $\mathbf{k}_{\perp}\to-\mathbf{k}_{\perp}$ (i.e., flipping the direction
of the wave vector relative to $\mathbf{b}_{\perp}$) in the two right
panels of Fig.~\ref{fig:consistent-tr-bx-by}.

The results for $\mathbf{k}\perp\mathbf{b}_\perp$ are presented in the two bottom panels of Fig.~\ref{fig:consistent-tr-bx-by}.
The corresponding
dispersion relations are qualitatively similar to those at nonzero $b_{\parallel}$. Indeed, this is evident from comparing the two
bottom panels in Fig.~\ref{fig:consistent-tr-bx-by} with those in Fig.~\ref{fig:consistent-tr-bz}.

\section{Summary and discussion}
\label{sec:Summary-Discussions}

By using the consistent chiral kinetic theory, we considered the spectrum of the gapped chiral \mbox{(pseudo-)magnetic}
plasmons in three-dimensional Weyl materials at nonzero wave vector $\mathbf{k}$ and in the linear order in background
magnetic and strain-induced pseudomagnetic fields. It came as a surprise that, even at the linear order,
the pseudomagnetic field affects the frequency of the longitudinal mode, i.e., the mode whose wave vector is parallel
to the external magnetic fields, as well as to the oscillating electric field. Despite being weak, this dependence is rather
uncommon and was not predicted before. We call this new type of collective excitations \emph{the chiral pseudomagnetic
plasmon}.

In the case of nonzero wave vector, the chiral nature of the chiral magnetic and pseudomagnetic plasmons is
manifested in the oscillations of the local chiral charge density, which are absent for ordinary electromagnetic
plasmons. These oscillations have a topological origin that stems from the dynamical version of the chiral electric
separation effect and chiral anomaly. Their topological nature is also manifested in
the absence of the temperature dependence. Interestingly, the oscillations of the electric charge density are also
unusual as they depend both on the pseudomagnetic field and the chiral shift parameter. Last but not least, we
found that contrary to the common belief, the oscillation amplitudes of the left- and right-handed fermion
densities in the longitudinal plasmon
(which is the CMW with the effects of dynamical electromagnetism taken into account) are different and
depend on the chiral shift, magnetic, and pseudomagnetic fields.

As we show, the presence of the magnetic (pseudomagnetic) field \emph{and} the electric chemical (chiral chemical) potential
results in a splitting of collective plasmon modes in Weyl materials. Moreover, not only all three plasmon frequencies
become completely nondegenerate, but also the dependence on the wave vector $\mathbf{k}$ in the dispersion relations
is affected by the background fields. Additionally, in Weyl materials with a nonzero energy separation between the Weyl nodes $b_0$, the two degenerate branches of the spectrum split even in the absence of external magnetic or pseudomagnetic field. It is worth noting that $\mu_5=-e b_0$ in the equilibrium state of such a system. This result is in agreement
with that obtained in Ref.~\cite{Akamatsu:2013} and can be explained by the dynamical version of the CME.
However, unlike the pseudomagnetic field, the chiral chemical potential does not affect the plasmon frequencies (gaps) in
the zero wave vector limit $\mathbf{k}=0$. At nonzero wave vectors, the size of the splitting increases approximately linearly
with $k$.

In this study, we also investigated the effects of the chiral shift $\mathbf{b}$ on the chiral plasmon properties.
Similarly to the external magnetic (pseudomagnetic) field with nonzero electric (chiral) chemical potential, the chiral shift
lifts the degeneracy of the plasmon modes. Moreover, it strongly affects the longitudinal plasmon mode by
mixing it with the transverse ones. It is worth noting that the dependence of the dispersion relations on
$\mathbf{b}$ is the characteristic feature of the consistent chiral kinetic theory since it originates exclusively
from the topological correction in the electric current.

In summary, the key results of this study relevant for experiments are as follows. (1) In Weyl materials with a finite electric chemical potential $\mu$, but without external magnetic field $\mathbf{B}_0$,
pseudomagnetic field $\mathbf{B}_{0,5}$, and chiral chemical potential $\mu_5$, there are three
gapped plasmon modes, two of which are degenerate.
(2) The chiral chemical potential $\mu_5$ leads to a splitting of the degenerate plasmon modes.
The amplitude of the splitting increases approximately linearly with wave vector $k$ and vanishes at $k=0$.
(3) An external magnetic field $\mathbf{B}_0$ (a pseudomagnetic field $\mathbf{B}_{0,5}$) applied to a
system with a finite $\mu$ ($\mu_5$) also leads to a splitting of the plasmon modes. Unlike the case
of a chiral chemical potential $\mu_5$ in result 2, this splitting remains nonzero even in the limit of vanishing wave vector.
(4) Similarly, a nonzero chiral shift $\mathbf{b}$ lifts the degeneracy of the plasmon modes. In the
case when $\mathbf{b}$ is parallel to $\mathbf{B}_{0}$ and $\mathbf{k}$, only the transverse modes are affected.

As is clear, all qualitative predictions of this study can be straightforwardly tested in future
experiments. We hope that such experiments will finally resolve the issue of the consistent
versus covariant implementations of the chiral anomaly in the chiral kinetic theory, as well as produce
abundant data in support of the quantum anomalies in condensed matter materials.

\begin{acknowledgments}
The work of E.V.G. was partially supported by the Program of Fundamental
Research of the Physics and Astronomy Division of the NAS of Ukraine.
The work of V.A.M. and P.O.S. was supported by the Natural Sciences and
Engineering Research Council of Canada.
The work of I.A.S. was supported by the U.S. National Science Foundation under Grant No.~PHY-1404232.
\end{acknowledgments}

\appendix

\section{Useful formulas and relations}
\label{sec:App-ref}

By making use of the short-hand notation $f^{(0)}_{\lambda} =1/[e^{(\epsilon_{\mathbf{p}}-\mu_{\lambda})/T}+1]$
with $\epsilon_{\mathbf{p}}=v_Fp$, where $p=|\mathbf{p}|$, it is straightforward to derive the following formulas:
\begin{eqnarray}
\int\frac{d^3p}{(2\pi)^3} p^{n-2}  f^{(0)}_{\lambda}
&=& -\frac{T^{n+1} \Gamma(n+1) }{2\pi^2 v_F^{n+1}}  \mbox{Li}_{n+1}\left(-e^{\mu_{\lambda}/T}\right),
\qquad n\geq 0,
\label{integral-3a} \\
\int\frac{d^3p}{(2\pi)^3} p^{n-2} \frac{\partial f^{(0)}_{\lambda}}{\partial \epsilon_{\mathbf{p}}}
&=& \frac{T^{n} \Gamma(n+1) }{2\pi^2 v_F^{n+1}}  \mbox{Li}_{n}\left(-e^{\mu_{\lambda}/T}\right),
\qquad n\geq 0,
\label{integral-3b}
\end{eqnarray}
where $T\partial f^{(0)}_{\lambda}/\partial \epsilon_{\mathbf{p}}
=-T\partial f^{(0)}_{\lambda}/\partial \mu_{\lambda}
=-e^{(\epsilon_{\mathbf{p}}-\mu_{\lambda})/T}/[e^{(\epsilon_{\mathbf{p}}-\mu_{\lambda})/T}+1]^2$ and $\mbox{Li}_{n}(x)$ is the polylogarithm function (see formula 1.1.14
in Ref.~\cite{Erdelyi:Vol1}). [Note that in the given reference $\mathrm{Li}_n(x) \equiv \mathrm{F}(x, n)$.] The polylogarithm function at
$n=0,1$ can be rewritten as follows:
\begin{eqnarray}
\mbox{Li}_{0}\left(-e^{x}\right) &=& -\frac{1}{1+e^{-x}}, \\
\mbox{Li}_{1}\left(-e^{x}\right) &=& -\ln{\left(1+e^{x}\right)}.
\label{App-polylog}
\end{eqnarray}
The following identities for the polylogarithm functions are useful when taking into account the antiparticles contributions:
\begin{eqnarray}
\ln(1+e^{x})-\ln(1+e^{-x}) &=& x,\\
\mbox{Li}_{2} (-e^{x}) +\mbox{Li}_{2} (-e^{-x})   &=& -\frac{x^2}{2}-\frac{\pi^2}{6},
\\
\mbox{Li}_{3} (-e^{x}) - \mbox{Li}_{3} (-e^{-x})   &=&  -\frac{x^3}{6}-\frac{\pi^2 x}{6}.
\end{eqnarray}

By integrating over the angular coordinates, one can derive the following general relations:
\begin{eqnarray}
\int \frac{d^3 p}{(2\pi)^3}  \mathbf{p} \, f(p^2)&=& 0,\\
\int \frac{d^3 p}{(2\pi)^3}  \mathbf{p} (\mathbf{p}\cdot \mathbf{a}) f(p^2) &=& \frac{\mathbf{a}}{3}
\int \frac{d^3 p}{(2\pi)^3}  p^2 f(p^2),\\
\int \frac{d^3 p}{(2\pi)^3}  \mathbf{p} (\mathbf{p}\cdot \mathbf{a}) (\mathbf{p}\cdot \mathbf{b})  f(p^2) &=& 0,
\end{eqnarray}
and, similarly,
\begin{eqnarray}
\label{App-integral-p-angle-be}
\int d\xi d\phi \left( \hat{\mathbf{p}} \cdot \mathbf{a} \right) e^{i\left( \hat{\mathbf{p}} \cdot \mathbf{k}\right)}
&=& 4\pi i \frac{\left(\mathbf{k} \cdot \mathbf{a} \right) }{k^2} \left(\frac{\sin k}{k}-\cos k \right),\\
\int d\xi d\phi \left( \hat{\mathbf{p}} \cdot \mathbf{a} \right) \left( \hat{\mathbf{p}} \cdot \mathbf{b} \right) e^{i\left( \hat{\mathbf{p}}
\cdot \mathbf{k}\right)}
&=&  4\pi \frac{\left(\mathbf{a} \cdot \mathbf{b} \right) }{k^2} \left(\frac{\sin k}{k}-\cos k \right)
+ 12\pi \frac{\left(\mathbf{a} \cdot \mathbf{k} \right) \left(\mathbf{b} \cdot \mathbf{k} \right)}{k^4}
\left(\cos k -\frac{\sin k}{k}+\frac{k}{3} \sin k \right) ,\\
\int d\xi d\phi \left( \hat{\mathbf{p}} \cdot \mathbf{a} \right) \left( \hat{\mathbf{p}} \cdot \mathbf{b} \right) \left( \hat{\mathbf{p}} \cdot
\mathbf{c} \right) e^{i\left( \hat{\mathbf{p}} \cdot \mathbf{k}\right)}
&=&  -4\pi i \frac{\left(\mathbf{a} \cdot \mathbf{k} \right)\left(\mathbf{b} \cdot \mathbf{c} \right) +\left(\mathbf{b} \cdot \mathbf{k} \right)
\left(\mathbf{a} \cdot \mathbf{c} \right) +\left(\mathbf{c} \cdot \mathbf{k} \right)\left(\mathbf{a} \cdot \mathbf{b} \right)}{k^3} \left(\sin k
+\frac{3\cos k}{k} -\frac{3\sin k}{k^2} \right) \nonumber\\
&-& 4\pi i \frac{\left(\mathbf{a} \cdot \mathbf{k} \right) \left(\mathbf{b} \cdot \mathbf{k} \right)\left(\mathbf{c} \cdot \mathbf{k} \right)}
{k^4}
\left(\cos k -\frac{6\sin k}{k} -\frac{15\cos k}{k^2} +\frac{15\sin k}{k^3}\right),
\end{eqnarray}
\begin{eqnarray}
\int d\xi d\phi  (\hat{\mathbf{p}}\cdot \mathbf{a}) (\hat{\mathbf{p}}\cdot \mathbf{b}) (\hat{\mathbf{p}}\cdot \mathbf{c})  (\hat{\mathbf{p}}\cdot \mathbf{d}) e^{i\left( \hat{\mathbf{p}} \cdot \mathbf{k}\right)}
&=&  -4\pi \frac{(\mathbf{a}\cdot \mathbf{b})(\mathbf{c}\cdot \mathbf{d})
+ (\mathbf{a}\cdot \mathbf{c})(\mathbf{b}\cdot \mathbf{d})
+(\mathbf{a}\cdot \mathbf{d})(\mathbf{b}\cdot \mathbf{c}) }{k^3} \left(\sin k +\frac{3\cos k}{k} -\frac{3\sin k}{k^2} \right) \nonumber\\
&+& 4\pi \frac{(\mathbf{a} \cdot \mathbf{b})(\mathbf{c}\cdot \mathbf{k})(\mathbf{d}\cdot \mathbf{k})
+(\mathbf{a} \cdot \mathbf{c})(\mathbf{b}\cdot \mathbf{k})(\mathbf{d}\cdot \mathbf{k})
+(\mathbf{a} \cdot \mathbf{d})(\mathbf{b}\cdot \mathbf{k})(\mathbf{c}\cdot \mathbf{k})}{k^4} \nonumber\\
&\times&\left(\cos k -\frac{6\sin k}{k} -\frac{15\cos k}{k^2} +\frac{15\sin k}{k^3}\right) \nonumber\\
&+& 4\pi \frac{(\mathbf{b} \cdot \mathbf{c})(\mathbf{a}\cdot \mathbf{k})(\mathbf{d}\cdot \mathbf{k})
+(\mathbf{b} \cdot \mathbf{d})(\mathbf{a}\cdot \mathbf{k})(\mathbf{c}\cdot \mathbf{k})
+(\mathbf{c} \cdot \mathbf{d})(\mathbf{a}\cdot \mathbf{k})(\mathbf{b}\cdot \mathbf{k})}{k^4}\nonumber\\
&\times&\left(\cos k -\frac{6\sin k}{k} -\frac{15\cos k}{k^2} +\frac{15\sin k}{k^3}\right)
+ 4\pi \frac{(\mathbf{a}\cdot \mathbf{k})(\mathbf{b}\cdot \mathbf{k})(\mathbf{c}\cdot \mathbf{k})(\mathbf{d}\cdot \mathbf{k})}{k^5} \nonumber\\
&\times&\left(\sin k +\frac{10\cos k}{k} -\frac{45\sin k}{k^2} -\frac{105\cos k}{k^3} +\frac{105\sin k}{k^3}\right),
\label{App-integral-p-angle-ee}
\end{eqnarray}
together with
\begin{eqnarray}
\int \frac{d^3 p}{(2\pi)^3}  \mathbf{p} \, {G}(p^2,\mathbf{p}\cdot \mathbf{B})&=&
 \frac{\mathbf{B} }{B}  \int \frac{dp d\xi d\phi}{(2\pi)^3}  \, p^3 \xi  \, {G}(p^2,p B \xi) ,\\
\int \frac{d^3 p}{(2\pi)^3}  \mathbf{p} (\mathbf{p}\cdot \mathbf{a}) {G}(p^2,\mathbf{p}\cdot \mathbf{B}) &=&
\int \frac{dp d\xi d\phi}{(2\pi)^3} p^4 \left[
\mathbf{a}  \frac{1-\xi^2}{2}
+\frac{\mathbf{B}(\mathbf{a}\cdot \mathbf{B}) }{B^2}  \frac{3\xi^2-1}{2}
\right]  {G}(p^2,p B \xi) ,\\
\int \frac{d^3 p}{(2\pi)^3}  (\mathbf{p}\cdot \mathbf{a}) (\mathbf{p}\cdot \mathbf{b}) (\mathbf{p}\cdot \mathbf{c})
\, {G}(p^2,\mathbf{p}\cdot \mathbf{B})&=& \int \frac{dp d\xi d\phi}{(2\pi)^3} p^5 \left[
\frac{(\mathbf{a}\cdot \mathbf{B})(\mathbf{b}\cdot \mathbf{c}) + (\mathbf{b}\cdot \mathbf{B})(\mathbf{a}\cdot \mathbf{c}) +(\mathbf{c}\cdot \mathbf{B})(\mathbf{a}\cdot \mathbf{b}) }{B} \frac{\xi(1-\xi^2)}{2} \right. \nonumber\\
&&
\left. +\frac{(\mathbf{a}\cdot \mathbf{B})(\mathbf{b}\cdot \mathbf{B})(\mathbf{c}\cdot \mathbf{B}) }{B^3}  \frac{\xi(5\xi^2-3)}{2}
\right]  {G}(p^2,p B \xi),
\end{eqnarray}
\begin{eqnarray}
&&\int \frac{d^3 p}{(2\pi)^3}  (\mathbf{p}\cdot \mathbf{a}) (\mathbf{p}\cdot \mathbf{b}) (\mathbf{p}\cdot \mathbf{c})  (\mathbf{p}\cdot \mathbf{d})
\, {G}(p^2,\mathbf{p}\cdot \mathbf{B}) = \int \frac{dp d\xi d\phi}{(2\pi)^3} \, p^6 \Bigg\{
\left[(\mathbf{a}\cdot \mathbf{b})(\mathbf{c}\cdot \mathbf{d})
+ (\mathbf{a}\cdot \mathbf{c})(\mathbf{b}\cdot \mathbf{d})
+(\mathbf{a}\cdot \mathbf{d})(\mathbf{b}\cdot \mathbf{c}) \right]\frac{(1-\xi^2)^2}{8}  \nonumber\\
&&
-\frac{(\mathbf{a} \cdot \mathbf{b})(\mathbf{c}\cdot \mathbf{B})(\mathbf{d}\cdot \mathbf{B})
+(\mathbf{a} \cdot \mathbf{c})(\mathbf{b}\cdot \mathbf{B})(\mathbf{d}\cdot \mathbf{B})
+(\mathbf{a} \cdot \mathbf{d})(\mathbf{b}\cdot \mathbf{B})(\mathbf{c}\cdot \mathbf{B})}{B^2}
\frac{(1-6\xi^2+5\xi^4)}{8} \nonumber\\
&&
-\frac{(\mathbf{b} \cdot \mathbf{c})(\mathbf{a}\cdot \mathbf{B})(\mathbf{d}\cdot \mathbf{B})
+(\mathbf{b} \cdot \mathbf{d})(\mathbf{a}\cdot \mathbf{B})(\mathbf{c}\cdot \mathbf{B})
+(\mathbf{c} \cdot \mathbf{d})(\mathbf{a}\cdot \mathbf{B})(\mathbf{b}\cdot \mathbf{B})
}{B^2}
\frac{(1-6\xi^2+5\xi^4)}{8} \nonumber\\
&&
+\frac{(\mathbf{a}\cdot \mathbf{B})(\mathbf{b}\cdot \mathbf{B})(\mathbf{c}\cdot \mathbf{B})(\mathbf{d}\cdot \mathbf{B}) }{B^4}
\frac{(3-30\xi^2+35\xi^4)}{8}
\Bigg\}  {G}(p^2,p B \xi) .
\end{eqnarray}
Here, $\xi=\cos\theta$ is the cosine of the angle between vectors $\mathbf{B}$ and $\mathbf{p}$, and $\hat{\mathbf{p}}=\mathbf{p}/p$.

In the limit of vanishing external effective magnetic field, $\mathbf{B}_{0,\lambda}\to 0$, the main contribution to the integrals over $\tau$ in
Sec.~\ref{sec:collective-B} comes from the region of small $\tau$. Expanding $\sin{\tau}$ and $\cos{\tau}$ in Eqs.~(\ref{collective-B-Omega-tau-be}) through (\ref{collective-B-K2-tau-ee}) in powers of $\tau$, one can obtain the following expressions:
\begin{eqnarray}
\label{collective-B-tensor-Ktau}
\mathbf{K}_{\tau} &\approx& \mathbf{k}\tau+(\hat{\mathbf{z}} \times \mathbf{k} )\frac{\tau^2}{2}
+\frac{\lambda \hbar \omega }{2c v_F p^2} e\mathbf{B}_{0,\lambda} \tau +O(\tau^3),\quad
\mathbf{K}_{\tau}^{(1)} \approx \mathbf{k}\left(1-\frac{\tau^2}{2}\right)+(\hat{\mathbf{z}} \times \mathbf{k})\tau
+\hat{\mathbf{z}}(\hat{\mathbf{z}} \cdot \mathbf{k})\frac{\tau^2}{2} +O(\tau^3),\\
\mathbf{E}_{\tau}&\approx& \mathbf{E}\left(1-\frac{\tau^2}{2}\right) + (\hat{\mathbf{z}} \times \mathbf{E} )\tau
+\hat{\mathbf{z}}(\mathbf{E}\cdot \hat{\mathbf{z}}) \frac{\tau^2}{2} +O(\tau^3),
\end{eqnarray}
and
\begin{eqnarray}
\label{collective-B-tensor-Ktau2}
K_{\tau}^2 &\approx&  \left(k_\parallel +\lambda\hbar \frac{\omega eB_{0,\lambda}}{2cv_Fp^2}\right)^2\tau^2+k_\perp^2 \tau^2+O(\tau^3) \approx K^2 \tau^2,\quad
K \approx  k\left(1+\lambda\hbar \frac{\omega eB_{0,\lambda}}{2cv_Fp^2} \frac{k_{\parallel}}{k^2}\right) +O(\tau^2),
\end{eqnarray}
\begin{eqnarray}
\label{collective-B-tensor-KEtau}
(\mathbf{K}_{\tau} \cdot \mathbf{E}_{\tau} )
&\approx& (\mathbf{k}\cdot \mathbf{E}) \tau
+(\hat{\mathbf{z}}\cdot [\mathbf{E}\times \mathbf{k}])\frac{\tau^2}{2} +(\hat{\mathbf{z}}\cdot \mathbf{E})\frac{\lambda \hbar\omega eB_{0,\lambda}}
{2cv_Fp^2} \tau +O(\tau^3),\\
\label{collective-B-tensor-KKtau}
\mathbf{K}_{\tau} (\mathbf{K}_{\tau} \cdot \mathbf{E}_{\tau} )
&\approx& \mathbf{k} (\mathbf{k}\cdot\mathbf{E})\tau^2
+\mathbf{k}(\hat{\mathbf{z}}\cdot[\mathbf{E}\times\mathbf{k}]) \frac{\tau^3}{2} + \frac{\lambda \hbar \omega eB_{0,\lambda}}{2cv_Fp^2}
\mathbf{k}(\hat{\mathbf{z}}\cdot\mathbf{E}) \tau^2 +[\hat{\mathbf{z}}\times\mathbf{k}](\mathbf{k}\cdot\mathbf{E})\frac{\tau^3}{2} \nonumber\\
&+& [\hat{\mathbf{z}}\times\mathbf{k}](\hat{\mathbf{z}}\cdot\mathbf{E})\frac{\lambda \hbar \omega eB_{0,\lambda}}{2cv_Fp^2} \frac{\tau^3}{2} +
\frac{\lambda \hbar \omega eB_{0,\lambda}}{2cv_Fp^2} \hat{\mathbf{z}}(\mathbf{k}\cdot\mathbf{E})\tau^2 +O(\tau^4),\\
\label{collective-B-tensor-kEtau}
\left[ \mathbf{k} \times \mathbf{E}_{\tau} \right] &\approx& [\mathbf{k}\times\mathbf{E}]\left(1-\frac{\tau^2}{2}\right) + \hat{\mathbf{z}}(\mathbf{k}\cdot
\mathbf{E}) \tau - \mathbf{E}(\hat{\mathbf{z}}\cdot \mathbf{k})\tau +O(\tau^3), \\
\label{collective-B-tensor-kKtau}
\left[\mathbf{k}\times\mathbf{K}_{\tau}\right](\mathbf{K}_{\tau} \cdot \mathbf{E}_{\tau}) &\approx& \hat{\mathbf{z}} (\mathbf{k}\cdot\mathbf{E})
k^2 \frac{\tau^3}{2} -\mathbf{k}(\hat{\mathbf{z}}\cdot\mathbf{k})(\mathbf{k}\cdot\mathbf{E}) \frac{\tau^3}{2}
+\frac{\lambda \hbar\omega eB_{0,\lambda}}{2cv_Fp^2} [\mathbf{k}\times\hat{\mathbf{z}}](\mathbf{k}\cdot\mathbf{E}) \tau^2 +O(\tau^4).
\end{eqnarray}
Then, the problem reduces to the calculation of table integrals, which read as
\begin{eqnarray}
\label{App-int-tau-be}
I_0&=&  \int_{0}^{-s_B\infty}\frac{d\tau}{\tau^2}   e^{-i  \tau \tilde{P}_{\lambda}\tilde{\omega} }
 \left(\frac{\sin(\tilde{P}_{\lambda}\tilde{K}\tau)}{\tilde{P}_{\lambda}\tilde{K}\tau}
 -\cos(\tilde{P}_{\lambda}\tilde{K}\tau)\right)
  = -\frac{i \tilde{P}_{\lambda}\tilde{\omega}}{2} \left[1+\frac{1}{2}\left(\frac{\tilde{K}}{\tilde{\omega}}-\frac{\tilde{\omega}}{\tilde{K}}
  \right) \ln{\left(\frac{\tilde{\omega}+\tilde{K}}{\tilde{\omega}-\tilde{K}}\right)}\right],\\
 I_1&=&  \int_{0}^{-s_B\infty} \frac{d\tau}{\tau} e^{-i  \tau \tilde{P}_{\lambda}\tilde{\omega} }
 \left(\frac{\sin(\tilde{P}_{\lambda}\tilde{K}\tau)}{\tilde{P}_{\lambda}\tilde{K}\tau}
 -\cos(\tilde{P}_{\lambda}\tilde{K}\tau)\right)
 =  \frac{i}{\tilde{P}_{\lambda}}\frac{\partial I_0}{\partial \tilde{\omega}}
= 1-\frac{\tilde{\omega}}{2\tilde{K}}\ln{\left(\frac{\tilde{\omega}+\tilde{K}}{\tilde{\omega}-\tilde{K}}\right)},\\
I_2&=& \int_{0}^{-s_B\infty} d\tau e^{-i  \tau \tilde{P}_{\lambda}\tilde{\omega}}
 \left(\frac{\sin(\tilde{P}_{\lambda}\tilde{K}\tau)}{\tilde{P}_{\lambda}\tilde{K}\tau}
 -\cos(\tilde{P}_{\lambda}\tilde{K}\tau)\right)
 =  \frac{i}{\tilde{P}_{\lambda}}\frac{\partial I_1}{\partial \tilde{\omega}}
 =-\frac{i\tilde{\omega}}{\tilde{P}_{\lambda}\left(\tilde{K}^2-\tilde{\omega}^2\right)}
 -\frac{i}{2\tilde{P}_{\lambda}\tilde{K}}\ln{\left(\frac{\tilde{\omega}+\tilde{K}}{\tilde{\omega}-\tilde{K}}\right)},\\
  I_3 &=&  \int_{0}^{-s_B\infty} \frac{d\tau}{\tau^2} e^{-i  \tau \tilde{P}_{\lambda}\tilde{\omega}}
 \left(
  \cos(\tilde{P}_{\lambda}\tilde{K}\tau)
  -\frac{\sin(\tilde{P}_{\lambda}\tilde{K}\tau)}{\tilde{P}_{\lambda}\tilde{K}\tau}
  +\frac{\tilde{P}_{\lambda}\tilde{K}\tau}{3}\sin(\tilde{P}_{\lambda}\tilde{K}\tau)
  \right) \nonumber\\
  &=&
  \frac{i \tilde{P}_{\lambda}\tilde{\omega}}{2} \left[1+\frac{1}{2}\left(\frac{\tilde{K}}{3\tilde{\omega}}-\frac{\tilde{\omega}}{\tilde{K}}\right) \ln{\left(\frac{\tilde{\omega}+\tilde{K}}{\tilde{\omega}-\tilde{K}}\right)} \right],\\
   I_4 &=&  \int_{0}^{-s_B\infty} \frac{d\tau}{\tau} e^{-i  \tau \tilde{P}_{\lambda}\tilde{\omega} }
 \left(
  \cos(\tilde{P}_{\lambda}\tilde{K}\tau)
  -\frac{\sin(\tilde{P}_{\lambda}\tilde{K}\tau)}{\tilde{P}_{\lambda}\tilde{K}\tau}
  +\frac{\tilde{P}_{\lambda}\tilde{K}\tau}{3}\sin(\tilde{P}_{\lambda}\tilde{K}\tau)
  \right)
 =  \frac{i}{\tilde{P}_{\lambda}}\frac{\partial I_3}{\partial \tilde{\omega}} \nonumber\\
 &=&-\frac{2\tilde{K}^2-3\tilde{\omega}^2}{3\left(\tilde{K}^2-\tilde{\omega}^2\right)}
 +\frac{\tilde{\omega}}{2\tilde{K}}\ln{\left(\frac{\tilde{\omega}+\tilde{K}}{\tilde{\omega}-\tilde{K}}\right)},
 \end{eqnarray}
\begin{eqnarray}
  I_5 &=&  \int_{0}^{-s_B\infty} d\tau e^{-i  \tau \tilde{P}_{\lambda}\tilde{\omega} }
 \left(
  \cos(\tilde{P}_{\lambda}\tilde{K}\tau)
  -\frac{\sin(\tilde{P}_{\lambda}\tilde{K}\tau)}{\tilde{P}_{\lambda}\tilde{K}\tau}
  +\frac{\tilde{P}_{\lambda}\tilde{K}\tau}{3}\sin(\tilde{P}_{\lambda}\tilde{K}\tau)
  \right)
 =  \frac{i}{\tilde{P}_{\lambda}}\frac{\partial I_4}{\partial \tilde{\omega}} \nonumber\\
 &=&i \frac{\tilde{\omega}\left(5\tilde{K}^2-3\tilde{\omega}^2\right)}{3\tilde{P}_{\lambda}\left(\tilde{K}^2-\tilde{\omega}^2\right)^2}
 +\frac{i}{2\tilde{K}\tilde{P}_{\lambda}}\ln{\left(\frac{\tilde{\omega}+\tilde{K}}{\tilde{\omega}-\tilde{K}}\right)},\\
  I_6 &=& \int_{0}^{-s_B\infty} \frac{d\tau}{\tau^2} e^{-i \tau \tilde{P}_{\lambda}\tilde{\omega} }
 \left(
  \sin(\tilde{P}_{\lambda}\tilde{K}\tau)
  +\frac{3\cos(\tilde{P}_{\lambda}\tilde{K}\tau)}{\tilde{P}_{\lambda}\tilde{K}\tau}
  -\frac{3\sin(\tilde{P}_{\lambda}\tilde{K}\tau)}{(\tilde{P}_{\lambda}\tilde{K}\tau)^2}
  \right) \nonumber\\
 &=&-\frac{1}{3\tilde{K}\tilde{P}_{\lambda}}\Bigg[1-\frac{3\tilde{\omega}^2}{2\tilde{K}^2} +\frac{3\tilde{\omega}}{\tilde{K}}\left(1-
 \frac{\tilde{\omega}^2}{\tilde{K}^2}\right) \ln{\left(\frac{\tilde{\omega}+\tilde{K}}{\tilde{\omega}-\tilde{K}}\right)} \Bigg],\\
I_7 &=& \int_{0}^{-s_B\infty} \frac{d\tau}{\tau} e^{-i  \tau \tilde{P}_{\lambda}\tilde{\omega} }
 \left(
  \cos(\tilde{P}_{\lambda}\tilde{K}\tau)
  -\frac{6\sin(\tilde{P}_{\lambda}\tilde{K}\tau)}{\tilde{P}_{\lambda}\tilde{K}\tau}
  -\frac{15\cos(\tilde{P}_{\lambda}\tilde{K}\tau)}{(\tilde{P}_{\lambda}\tilde{K}\tau)^2} +\frac{15\sin(\tilde{P}_{\lambda}\tilde{K}\tau)}
  {(\tilde{P}_{\lambda}\tilde{K}\tau)^3}
  \right) \nonumber\\
 &=&\frac{2}{3\tilde{K}\tilde{P}_{\lambda}}\Bigg[1-\frac{15\tilde{\omega}^2}{4\tilde{K}^2} -\frac{9\tilde{\omega}}{8\tilde{K}}\left(1-
 \frac{5\tilde{\omega}^2}{3\tilde{K}^2}\right) \ln{\left(\frac{\tilde{\omega}+\tilde{K}}{\tilde{\omega}-\tilde{K}}\right)} \Bigg],\\
I_8 &=& \int_{0}^{-s_B\infty} \frac{d\tau}{\tau^2} e^{-i \tau \tilde{P}_{\lambda}\tilde{\omega} }
 \left(
  \sin(\tilde{P}_{\lambda}\tilde{K}\tau)
  +\frac{3\cos(\tilde{P}_{\lambda}\tilde{K}\tau)}{\tilde{P}_{\lambda}\tilde{K}\tau}
  -\frac{3\sin(\tilde{P}_{\lambda}\tilde{K}\tau)}{(\tilde{P}_{\lambda}\tilde{K}\tau)^2}
  \right) \nonumber\\
 &=&-\frac{\tilde{P}_{\lambda}\tilde{K}}{3}\Bigg\{1-\frac{3\tilde{\omega}^2}{2\tilde{K}^2} -\frac{3\tilde{\omega}}{4\tilde{K}}\left(1-\frac{\tilde{\omega}^2}{\tilde{K}^2}\right) \ln{\left(\frac{\tilde{\omega}+\tilde{K}}{\tilde{\omega}-\tilde{K}}\right)} \Bigg\}, \\
I_9 &=& \int_{0}^{-s_B\infty} \frac{d\tau}{\tau} e^{-i  \tau \tilde{P}_{\lambda}\tilde{\omega} }
 \left(
  \cos(\tilde{P}_{\lambda}\tilde{K}\tau)
  -\frac{6\sin(\tilde{P}_{\lambda}\tilde{K}\tau)}{\tilde{P}_{\lambda}\tilde{K}\tau}
  -\frac{15\cos(\tilde{P}_{\lambda}\tilde{K}\tau)}{(\tilde{P}_{\lambda}\tilde{K}\tau)^2} +\frac{15\sin(\tilde{P}_{\lambda}\tilde{K}\tau)}{(\tilde{P}_{\lambda}\tilde{K}\tau)^3}
  \right) \nonumber\\
 &=&\frac{2}{3}\Bigg\{1-\frac{15\tilde{\omega}^2}{4\tilde{K}^2} -\frac{9\tilde{\omega}}{8\tilde{K}}\left(1-\frac{5\tilde{\omega}^2}{3\tilde{K}^2}\right) \ln{\left(\frac{\tilde{\omega}+\tilde{K}}{\tilde{\omega}-\tilde{K}}\right)} \Bigg\},\\
  I_{10} &=& \int_{0}^{-s_B\infty} \frac{d\tau}{\tau} e^{-i \tau \tilde{P}_{\lambda}\tilde{\omega} }
 \left(
  \sin(\tilde{P}_{\lambda}\tilde{K}\tau)
  +\frac{3\cos(\tilde{P}_{\lambda}\tilde{K}\tau)}{\tilde{P}_{\lambda}\tilde{K}\tau}
  -\frac{3\sin(\tilde{P}_{\lambda}\tilde{K}\tau)}{(\tilde{P}_{\lambda}\tilde{K}\tau)^2}
  \right) \nonumber\\
 &=&  \frac{i}{\tilde{P}_{\lambda}}\frac{\partial I_8}{\partial \tilde{\omega}} =\frac{3i\tilde{\omega}}{2\tilde{K}}\Bigg\{1 -\frac{\tilde{\omega}}{2\tilde{K}}\left(1-\frac{\tilde{K}^2}{3\tilde{\omega}^2}\right) \ln{\left(\frac{\tilde{\omega}+\tilde{K}}{\tilde{\omega}-\tilde{K}}\right)} \Bigg\},\\
 I_{11} &=& \int_{0}^{-s_B\infty} d\tau e^{-i  \tau \tilde{P}_{\lambda}\tilde{\omega} }
 \left(
  \cos(\tilde{P}_{\lambda}\tilde{K}\tau)
  -\frac{6\sin(\tilde{P}_{\lambda}\tilde{K}\tau)}{\tilde{P}_{\lambda}\tilde{K}\tau}
  -\frac{15\cos(\tilde{P}_{\lambda}\tilde{K}\tau)}{(\tilde{P}_{\lambda}\tilde{K}\tau)^2} +\frac{15\sin(\tilde{P}_{\lambda}\tilde{K}\tau)}{(\tilde{P}_{\lambda}\tilde{K}\tau)^3}
  \right) \nonumber\\
 &=&\frac{i}{\tilde{P}_{\lambda}}\frac{\partial I_{9}}{\partial \tilde{\omega}}=i\frac{13\tilde{\omega}}{2\tilde{P}_{\lambda}(\tilde{\omega}^2-\tilde{K}^2)} \Bigg\{1-\frac{15\tilde{\omega}^2}{13\tilde{K}^2} -\frac{9\tilde{\omega}}{13\tilde{K}}\left[1-\frac{5\tilde{\omega}^2}{6\tilde{K}^2}-\frac{\tilde{K}^2}{6\tilde{\omega}^2}\right] \ln{\left(\frac{\tilde{\omega}+\tilde{K}}{\tilde{\omega}-\tilde{K}}\right)} \Bigg\},
\\
 I_{12} &=& \int_{0}^{-s_B\infty} \frac{d\tau}{\tau^3} e^{-i \tau \tilde{P}_{\lambda}\tilde{\omega} }
 \left(
  \sin(\tilde{P}_{\lambda}\tilde{K}\tau)
  +\frac{3\cos(\tilde{P}_{\lambda}\tilde{K}\tau)}{\tilde{P}_{\lambda}\tilde{K}\tau}
  -\frac{3\sin(\tilde{P}_{\lambda}\tilde{K}\tau)}{(\tilde{P}_{\lambda}\tilde{K}\tau)^2}
  \right) \nonumber\\
 &=&i\frac{5\tilde{\omega}\tilde{P}_{\lambda}^2\tilde{K}}{24} \Bigg\{1-\frac{3\tilde{\omega}^2}{5\tilde{K}^2} -\frac{3\tilde{\omega}}{5\tilde{K}}\left(1-\frac{\tilde{\omega}^2}{2\tilde{K}^2}+\frac{\tilde{K}^2}{2\tilde{\omega}^2}\right) \ln{\left(\frac{\tilde{\omega}+\tilde{K}}{\tilde{\omega}-\tilde{K}}\right)} \Bigg\},\\
  I_{13} &=& \int_{0}^{-s_B\infty} \frac{d\tau}{\tau^2} e^{-i  \tau \tilde{P}_{\lambda}\tilde{\omega} }
 \left(
  \cos(\tilde{P}_{\lambda}\tilde{K}\tau)
  -\frac{6\sin(\tilde{P}_{\lambda}\tilde{K}\tau)}{\tilde{P}_{\lambda}\tilde{K}\tau}
  -\frac{15\cos(\tilde{P}_{\lambda}\tilde{K}\tau)}{(\tilde{P}_{\lambda}\tilde{K}\tau)^2} +\frac{15\sin(\tilde{P}_{\lambda}\tilde{K}\tau)}{(\tilde{P}_{\lambda}\tilde{K}\tau)^3}
  \right) \nonumber\\
 &=&-i\frac{13\tilde{\omega}\tilde{P}_{\lambda}}{24} \Bigg\{1-\frac{15\tilde{\omega}^2}{13\tilde{K}^2} -\frac{9\tilde{\omega}}{13\tilde{K}} \left[1-\frac{5\tilde{\omega}^2}{6\tilde{K}^2}-\frac{\tilde{K}^2}{6\tilde{\omega}^2}\right] \ln{\left(\frac{\tilde{\omega}+\tilde{K}}{\tilde{\omega}-\tilde{K}}\right)} \Bigg\},\\
  I_{14} &=& \int_{0}^{-s_B\infty} \frac{d\tau}{\tau} e^{-i  \tau \tilde{P}_{\lambda}\tilde{\omega} }
 \left(
  \sin(\tilde{P}_{\lambda}\tilde{K}\tau)
  +\frac{10\cos(\tilde{P}_{\lambda}\tilde{K}\tau)}{\tilde{P}_{\lambda}\tilde{K}\tau}
  -\frac{45\sin(\tilde{P}_{\lambda}\tilde{K}\tau)}{(\tilde{P}_{\lambda}\tilde{K}\tau)^2} -\frac{105\cos(\tilde{P}_{\lambda}\tilde{K}\tau)}{(\tilde{P}_{\lambda}\tilde{K}\tau)^3}
  +\frac{105\sin(\tilde{P}_{\lambda}\tilde{K}\tau)}{(\tilde{P}_{\lambda}\tilde{K}\tau)^3}
  \right) \nonumber\\
 &=&-i\frac{55\tilde{\omega}}{24\tilde{K}} \Bigg\{1-\frac{21\tilde{\omega}^2}{11\tilde{K}^2} -\frac{9\tilde{\omega}}{11\tilde{K}} \left[1-\frac{7\tilde{\omega}^2}{6\tilde{K}^2}-\frac{\tilde{K}^2}{10\tilde{\omega}^2}\right] \ln{\left(\frac{\tilde{\omega}+\tilde{K}}{\tilde{\omega}-\tilde{K}}\right)} \Bigg\},
\label{App-int-tau-ee}
\end{eqnarray}
where we used the following notations:
\begin{equation}
s_B\equiv\sign{eB_{0,\lambda}}, \quad \tilde{P}_{\lambda} \equiv \frac{cp\Omega_e}{v_F eB_{0,\lambda}}, \quad \tilde{\omega}\equiv
\frac{\omega}{\Omega_e}, \quad \tilde{K}\equiv\frac{v_FK}{\Omega_e}, \quad \Omega_e \equiv \sqrt{\frac{4\alpha}{3\pi \hbar^2}\left(\mu^2+\mu_5^2 +
\frac{1}{3}\pi^2 T^2\right)}.
\end{equation}
Here $\Omega_e$ is the Langmuir (plasma) frequency, $\alpha=e^2/(\hbar v_Fn_0^2)$ denotes the fine structure constant in Weyl
materials, and $K\equiv|\mathbf{K}_{\tau}|/\tau$.

\section{Coefficients $\mathbf{X}_2$ and $\mathbf{X}_3$}
\label{sec:App-X-j}

In this appendix we present the explicit form of the coefficients $\mathbf{X}_2^{\rm (a)}$ through $\mathbf{X}_2^{\rm (d)}$ and
$\mathbf{X}_3^{\rm (a)}$ through $\mathbf{X}_3^{\rm (d)}$ in the limit of small $\mathbf{B}_{0,\lambda}$.
Using expressions (\ref{collective-B-tensor-Ktau}) through (\ref{collective-B-tensor-kKtau}) and formulas from
Appendix~\ref{sec:App-ref}, one can show that coefficients  $\mathbf{X}_2^{\rm (a)}$ through
$\mathbf{X}_2^{\rm (d)}$ equal
\begin{eqnarray}
\mathbf{X}_{2}^{\rm (a)} &=& \frac{ie}{\omega}\frac{cv_F}{B_{0,\lambda}}
 \int_{0}^{-s_B\infty}d\tau \int\frac{p^3 dp d\xi d\phi}{(2\pi \hbar)^3}
 \hat{\mathbf{p}} \left( \hat{\mathbf{p}} \cdot \mathbf{E}_{\tau} \right) \frac{\partial f^{(0)}_{\lambda}}{\partial \epsilon_{\mathbf{p}}} e^{-i p\frac{\Omega_{\tau}- c\hat{\mathbf{p}} \cdot \mathbf{K}_{\tau}}{ eB_{0,\lambda}}}\nonumber\\
&\simeq& \mathbf{E}\frac{e^2 T^2}{4\pi^2\hbar^3v_F^3k^2} \left[1-\frac{\omega}{2v_Fk} \left(1-\frac{v_F^2k^2}{\omega^2}\right) \ln{\left|\frac{\omega+v_Fk}{\omega-v_Fk}\right|} \right] \mathrm{Li}_2\left(-e^{\mu_{\lambda}/T}\right) \nonumber\\
&+& \mathbf{E}(\hat{\mathbf{z}}\cdot\hat{\mathbf{k}}) \frac{3e^3B_{0,\lambda} \omega}{8\pi^2 \hbar^2 c v_F^2 k^3} \left[1-\frac{\omega}{2v_Fk}\left(1-\frac{v_F^2k^2}{3\omega^2}\right)\ln{\left|\frac{\omega+v_Fk}{\omega-v_Fk}\right|} \right] \frac{1}{1+e^{-\mu_{\lambda}/T}}  \nonumber\\
&-& i[\hat{\mathbf{z}}\times\mathbf{E}] \frac{e^3B_{0,\lambda} T}{2\pi^2 \hbar^3 cv_F k^2 \omega} \left[1-\frac{\omega}{2v_Fk}\ln{\left|\frac{\omega+v_Fk}{\omega-v_Fk}\right|}\right] \ln{\left(1+e^{\mu_{\lambda}/T}\right)} \nonumber\\
&-&\hat{\mathbf{k}}(\mathbf{E}\cdot\hat{\mathbf{k}}) \frac{3e^2 T^2}{4\pi^2 \hbar^3v_F^3k^2} \left[1-\frac{\omega}{2v_Fk}\left(1-\frac{v_F^2k^2}{3\omega^2}\right)\ln{\left|\frac{\omega+v_Fk}{\omega-v_Fk}\right|} \right] \mathrm{Li}_2\left(-e^{\mu_{\lambda}/T}\right)  \nonumber\\
&+&\hat{\mathbf{k}}(\mathbf{E}\cdot\hat{\mathbf{k}})(\hat{\mathbf{k}}\cdot\hat{\mathbf{z}}) \frac{15e^3B_{0,\lambda} \omega^3}{8\pi^2\hbar^2c v_F^2 k^3(v_F^2k^2-\omega^2)} \left[1-\frac{13v_F^2k^2}{15\omega^2} -\frac{\omega}{2v_Fk}\left(1-\frac{6v_F^2k^2}{5\omega^2} +\frac{v_F^4k^4}{5\omega^4}\right)\ln{\left|\frac{\omega+v_Fk}{\omega-v_Fk}\right|} \right] \frac{1}{1+e^{-\mu_{\lambda}/T}} \nonumber\\
&+&\hat{\mathbf{k}}(\mathbf{E}\cdot\hat{\mathbf{z}}) \frac{3e^3B_{0,\lambda} \omega}{8\pi^2\hbar^2cv_F^2k^3} \left[1-\frac{\omega}{2v_Fk}\left(1-\frac{v_F^2k^2}{3\omega^2}\right)\ln{\left|\frac{\omega+v_Fk}{\omega-v_Fk}\right|} \right] \frac{1}{1+e^{-\mu_{\lambda}/T}} \nonumber\\
&+&\hat{\mathbf{z}}(\mathbf{E}\cdot\hat{\mathbf{k}}) \frac{3e^3B_{0,\lambda} \omega}{8\pi^2\hbar^2cv_F^2k^3} \left[1-\frac{\omega}{2v_Fk}\left(1-\frac{v_F^2k^2}{3\omega^2}\right)\ln{\left|\frac{\omega+v_Fk}{\omega-v_Fk}\right|} \right] \frac{1}{1+e^{-\mu_{\lambda}/T}} +O(B_{0,\lambda}^2),
\label{collective-B-X2-a}
\end{eqnarray}
\begin{eqnarray}
\mathbf{X}_{2}^{\rm (b)}&=& \frac{ie^2}{\omega}\frac{\lambda \hbar v_F}{B_{0,\lambda}}
 \int_{0}^{-s_B\infty}d\tau \int\frac{p dp d\xi d\phi}{(2\pi \hbar)^3}
 \hat{\mathbf{p}} \left( \hat{\mathbf{p}} \cdot \mathbf{E}_{\tau} \right)
(\mathbf{B}_{0,\lambda}\cdot\hat{\mathbf{p}}) e^{-i p\frac{\Omega_{\tau}- c\hat{\mathbf{p}} \cdot \mathbf{K}_{\tau}}{ eB_{0,\lambda}}} \left[\frac{\partial f^{(0)}_{\lambda}}{\partial \epsilon_{\mathbf{p}}} -\frac{v_F p }{2} \frac{\partial^2 f^{(0)}_{\lambda}}{\partial \epsilon_{\mathbf{p}}^2} \right]\nonumber\\
&\simeq& -\frac{\lambda e^3B_{0,\lambda}}{4\pi^2\hbar^2 ck \omega} \frac{1}{1+e^{-\mu_{\lambda}/T}} \Bigg\{- \left[\hat{\mathbf{k}}(\mathbf{E}\cdot\hat{\mathbf{z}}) +\mathbf{E}(\hat{\mathbf{z}}\cdot\hat{\mathbf{k}}) +\hat{\mathbf{z}}(\hat{\mathbf{k}}\cdot\mathbf{E})\right] \left[1-\frac{3\omega^2}{2v_F^2k^2} -\frac{3\omega}{4v_Fk}\left(1-\frac{\omega^2}{v_F^2k^2}\right) \ln{\left|\frac{\omega+v_Fk}{\omega-v_Fk}\right|} \right] \nonumber\\ &+&2\hat{\mathbf{k}}(\mathbf{k}\cdot\hat{\mathbf{z}})(\mathbf{k}\cdot\mathbf{E}) \left[1-\frac{15\omega^2}{4v_F^2k^2} -\frac{9\omega}{8v_Fk}\left(1-\frac{5\omega^2}{3v_F^2k^2}\right) \ln{\left|\frac{\omega+v_Fk}{\omega-v_Fk}\right|} \right]\Bigg\}
+O(B_{0,\lambda}^2),
\label{collective-B-X2-b}
\end{eqnarray}
\begin{eqnarray}
\mathbf{X}_{2}^{\rm (c)}&=& -\frac{e}{\omega}\int\frac{p^2 dp d\xi d\phi}{(2\pi \hbar)^3} \int_{0}^{-s_B\infty}d\tau
 v_F\hat{\mathbf{p}} \frac{\lambda \hbar v_F c}{2\omega B_{0,\lambda}} \left(\hat{\mathbf{p}}\cdot [\mathbf{K}_{\tau}^{(1)}\times \mathbf{E}_{\tau}]\right) (\hat{\mathbf{p}}\cdot\mathbf{K}_{\tau}^{(1)})
\frac{\partial f^{(0)}_{\lambda}}{\partial \epsilon_{\mathbf{p}}} e^{-i p\frac{\Omega_{\tau}- c\hat{\mathbf{p}} \cdot \mathbf{K}_{\tau}}{ eB_{0,\lambda}}}\nonumber\\
&\simeq& -i\frac{\lambda v_F^2 e^2}{4\pi^2\hbar^2 \omega^2 c k^2} \Bigg\{ -\frac{c}{3} \left[1-\frac{3\omega^2}{2v_F^2k^2} -\frac{3\omega}{4v_Fk}\left(1-\frac{\omega^2}{v_F^2k^2}\right) \ln{\left|\frac{\omega+v_Fk}{\omega-v_Fk}\right|} \right] \nonumber\\
&\times& \left[k(\hat{\mathbf{k}}\times\mathbf{E}) \frac{T}{v_F^2}\ln{\left(1+e^{\mu_{\lambda}/T}\right)} -\frac{\lambda \hbar \omega eB_{0,\lambda}}{2cv_F}\left[(\hat{\mathbf{k}}\cdot\hat{\mathbf{z}})(\hat{\mathbf{k}}\times\mathbf{E}) +\hat{\mathbf{k}}(\hat{\mathbf{z}}\cdot[\mathbf{k}\times\mathbf{E}])\right] \int \frac{dp}{p} \frac{\partial f^{(0)}_{\lambda}}{\partial \epsilon_{\mathbf{p}}} \right]  \nonumber\\
&-&(\hat{\mathbf{k}}\cdot\hat{\mathbf{z}})(\hat{\mathbf{k}}\times\mathbf{E}) \frac{\lambda eB_{0,\lambda}\hbar \omega}{3v_F} \left[1-\frac{15\omega^2}{4v_F^2k^2} -\frac{9\omega}{8v_Fk}\left(1-\frac{5\omega^2}{3v_F^2k^2}\right) \ln{\left|\frac{\omega+v_Fk}{\omega-v_Fk}\right|} \right] \int \frac{dp}{p} \frac{\partial f^{(0)}_{\lambda}}{\partial \epsilon_{\mathbf{p}}} \nonumber\\
&+& \frac{3ieB_{0,\lambda}\omega}{4v_F^2k}\left[1 +\frac{v_Fk}{6\omega}\left(1-\frac{3\omega^2}{v_F^2k^2}\right) \ln{\left|\frac{\omega+v_Fk}{\omega-v_Fk}\right|} \right]
\left[3\hat{\mathbf{k}}(\hat{\mathbf{z}}\cdot\mathbf{E}) -2\mathbf{E}(\hat{\mathbf{z}}\cdot\hat{\mathbf{k}}) -\hat{\mathbf{k}}(\hat{\mathbf{k}}\cdot\mathbf{E})(\hat{\mathbf{z}}\cdot\hat{\mathbf{k}})\right]\frac{1}{1+e^{-\mu_{\lambda}/T}} \nonumber\\
&-&\frac{\lambda \hbar \omega eB_{0,\lambda}}{3v_F}\left[1-\frac{15\omega^2}{4v_F^2k^2} -\frac{9\omega}{8v_Fk}\left(1-\frac{5\omega^2}{3v_F^2k^2}\right) \ln{\left|\frac{\omega+v_Fk}{\omega-v_Fk}\right|} \right]  \hat{\mathbf{k}}(\hat{\mathbf{z}}\cdot[\hat{\mathbf{k}}\times\mathbf{E}]) \int \frac{dp}{p} \frac{\partial f^{(0)}_{\lambda}}{\partial \epsilon_{\mathbf{p}}}\nonumber\\
&+&i\frac{13\omega k eB_{0,\lambda}}{4(\omega^2-v_F^2k^2)} \left[1-\frac{15\omega^2}{13v_F^2k^2} -\frac{9\omega}{13v_Fk}\left(1-\frac{5\omega^2}{6v_F^2k^2}-\frac{v_F^2k^2}{6\omega^2}\right) \ln{\left|\frac{\omega+v_Fk}{\omega-v_Fk}\right|} \right] \frac{\hat{\mathbf{k}} \left[(\hat{\mathbf{z}}\cdot\mathbf{E})-(\hat{\mathbf{k}}\cdot\mathbf{E})(\hat{\mathbf{z}}\cdot\hat{\mathbf{k}})\right]}{ 1+e^{-\mu_{\lambda}/T}}\Bigg\}\nonumber\\
&+&O(B_{0,\lambda}^2),
\label{collective-B-X2-c}
\end{eqnarray}
and
\begin{eqnarray}
\mathbf{X}_{2}^{\rm (d)}&=& -\frac{v_F^2e^2}{16\pi^3\hbar\omega^2 B_{0,\lambda}} \int dp d\xi d\phi \int_{0}^{-s_B\infty}d\tau \hat{\mathbf{p}} (\hat{\mathbf{p}}\cdot\mathbf{B}_{0,\lambda}) \left(\hat{\mathbf{p}}\cdot [\mathbf{K}_{\tau}^{(1)}\times \mathbf{E}_{\tau}]\right) (\hat{\mathbf{p}}\cdot\mathbf{K}_{\tau}^{(1)}) \left[ \frac{\partial f^{(0)}_{\lambda}}{\partial \epsilon_{\mathbf{p}}}  -\frac{v_F p }{2} \frac{\partial^2 f^{(0)}_{\lambda}}{\partial \epsilon_{\mathbf{p}}^2} \right]\nonumber\\
&\times&
e^{-i p\frac{\Omega_{\tau}- c\hat{\mathbf{p}} \cdot \mathbf{K}_{\tau}}{ eB_{0,\lambda}}}
\simeq i\frac{3v_F e^3B_{0,\lambda}}{16\pi^2 \hbar c \omega} \Big[(\hat{\mathbf{z}}\cdot\hat{\mathbf{k}})(\hat{\mathbf{k}}\times \mathbf{E}) +\hat{\mathbf{k}}(\hat{\mathbf{z}}\cdot[\hat{\mathbf{k}}\times\mathbf{E}])\Big] \int\frac{dp}{p} \Bigg[\frac{\partial f^{(0)}_{\lambda}}{\partial \epsilon_{\mathbf{p}}} -\frac{v_F p }{2} \frac{\partial^2 f^{(0)}_{\lambda}}{\partial \epsilon_{\mathbf{p}}^2} \Bigg] \nonumber\\
&\times&\left[1-\frac{\omega^2}{v_F^2k^2} -\frac{2\omega}{3v_Fk} \left(1-\frac{3\omega^2}{4v_F^2k^2}\right) \ln{\left|\frac{\omega+v_Fk}{\omega-v_Fk}\right|}\right] +O(B_{0,\lambda}^2).
\label{collective-B-X2-d}
\end{eqnarray}
Here, $\hat{\mathbf{z}}=\mathbf{B}_{0,\lambda}/B_{0,\lambda}$, $B_{0,\lambda}=B_{0}+\lambda B_{0,5}$, and $\mu_{\lambda}=\mu+\lambda \mu_{5}$. Further, we present the coefficients $\mathbf{X}_{3}^{\rm (a)}$ through $\mathbf{X}_{3}^{\rm (d)}$, i.e.,
\begin{eqnarray}
\mathbf{X}_{3}^{\rm (a)}&=&-e\frac{\lambda\hbar v_F}{\omega}\int\frac{d^3p}{(2\pi \hbar)^3} \frac{(\mathbf{k}\times\hat{\mathbf{p}})}{2p} \int_{0}^{-s_B\infty}d\tau \frac{c (\mathbf{p} \cdot \mathbf{E}_{\tau})}{B_{0,\lambda}} \frac{\partial f^{(0)}_{\lambda}}{\partial \epsilon_{\mathbf{p}}}
e^{-i p\frac{\Omega_{\tau}- c\hat{\mathbf{p}} \cdot \mathbf{K}_{\tau}}{ eB_{0,\lambda}}} \nonumber\\
&\simeq&-i(\hat{\mathbf{k}}\times\mathbf{E}) \frac{\lambda e^2 T}{8\pi^2\hbar^2 v_F^2k} \left[1-\frac{\omega}{2v_Fk}\left(1-\frac{v_F^2k^2}{\omega^2}\right)\ln{\left|\frac{\omega+v_Fk}{\omega-v_Fk}\right|} \right] \ln{\left(1+e^{\mu_{\lambda}/T}\right)} \nonumber\\
&-&i(\hat{\mathbf{k}}\times\mathbf{E})(\hat{\mathbf{z}}\cdot\hat{\mathbf{k}}) \frac{3 \omega e^3 B_{0,\lambda}}{16\pi^2\hbar cv_F k^2} \left[1-\frac{\omega}{2v_Fk}\left(1-\frac{v_F^2k^2}{3\omega^2}\right)\ln{\left|\frac{\omega+v_Fk}{\omega-v_Fk}\right|} \right] \int \frac{dp}{p} \frac{\partial f_{\lambda}^{(0)}}{\partial \epsilon_{\mathbf{p}}} \nonumber\\
&+&\hat{\mathbf{z}}(\hat{\mathbf{k}}\cdot\mathbf{E}) \frac{3 e^3\lambda B_{0,\lambda}}{8\pi^2\hbar^2 c k \omega} \left[-\frac{2v_F^2k^2-3\omega^2}{3(v_F^2k^2-\omega^2)} +\frac{\omega}{2v_Fk}\ln{\left|\frac{\omega+v_Fk}{\omega-v_Fk}\right|} \right] \frac{1}{1+e^{-\mu_{\lambda}/T}} \nonumber\\
&-&\hat{\mathbf{k}}(\hat{\mathbf{k}}\cdot\mathbf{E})(\hat{\mathbf{z}}\cdot\hat{\mathbf{k}}) \frac{3 e^3\lambda B_{0,\lambda}}{8\pi^2\hbar^2 c k \omega} \left[-\frac{2v_F^2k^2-3\omega^2}{3(v_F^2k^2-\omega^2)} +\frac{\omega}{2v_Fk}\ln{\left|\frac{\omega+v_Fk}{\omega-v_Fk}\right|} \right] \frac{1}{1+e^{-\mu_{\lambda}/T}} \nonumber\\
&-&i(\hat{\mathbf{k}}\times \hat{\mathbf{z}})(\mathbf{E}\cdot\hat{\mathbf{k}}) \frac{3\omega e^3B_{0,\lambda}}{16\pi^2\hbar c v_F k^2} \left[1-\frac{\omega}{2v_Fk}\left(1-\frac{v_F^2k^2}{3\omega^2}\right)\ln{\left|\frac{\omega+v_Fk}{\omega-v_Fk}\right|} \right] \int \frac{dp}{p} \frac{\partial f_{\lambda}^{(0)}}{\partial \epsilon_{\mathbf{p}}} +O(B_{0,\lambda}^2),
\label{collective-B-magnetization-a}
\end{eqnarray}
\begin{eqnarray}
\mathbf{X}_{3}^{\rm (b)}&=&\frac{e^2\hbar^2 v_F}{2c\omega}\int\frac{d^3p}{(2\pi \hbar)^3} \frac{(\mathbf{k}\times\hat{\mathbf{p}})}{2p^2} \int_{0}^{-s_B\infty}d\tau \frac{c (\hat{\mathbf{p}} \cdot \mathbf{E}_{\tau})(\hat{\mathbf{p}}\cdot\mathbf{B}_{0,\lambda})}{B_{0,\lambda}} \Bigg[\frac{\partial f^{(0)}_{\lambda}}{\partial \epsilon_{\mathbf{p}}} +v_F p \frac{\partial^2 f^{(0)}_{\lambda}}{\partial \epsilon_{\mathbf{p}}^2} \Bigg]e^{-i p\frac{\Omega_{\tau}- c\hat{\mathbf{p}} \cdot \mathbf{K}_{\tau}}{B_{0,\lambda}}} \nonumber\\
&\simeq& i\frac{v_Fe^3B_{0,\lambda}}{24\pi^2\hbar c \omega} \int\frac{dp}{p} \left[\frac{\partial f^{(0)}_{\lambda}}{\partial \epsilon_{\mathbf{p}}} +v_Fp \frac{\partial^2 f^{(0)}_{\lambda}}{\partial \epsilon_{\mathbf{p}}^2}\right]
\left[1-\frac{3\omega^2}{2v_F^2k^2} -\frac{3\omega}{4v_Fk}\left(1-\frac{\omega^2}{v_F^2k^2}\right) \ln{\left|\frac{\omega+v_Fk}{\omega-v_Fk}\right|} \right] \nonumber\\
&\times&\left[(\hat{\mathbf{k}}\times\mathbf{E})(\hat{\mathbf{k}}\cdot\hat{\mathbf{z}}) +(\hat{\mathbf{k}}\times\hat{\mathbf{z}})(\hat{\mathbf{k}}\cdot\mathbf{E})\right]+O(B_{0,\lambda}^2),
\label{collective-B-magnetization-b}
\end{eqnarray}
\begin{eqnarray}
\mathbf{X}_{3}^{\rm (c)}&=&-i\frac{e\hbar^2 v_F^2c}{4\omega^2 B_{0,\lambda}}\int\frac{d^3p}{(2\pi \hbar)^3} \frac{(\mathbf{k}\times\hat{\mathbf{p}})}{p} \int_{0}^{-s_B\infty}d\tau  \frac{\partial f^{(0)}_{\lambda}}{\partial \epsilon_{\mathbf{p}}} \left(\hat{\mathbf{p}}\cdot [\mathbf{K}_{\tau}^{(1)}\times \mathbf{E}_{\tau}]\right) (\hat{\mathbf{p}}\cdot\mathbf{K}_{\tau}^{(1)}) e^{-i p\frac{\Omega_{\tau}- c\hat{\mathbf{p}} \cdot \mathbf{K}_{\tau}}{ eB_{0,\lambda}}} \nonumber\\
&\simeq&\frac{e^2v_F^2}{8\pi^2\hbar \omega^2 c} \Bigg\{
-\frac{c k^2}{3v_F} \left[1-\frac{3\omega^2}{2v_F^2k^2} -\frac{3\omega}{4v_Fk}\left(1-\frac{\omega^2}{v_F^2k^2}\right) \ln{\left|\frac{\omega+v_Fk}{\omega-v_Fk}\right|} \right] \left[\hat{\mathbf{k}}(\hat{\mathbf{k}}\cdot\mathbf{E})-\mathbf{E}\right] \frac{1}{1+e^{-\mu_{\lambda}/T}} \nonumber\\
&-&\frac{\lambda \hbar eB_{0,\lambda}\omega k}{6v_F} (\hat{\mathbf{k}}\cdot\hat{\mathbf{z}}) \left[\hat{\mathbf{k}}(\hat{\mathbf{k}}\cdot\mathbf{E})-\mathbf{E}\right] \left[1-\frac{6\omega^2}{v_F^2k^2} -\frac{3\omega}{2v_Fk}\left(1-\frac{2\omega^2}{v_F^2k^2}\right) \ln{\left|\frac{\omega+v_Fk}{\omega-v_Fk}\right|} \right] \int \frac{dp}{p^2} \frac{\partial f^{(0)}_{\lambda}}{\partial \epsilon_{\mathbf{p}}} \nonumber\\
&+& \frac{3i\omega eB_{0,\lambda}}{2v_F}\left[1 +\frac{v_Fk}{6\omega}\left(1-\frac{3\omega^2}{v_F^2k^2}\right) \ln{\left|\frac{\omega+v_Fk}{\omega-v_Fk}\right|} \right] (\hat{\mathbf{k}}\times\mathbf{E})(\hat{\mathbf{k}}\cdot\hat{\mathbf{z}}) \int \frac{dp}{p} \frac{\partial f^{(0)}_{\lambda}}{\partial \epsilon_{\mathbf{p}}} \Bigg\} +O(B_{0,\lambda}^2),
\label{collective-B-magnetization-c}
\end{eqnarray}
\begin{eqnarray}
\mathbf{X}_{3}^{\rm (d)}&=&i\frac{\lambda\hbar^3 e^2 v_F^2}{16\omega^2 B_{0,\lambda}} \int\frac{d^3p}{(2\pi \hbar)^3} \int_{0}^{-s_B\infty}d\tau \left[\frac{\partial f^{(0)}_{\lambda}}{\partial \epsilon_{\mathbf{p}}} +v_F p\frac{\partial^2 f^{(eq)}_{\lambda}}{\partial \epsilon_{\mathbf{p}}^2} \right] \frac{(\mathbf{k}\times\hat{\mathbf{p}})}{p^3} \left(\hat{\mathbf{p}}\cdot [\mathbf{K}_{\tau}^{(1)}\times \mathbf{E}_{\tau}]\right) (\hat{\mathbf{p}}\cdot\mathbf{K}_{\tau}^{(1)}) (\hat{\mathbf{p}}\cdot\mathbf{B}_{0,\lambda})\nonumber\\
&\times&e^{-i p\frac{\Omega_{\tau}- c\hat{\mathbf{p}} \cdot \mathbf{K}_{\tau}}{ eB_{0,\lambda}}} = \frac{3\lambda e^3 B_{0,\lambda}v_F k}{64\pi^2c\omega} (\hat{\mathbf{z}}\cdot\hat{\mathbf{k}})(\hat{\mathbf{k}}(\hat{\mathbf{k}}\cdot\mathbf{E})-\mathbf{E}) \int\frac{dp}{p^2} \Bigg[\frac{\partial f^{(0)}_{\lambda}}{\partial \epsilon_{\mathbf{p}}} +v_F p\frac{\partial^2 f^{(0)}_{\lambda}}{\partial \epsilon_{\mathbf{p}}^2} \Bigg] \nonumber\\
&\times&
\left[1-\frac{\omega^2}{v_F^2k^2} -\frac{2\omega}{3v_Fk}\left(1-\frac{3\omega^2}{4v_F^2k^2}\right) \ln{\left|\frac{\omega+v_Fk}{\omega-v_Fk}\right|}\right] +O(B_{0,\lambda}^2).
\label{collective-B-magnetization-d}
\end{eqnarray} 

\section{Coefficients $A_i$}
\label{sec:App-Ai}

In this appendix, we present the coefficients $A_i$, where $i=\overline{0,11}$, up to the linear order in $B_{0}$ and $B_{0,5}$. They are
\begin{eqnarray}
\label{collective-B-tensor-A0}
A_0&\simeq& -\frac{3n_0^2\Omega_e^2}{8\pi v_F^2k^2} \left[1-\frac{\omega}{2v_Fk}\left(1-\frac{v_F^2k^2}{\omega^2}\right) \ln{\left|\frac{\omega+v_Fk}{\omega-v_Fk}\right|}\right]
+\frac{e^2v_F k^2}{12\pi^2\hbar \omega^2}  \left[1-\frac{3\omega^2}{2v_F^2k^2} -\frac{3\omega}{4v_Fk}\left(1-\frac{\omega^2}{v_F^2k^2}\right) \ln{\left|\frac{\omega+v_Fk}{\omega-v_Fk}\right|} \right] \nonumber\\
&+&O(B_{0}^2, B_{0,5}^2),\\
\label{collective-B-tensor-A1}
A_1&\simeq&
i\frac{v_Fe^3}{24\pi^2\hbar c\omega T} \sum_{\lambda=\pm}B_{0, \lambda} F \left(\frac{\mu_{\lambda}}{T}\right)
+i\frac{e^3 (\mu B_{0}+\mu_5B_{0,5})}{\pi^2\hbar^3 v_Fc\omega k^2} \left[1-\frac{\omega}{2v_Fk}\ln{\left|\frac{\omega+v_Fk}{\omega-v_Fk}\right|}\right]
+O(B_{0}^2, B_{0,5}^2),\\
\label{collective-B-tensor-A2}
A_2&\simeq&-i\frac{2\mu_{5} ke^2}{3\pi^2\hbar^2\omega^2} \left[1-\frac{3\omega^2}{4v_F^2k^2} -\frac{3\omega}{8v_Fk}\left(1-\frac{\omega^2}{v_F^2k^2}\right) \ln{\left|\frac{\omega+v_Fk}{\omega-v_Fk}\right|}\right] -i \frac{e^3 b_0 k}{2\pi^2\omega^2 \hbar^2}+O(B_{0}^2,B_{0,5}^2),\\
\label{collective-B-tensor-A3}
A_3 &\simeq&\frac{9n_0^2\Omega_e^2}{8\pi v_F^2k^2} \left[1-\frac{\omega}{2v_Fk}\left(1-\frac{v_F^2k^2}{3\omega^2}\right) \ln{\left|\frac{\omega+v_Fk}{\omega-v_Fk}\right|}\right]
-\frac{e^2v_F k^2}{12\pi^2\hbar \omega^2}  \left[1-\frac{3\omega^2}{2v_F^2k^2} -\frac{3\omega}{4v_Fk}\left(1-\frac{\omega^2}{v_F^2k^2}\right) \ln{\left|\frac{\omega+v_Fk}{\omega-v_Fk}\right|} \right] \nonumber\\
&+&O(B_0^2, B_{0,5}^2),\\
\label{collective-B-tensor-A4}
A_4&\simeq& -\frac{\omega e^3B_{0,5}}{4\pi^2\hbar^2 ck(\omega^2-v_F^2k^2)}\left[1+ \frac{v_Fk}{2\omega}\left(1-\frac{\omega^2}{v_F^2k^2}\right) \ln{\left|\frac{\omega+v_Fk}{\omega-v_Fk}\right|}\right] +O(B_0^2, B_{0,5}^2),\\
\label{collective-B-tensor-A5}
A_5&\simeq&-\frac{e^3B_{0,5} v_F^2k}{2\pi^2 \hbar^2c\omega(\omega^2 -v_F^2k^2)} \left[1-\frac{\omega^2}{2v_F^2k^2}  -\frac{\omega}{4v_Fk}\left(1-\frac{\omega^2}{v_F^2k^2}\right) \ln{\left|\frac{\omega+v_Fk}{\omega-v_Fk}\right|} \right] +O(B_0^2, B_{0,5}^2),\\
\label{collective-B-tensor-A6}
A_6&\simeq& -\frac{3e^3B_{0,5}}{4\pi^2\hbar^2c\omega k} \left[1 -\frac{\omega}{2v_Fk}\left(1-\frac{v_F^2k^2}{3\omega^2}\right) \ln{\left|\frac{\omega+v_Fk}{\omega-v_Fk}\right|}\right]
+ \frac{5e^2v_Fk}{384C\pi^2 \omega T \sqrt{\hbar c}} \left[1+\frac{3\omega^2}{v_F^2k^2} -\frac{3\omega^3}{2v_F^3k^3} \ln{\left|\frac{\omega+v_Fk}{\omega-v_Fk}\right|}\right] \nonumber\\
&\times&\sum_{\lambda=\pm}\lambda\frac{\sign{eB_{0,\lambda}} \sqrt{|eB_{0,\lambda}|}}{\cosh^2{\left(\frac{\mu_{\lambda}}{2T}\right)}}
-\frac{7e^3v_F^2k}{96\pi^2 \omega c T^2} \left[1+\frac{3\omega^2}{7v_F^2k^2} -\frac{3\omega}{7v_Fk}\left(1+\frac{\omega^2}{2v_F^2k^2}\right) \ln{\left|\frac{\omega+v_Fk}{\omega-v_Fk}\right|}\right] \nonumber\\
&\times&\sum_{\lambda=\pm}\lambda B_{0, \lambda} T\frac{\partial F\left(\mu_{\lambda}/T\right)}{\partial \mu_{\lambda}} +O(B_0^2, B_{0,5}^2),\\
\label{collective-B-tensor-A7}
A_7&\simeq&i\frac{2v_Fe^3(\mu B_{0}+\mu_5B_{0,5})}{\pi^2 c \hbar^3 \omega (\omega^2-v_F^2k^2)} \left[1-\frac{3\omega^2}{2v_F^2k^2}\right] +i\frac{3e^3(\mu B_{0}+\mu_5B_{0,5})}{2\pi^2 c v_F^2\hbar^3 k^3} \ln{\left|\frac{\omega+v_Fk}{\omega-v_Fk}\right|} \nonumber\\
&-&i\frac{e^3v_F}{24\pi^2c \hbar \omega T} \left[1-\frac{6\omega^2}{v_F^2k^2} -\frac{3\omega}{2v_Fk}\left(1-\frac{2\omega^2}{v_F^2k^2}\right) \ln{\left|\frac{\omega+v_Fk}{\omega-v_Fk}\right|}\right] \sum_{\lambda=\pm} B_{0, \lambda} F \left(\frac{\mu_{\lambda}}{T}\right)
+O(B_0^2, B_{0,5}^2),
 \end{eqnarray}
\begin{eqnarray}
\label{collective-B-tensor-A8}
A_8&\simeq&i\frac{11v_Fe^3}{48\pi^2\hbar \omega c T}\left[1-\frac{21\omega^2}{11v_F^2k^2} -\frac{9\omega}{11v_Fk}\left(1-\frac{7 \omega^2}{6 v_F^2k^2}\right)\ln{\left|\frac{\omega+v_Fk}{\omega-v_Fk}\right|}\right] \sum_{\lambda=\pm} B_{0, \lambda} F \left(\frac{\mu_{\lambda}}{T}\right) +O(B_0^2, B_{0,5}^2),\\
\label{collective-B-tensor-A9}
A_{9}&\simeq&  -A_6 +O(B_{0}^2,B_{0,5}^2),\\
\label{collective-B-tensor-A10}
A_{10}&\simeq& -i \frac{7v_Fe^3}{48\pi^2\hbar c\omega T} \left[1-\frac{15\omega^2}{7v_F^2k^2} -\frac{6\omega}{7v_Fk}\left(1 -\frac{5\omega^2}{4v_F^2k^2}\right) \ln{\left|\frac{\omega+v_Fk}{\omega-v_Fk}\right|}\right] \sum_{\lambda=\pm} B_{0, \lambda} F \left(\frac{\mu_{\lambda}}{T}\right)
+O(B_0^2, B_{0,5}^2), \\
\label{collective-B-tensor-A11}
A_{11}&=&-i\frac{e^3}{2\pi^2\omega c \hbar^2},
\end{eqnarray}
where the function $F(x)$ is defined in Eq.~(\ref{collective-B-F1-def}) and regularization (\ref{collective-B-tensor-Lambda-IR}) was used.

Finally, let us consider characteristic equation (\ref{collective-B-tensor-dispersion-relation-general}) when
the wave vector $\mathbf{k}$ has an arbitrary orientation with respect to the effective magnetic field $\mathbf{B}_{0,\lambda}$.
Without loss of generality, it is convenient to use the parametrization $\mathbf{k} = (k\sin{\theta}, 0, k\cos{\theta})$, where $\theta$
is the angle between vectors $\mathbf{B}_{0,\lambda}$ and $\mathbf{k}$. Then, Eq.~(\ref{collective-B-tensor-dispersion-relation-general})
takes the following form:
\begin{eqnarray}
&&\left\{4 \pi \omega^2 \left[\sin{\theta} \cos{\theta} \left(A_3+A_9 \cos{\theta}\right)+\sin{\theta} A_5-A_{11} b_y\right]+c^2 k^2 \sin{\theta} \cos{\theta}\right\}
\Big\{16 \pi^2 \omega^4 \left[\sin{\theta} A_2-\left(A_{10}+A_7\right) \sin{\theta}\cos{\theta}-A_{11} b_x\right] \nonumber\\
&&\times\left(A_1- A_{10} \cos^2{\theta}-A_{11} b_{\parallel}+A_{2} \cos{\theta}-A_8 \sin^2{\theta}\right)
-\left[4\pi\omega^2 (A_0+A_6 \cos{\theta})-c^2 k^2+ n_0^2 \omega^2\right] \nonumber\\
&&\times\left[4\pi\omega^2 \left(A_{11} b_y+\sin{\theta}\cos{\theta} (A_3+A_9\cos{\theta})+\sin{\theta}A_4\right)+c^2 k^2 \sin{\theta} \cos{\theta}\right]\Big\} \nonumber\\
&&+\left\{4\pi\omega^2 \left[A_0+\cos{\theta} \left(A_3 \cos{\theta}+A_4+A_5+A_6+A_9 \cos^2{\theta}\right)\right] -c^2k^2 \sin^2{\theta}+ n_0^2 \omega^2\right\} \nonumber\\
&&\times\Big\{\left[4\pi\omega^2 (A_0+A_6 \cos{\theta})-c^2 k^2+ n_0^2 \omega^2\right] \left[4\pi\omega^2 \left(A_0+A_3 \sin^2{\theta}+\cos {\theta} \left(A_6+A_9 \sin^2{\theta}\right)\right) -c^2 k^2 \cos^2{\theta}+ n_0^2 \omega^2\right] \nonumber\\
&&+16 \pi^2 \omega^4 \left[A_1-A_{10} \cos^2{\theta}-A_{11}b_{\parallel}+A_2\cos{\theta} +A_7 \sin^2{\theta}\right]
\left[A_1-A_{10} \cos^2{\theta}-A_{11}b_{\parallel}+A_2 \cos{\theta}-A_8 \sin^2{\theta}\right]\Big\} \nonumber\\
&&-16 \pi^2 \omega^4 \left[A_{11}b_x- (A_8-A_{10}) \sin{\theta}\cos{\theta}-A_2\sin{\theta}\right]
\Big\{\left[A_2\sin{\theta}-(A_{10}+A_7) \cos{\theta}\sin{\theta}-A_{11}b_x\right] \nonumber\\
&&\times\left[4\pi\omega^2 \left(A_0+A_3 \sin^2{\theta }+\cos{\theta} \left(A_6+A_9 \sin^2{\theta}\right)\right)-c^2 k^2 \cos^2{\theta}+ n_0^2 \omega^2\right]
+\left[A_1-A_{10} \cos^2{\theta }-A_{11}b_{\parallel}+A_2 \cos{\theta}+A_7 \sin^2{\theta}\right] \nonumber\\
&&\times\left[4\pi\omega^2 \left(A_{11}b_y+\sin{\theta}\cos{\theta} (A_3+A_9 \cos{\theta})+A_4\sin{\theta}\right)+c^2 k^2 \sin{\theta} \cos{\theta}\right]\Big\}=0,
\label{collective-B-tensor-dispersion-relation-theta}
\end{eqnarray}
where $b_{\parallel}\equiv b_z$.

\end{document}